\documentclass[12pt, onecolumn,draftcls]{IEEEtran}
\usepackage{amsmath}
\usepackage{amsfonts}
\usepackage{amssymb}
\usepackage{wrapfig}
\usepackage{psfrag}
\usepackage[ ] {graphicx}
\usepackage{subfigure}
\usepackage{cite}
\usepackage{color}
\input{TNDs_IEEEtran_v16_mods}
\newcommand{\paperstatus}{Submitted to \textit{IEEE Trans. on Information Theory}}

\newtheorem{theorem}{Theorem}
\newtheorem{lemma}{Lemma}

\newtheorem{problem}{Problem}
\newtheorem{example}{Example}

\newtheorem{proposition}{Proposition}

\newtheorem{definition}{Definition}
\newtheorem{property}{Property}

\makeatletter
\newcommand{\smatlabaxislabel}[1]{\fontsize{12}{\f@baselineskip}%
\textsf{#1}}
\newcommand{\matlabaxislabel}[1]{\fontsize{14.4}{\f@baselineskip}%
\textsf{#1}}
\newcommand{\mmatlabaxislabel}[1]{\fontsize{17.28}{\f@baselineskip}%
\textsf{#1}}
\newcommand{\bmatlabaxislabel}[1]{\fontsize{20.74}{\f@baselineskip}%
\textsf{#1}}
\newcommand{\bbmatlabaxislabel}[1]{\fontsize{24.88}{\f@baselineskip}%
\textsf{#1}} \makeatother

\allowdisplaybreaks

\centerfigcaptionstrue

\newcommand{\half}%
{\raisebox{2.5pt}{\scriptsize 1}{\small
/}\raisebox{-1pt}{\scriptsize 2}}

\newcommand{\beq}{\begin{equation}}
\newcommand{\eeq}{\end{equation}}

\title{Energy-Efficient Full Diversity Collaborative Unitary Space-Time Block Code Design\\ via Unique Factorization of Signals}
\author{Dong Xia, Jian-Kang Zhang and Sorina Dumitrescu \\
Department of Electrical and Computer Engineering, \\McMaster
University\\
Hamilton, Ontario, Canada.
\thanks{The authors are with the
Department of Electrical and Computer Engineering, McMaster
University, 1280 Main Street West, Hamilton, Ontario, L8S 4K1, Canada. Emails:
(jkzhang, sorina)@mail.ece.mcmaster.ca, askxd@grads.ece.mcmaster.ca.}
}
\begin{document}
\maketitle

\begin{abstract}
In this paper, a novel concept called a \textit{uniquely factorable constellation pair} (UFCP) is proposed for the systematic design of a noncoherent full diversity collaborative unitary space-time block code by normalizing two Alamouti codes for a wireless communication
system having two transmitter antennas and a single receiver
antenna. It is proved that such a unitary UFCP code assures the unique identification of both channel coefficients and transmitted signals in a noise-free case as well as full
diversity for the noncoherent maximum likelihood (ML) receiver in a noise case. To  further improve error performance, an optimal unitary UFCP code is designed by appropriately and uniquely factorizing  a pair of energy-efficient cross quadrature amplitude modulation (QAM) constellations to maximize the coding gain subject to a transmission bit rate constraint. After a deep investigation of the fractional coding gain  function,  
a technical approach developed in this paper to maximizing the coding gain is to carefully design an energy scale to compress the first three largest energy points in the corner of the QAM constellations in the denominator of the objective as well as carefully design a constellation triple forming two UFCPs, with one collaborating with the other two  so as to make the accumulated minimum Euclidean distance along the two transmitter antennas in the numerator of the objective as large as possible and at the same time, to avoid as many corner points of the QAM constellations with the largest energy as possible to achieve the minimum of the numerator. 
In other words, the optimal coding gain is attained by intelligent constellations collaboration and efficient energy compression. Computer simulations demonstrate that error performance of the optimal unitary UFCP code presented in this paper outperforms those of the differential code and the SNR-efficient training code, which is the best code in current literatures for the system. 
\end{abstract}


\section{Introduction}\label{sec:Intro}
At present, the technology intelligently  combining multiple
antennas~\cite{teletar95,foschini98, moustakas2000} with space-time block
coding~\cite{tarokh98, hassibi02, heath02, gamal03, ma03, sethuraman03,
jkz06, yao03,
dayal05, belfiore03, belfiore05, rekaya04, wang-it05, wang-it04,
jkz-icassp05, wang05, rajan05, oggier06, elia06,liao07,jkz-it06,
shang08 } has been well developed to improve 
the spectral efficiency of a coherent wireless communication system. Lately, 
simple space-time block code designs with low complexity decoding~\cite{alamouti98,tarokh99,tirkkonen02,liang03, yuen05, karmakar-it09, dao08, rajan10, sezginer07, paredes08, biglieri09, sirinaunpiboon-it11, 
jkz-it06, shang08, guo-it10, xu-it11} have attracted much attention. 
In this
paper, we are specifically interested in  a flat fading wireless
communication system with two transmitter antennas and a single
receiver antenna. This system is often encountered in mobile
down-link communications for which the mobile receiver may not be
able to deploy multiple antennas. 
For such a
system, if the exact knowledge of the
channel coefficients is available at the receiver, the orthogonal Alamouti~\cite{alamouti98} space-time block code is particularly appealing, 
since it enables the coherent ML receiver to extract full diversity not only with linear processing complexity, but with information losslessness as well~\cite{sandhu00}. Unfortunately,
perfect channel state information at the receiver,
in practice, is not easily obtainable. If the channel changes slowly, then, the transmitter may have sufficiently long 
channel coherence time and send training signals enabling the
channel coefficients to be estimated accurately. However,  the fading coefficients in mobile wireless
communications may vary rapidly and  the coherence time may be too short to allow reliable
estimation of the coefficients. Therefore, the time cost on
sending training signals cannot be ignored since more training signals need sending for the accurate
estimation of the channel~\cite{marzetta-blast99, hassibi03,
zheng02}. In this paper, we consider the communication scenario where channel fading changes very promptly, assuming that the channel gains are completely unknown
at both the transmitter and the receiver, but remain unchanged within four transmission time slots, after which
they change to new independent values that are fixed for next four time slots, and so
on. Such fast varying channel with flat fading for a single transmitter and single receiver antenna was first considered for determining the capacity-achieving input distribution~\cite{richters69, faycal-it01, gursoy-wire05-1, gursoy-wire05-2, huang-it05}. In order to make communication as reliable as possible under this severe environment and avoid sending the training signals for estimation of the channel, using differential space-time
block coding~\cite{hochwald00, hughes00,
sweldens00, Tarokh00, shokrollahi01, hass-hoch02, hoch-marze00,
marze-hass02, ganesan02, jing-hass03, liang05} is one of the possible solutions.  Unfortunately, this approach results in an
approximate loss of 3dB in performance compared to coherent
detection. Recently, some techniques of blind signal processing such as the subspace
method based on the
second order statistics have been utilized to blindly identify the space-time block coded
channel~\cite{Swindlehurst02,Stoica03,Larsson03,
Shahbazpanahi05,Ma06}. However, phase ambiguity incurs the channel not being able to be identified uniquely, even in
the noise-free case. In addition, even if there were no phase ambiguity, the subspace method could not  be successfully applicable to our case, since the 4 coherence time slots are 
too short to allow the second-order statistics to be estimated accurately. 
Therefore, in order to attain a more satisfactory solution, 
noncoherent space-time block coding techniques~\cite{hochwald00, tarokh02,
mccloud02, warrier02, zhao04, kammoun03} have been developed. It has been proved that either at high signal-to-noise ratio
(SNR) or for long coherence time, the unitary code is
optimal~\cite{marzetta99, hochwald00, brehler01, zheng02}. Hence, most of the noncoherent space-time block code designs have been primarily concentrated on unitary designs~\cite{hochwald00, tarokh02, mccloud02,
warrier02, kammoun03, jing-hass03, zhao04, gohary09}. The Cayley~\cite{hass-hoch02, jing-hass03} transform and the exponential
transform~\cite{kammoun03} are now two well-established transforms which convert a linear dispersion code and a linear
space-time block code into unitary codes. The exponential
transform~\cite{kammoun03} requires that the number of the receiver antennas is greater than or equal to that of the transmitter antennas. In general, it cannot assure full diversity for the noncoherent ML receiver. 
The unitary design using  the Cayley transform aimed mainly at differential modulation and a differential receiver. Recently, the original non-full diversity design~\cite{hass-hoch02, jing-hass03} based on the Cayley transform has been improved into a full diversity design utilizing division algebra and algebraic number theory~\cite{oggier07}. More recently, the systematic design of nocoherent unitary space-time block codes with full diversity and a high transmission rate for an
arbitrary number of the transmitter antennas and the receiver
antennas has been established by using a pair of coprime PSK constellations and the QR decomposition~\cite{jkz-it09}. Particularly for the system with two transmitter antennas and a single receiver antenna, the phase ambiguity and full diversity issue for the noncoherent Alamouti
space-time block code has been completely resolved~\cite{cui07, jkz08}.

However, the PSK constellation is not as energy-efficient as the QAM constellation. Therefore, our primary target in this paper is to design a full diversity unitary space-time block code for the system by using the two Alamouti codes and the energy-efficient cross QAM constellations such that the noncoherent coding gain is maximized subject a transmission bit rate constraint. Despite the fact that recent research on coherent MIMO communications has told us that the Alamouti code enables coherent full diversity for any constellations and with any receivers, this is no longer true for noncoherent communications, even if the commonly-used QAM constellations are transmitted and even if the noncoherent ML receiver is employed, since the likelihood function is invariant under certain rotation of some QAM constellation points~\cite{Ma06, zhou07}. Hence, signals must be carefully designed to combat against fading. 
In the noncoherent wireless communication scenario, the unknown of the fading channel at both the transmitter and the receiver requires that the transmitted signals emitting from different time slots must be more correlated than in the coherent environment so that reliable communications with noncoherent full diversity are made possible under a maximum allowable transmission date rate. 
 
All the aforementioned factors greatly motivate us  
to proposing a novel concept, a uniquely factorable constellation pair (UFCP), for the systematic design of an optimal unitary constellation for the system. 
The main idea of the UFCP design essentially comprises the following two major steps.
\begin{enumerate}
\item \textit{Intelligent constellations collaboration}: From a pair of the energy-efficient cross QAM constellations, a constellation triple constituting two UFCPs will be carefully designed, with the one collaboratively shared with the other two through the two transmitter antennas,  
so that the minimum of the numerator in the factional objection function is made as large as possible and at the same time, the largest energy points of the QAM constellations as many as possible are avoided to reach the minimum of the numerator. 
\item \textit{Efficient energy compression}: An energy scale will be carefully designed  to compress the first three largest energy points of the QAM constellations in the denominator of the objective. 
\end{enumerate}
 
{\bf Notation}: Most notations used throughout this paper are
standard: column vectors and matrices are boldface lowercase and
uppercase letters, respectively; the matrix transpose, the complex
conjugate, the Hermitian are denoted by $(\cdot)^T,(\cdot)^*,
(\cdot)^H$, respectively; 
$\mathbf{I}_N$ denotes the $N\times N$ identity
matrix; Notation ${\rm Tr}({\mathbf M})$ denotes the trace of an $K\times K$ matrix ${\mathbf M}$, i.e., ${\rm Tr}({\mathbf M})=\sum_{i=1}^K m_{ii}$, whereas notation $\det({\mathbf M})$ denotes the determinant of ${\mathbf M}$; Notation $\|{\mathbf M}\|_{\rm F}$ denotes the Frobenius norm of ${\mathbf M}$; Notation $f(x)=o(g(x))$ denotes $\lim_{x\rightarrow\infty}\frac{f(x)}{g(x)}=0$; ${\Phi}$ denotes an empty set.  
\section{Channel Model and Unitary Space-time Block Coding}\label{sec:code}
In this section, we first briefly review the channel model in which we are interested in this paper. Then, we propose our transmission scheme and unitary code structure.  
\subsection{Channel Model}\label{subsec:channel}
Let us consider a wireless communication system having two transmitter antennas and a
single receiver antenna. The transmitted symbols from the two
transmitter antennas arrive at the receiver via two different
channels $h_1$ and $h_2$. Then, the discrete baseband received signal $r$ can be represented as
\begin{eqnarray}\label{channel-model}
    r = h_1 s_1+h_2s_2+\xi.
\end{eqnarray}
Throughout this paper we assume $h_1$ and $h_2$
are samples of independent circularly symmetric zero-mean complex white Gaussian
random variables with unit variances and remain constant for the first 4 time slots, after which
they change to new independent values that are fixed for the next 4 time slots, and so on, the
explanation for which will be given in the ensuing subsection.  $s_1$ and $s_2$ are two corresponding
transmitted symbols from these two antennas, and $\xi$ is a
circularly symmetric complex Gaussian noise with zero mean and variance~$\sigma^2$.
\subsection{Unitary Space-Time Block Codes}\label{subsec:blind-code}
Let ${\mathcal A}, {\mathcal B}_1$ and ${\mathcal B}_2$ be three constellations to be designed. Then,  our unitary space-time block code for the channel model~\eqref{channel-model} is basically 
generated by normalizing two Alamouti codes and is  described as follows:
First,  randomly, independently  and equally likely choose three symbols $a\in{\mathcal A}, b_1\in{\mathcal B}_1$ and $b_2\in{\mathcal B}_2$ and then, transmit their normalized version from the two transmitter antennas within four time slots.
During the first time slot, we transmit the signals $s_1=a/\sqrt{|a|^2+|b_1|^2+|b_2|^2}$ and $s_2=0$ in~\eqref{channel-model} at the same time from the respective Antennas 1 and 2. During the second time slot, the signals $s_1=0$ and $s_2=a^*/\sqrt{|a|^2+|b_1|^2+|b_2|^2},$ are simultaneously transmitted from Antennas 1 and 2, respectively. Collecting these two received signals yields 
\begin{subequations}\label{stbc}
\begin{eqnarray}\label{eq:code1}
\left(%
\begin{array}{c}
  r_1 \\
  r_2 \\
\end{array}%
\right) =\frac{1}{\sqrt{|a|^2+|b_1|^2+|b_2|^2}}\left(%
\begin{array}{cc}
  a & 0 \\
  0 & a^* \\
\end{array}%
\right)\left(%
\begin{array}{c}
  h_{1} \\
  h_{2} \\
\end{array}%
\right) +\left(%
\begin{array}{c}
  \xi_1 \\
  \xi_2  \\
\end{array}%
\right),
\end{eqnarray}
or equivalently,
\begin{equation}\label{eq:simple-code-x}
     \mathbf{r}_a =\frac{1}{\sqrt{|a|^2+|b_1|^2+|b_2|^2}} \mathbf{A}\mathbf{h} + \boldsymbol{\xi}_a,
\end{equation}
where ${\mathbf r}_a=(r_1, r_2)^T, {\mathbf A}={\rm diag}(a, a^*), {\mathbf h}=(h_1, h_2)^T$ and ${\boldsymbol\xi}_a=(\xi_1, \xi_2)^T$. In the rest two time slots, the second and third symbols are transmitted using the Alamouti coding scheme, i.e.,
\begin{eqnarray}\label{eq:code2}
\left(%
\begin{array}{c}
  r_3 \\
  r_4 \\
\end{array}%
\right) =\frac{1}{\sqrt{|a|^2+|b_1|^2+|b_2|^2}}\left(%
\begin{array}{cc}
  b_1 & b_2 \\
  -b_2^* & b^*_1 \\
\end{array}%
\right)\left(%
\begin{array}{c}
  h_1 \\
  h_2 \\
\end{array}%
\right) +\left(%
\begin{array}{c}
  \xi_3 \\
  \xi_4  \\
\end{array}%
\right),
\end{eqnarray}
or equivalently,
\begin{equation}\label{eq:simple-code-y}
    \mathbf{r}_b =\frac{1}{\sqrt{|a|^2+|b_1|^2+|b_2|^2}} \mathbf{B}\mathbf{h} + \boldsymbol{\xi}_b,
\end{equation}
where ${\mathbf r}_b=(r_3, r_4)^T, {\boldsymbol\xi}_b=(\xi_3, \xi_4)^T$ and 
\begin{eqnarray}
 \mathbf{B}&=&\left(%
\begin{array}{cc}
  b_1 & b_2 \\
  -b^*_2 & b_1^* \\
\end{array}%
\right),\qquad b_1\in{\mathcal B}_1,\,b_2\in{\mathcal B}_2.\nonumber
\end{eqnarray}
By stacking the above four received signals~\eqref{eq:simple-code-x} and~\eqref{eq:simple-code-y}, the relationship between the transmitted and
received signals within the four time slots can be represented in a more compact matrix form as  
\end{subequations}
\begin{subequations}\label{eq:code}
\begin{eqnarray}
{\mathbf r}&=&\mathbf{U}\mathbf{h} + \boldsymbol{\xi},\label{eq:code-a}
\end{eqnarray}
where $\mathbf{r}=(r_1, r_2,r_3,r_4)^T,
\boldsymbol{\xi}=(\xi_1, \xi_2, \xi_3, \xi_4)^T$ and
\begin{eqnarray}
{\mathbf U}=\frac{1}{\sqrt{|a|^2+|b_1|^2+|b_2|^2}}\left(%
\begin{array}{cc}
  {\mathbf A}\\
  {\mathbf B}
\end{array}%
\right)
=\frac{1}{\sqrt{|a|^2+|b_1|^2+|b_2|^2}}\left(%
\begin{array}{cc}
  a & 0\\
  0 & a^*\\
  b_1 & b_2  \\
  -b^*_2  & b_1^*
\end{array}%
\right)\label{eq:code-b}
\end{eqnarray}
for $a\in{\mathcal A}, b_1\in{\mathcal B}_1, b_2\in{\mathcal B}_2$.
\end{subequations}
We would like to make the following comments on the unitary space-time block coded channel model~\eqref{eq:code}:
\begin{enumerate}
\item \underline{\textit{Why do we need four time slots}}? For a noncoherent MIMO system with $M$ transmitter antennas, it has been proved~\cite{brehler01} that a necessary condition for a space-time block code to enable full diversity for the noncoherent ML receiver is coherent time $T\ge 2M$. Particularly for $M=2$, we should need at least 4 time slots. Actually, in this paper, we consider the shortest coherent time slots enabling the unique identification and full diversity. See more details in Section~\ref{sec:gain}.  
\item \underline{\textit{Why is the symbol rate of $\frac{3}{4}$ per channel use reasonable}}? The answer to this question is mainly motivated from the following observation: For a noncoherent MIMO communication system with $M$ transmitter antennas and $N$ receiver antennas, Zheng and Tse~[4] have proved that in a high SNR regime and for the Rayleigh-faded channel, the average channel capacity is given by
\begin{eqnarray}
C=M^*\left(1-\frac{M^*}{T}\right)\log{\rm SNR}+O(1),
\end{eqnarray}
where $M^*=\min\{M, N, \lfloor\frac{T}{2}\rfloor\}$ and $T$ is coherent time. 
This benchmark result tells us that the original noncoherent MIMO system can be asymptotically  regarded as $M^*\left(1-\frac{M^*}{T}\right)$ parallel spatial channels and thus, the number $M^*\left(1-\frac{M^*}{T}\right)$ is the total number of degrees of freedom to communication. The result also suggests us that the symbol rate of a space-time block code for the noncoherent MIMO channel should be $M^*\left(1-\frac{M^*}{T}\right)$. Especially for the noncoherent system with $T=4, M=2$ and $N=1$, the symbol rate should be $\frac{3}{4}$.   
\end{enumerate}
\subsection{Problem Statement}\label{subsec:prob}
To formally state our design problem, we make the following assumptions throughout this paper:
\begin{enumerate}
 \item The channel coefficients $h_1$ and $h_2$ are samples of independent circularly-symmetric white
Gaussian random variables with zero mean and unit variances, and remain constant
for the first $4$ time slots, after which they change to new
independent values that are fixed for the next $4$ time slots, and
so on. 
 \item
The elements of $\boldsymbol\xi$ are circularly-symmetric zero-mean complex Gaussian samples with covariance matrix $\sigma^2{\mathbf
    I}_4$;
 \item During $4$ observable time slots, the space-time block coding matrix ${\mathbf U}$ is
transmitted with $a, b_1$ and $b_2$  being independently and equally likely chosen from the respective
constellations ${\mathcal A}, {\mathcal B}_1$ and ${\mathcal B}_2$.
 \item Channel state
information is not available at either the transmitter or the
receiver.
\end{enumerate}
Under the above assumptions, our primary purpose in this paper is to solve the following problem.
\begin{problem}\label{prob:start} Design the constellation triple ${\mathcal A}, {\mathcal B}_1$ and ${\mathcal B}_2$ for the unitary space-time block coded channel~\eqref{eq:code} such that
\begin{enumerate}
 \item in the noise-free case, for any given nonzero received signal vector 
${\mathbf r}\ne {\mathbf 0}$, the equation reduced from~\eqref{eq:code-a}
\begin{eqnarray}\label{eq:blind}
{\mathbf r}={\mathbf U}{\mathbf h}
\end{eqnarray}
with respect to the transmitted symbol variables $a$, $b_1$ and $b_2$, and the channel vector ${\mathbf h}$ has a unique solution, and
 \item in the noisy environment, full diversity and the optimal coding gain are enabled for
 the noncoherent ML receiver.~\hfill\QED
 \end{enumerate}
\end{problem}
\section{Uniquely Factorable Constellation Pair }\label{sec:ufcp}
In order to systematically design the constellation triple, ${\mathcal A}, {\mathcal B}_1$ and ${\mathcal B}_2$ in Problem~\ref{prob:start}, in this section we propose a novel concept called Uniquely Factorable Constellation Pair (UFCP). 
\subsection{UFCP}
\begin{definition}\label{def:UFCP}
A pair of constellations $\mathcal{X}$ and $\mathcal{Y}$ is said to be a UFCP, which is denoted by ${\mathcal Y}\sim{\mathcal X}$, if there exist
$x,\tilde{x}\in \mathcal{X}$ and $y,\tilde{y}\in \mathcal{Y}$ such that $x\tilde{y}=\tilde{x}y$,
then $x=\tilde{x}$, $y=\tilde{y}$.
~\hfill\QED
\end{definition}
In Section~\ref{sec:identi}, we will see that it is this kind of the unique factorization of constellations that enables the unique identification of the channel and transmitted signals as well as full diversity. The following example provides us with a trivial UFCP. 
\begin{example}\label{exam:train} For any set ${\mathcal Y}$, if we take ${\mathcal X}=\{1\}$, then, ${\mathcal X}$ and ${\mathcal Y}$ form a UFCP.~\hfill\QED  
\end{example}
Example~\ref{exam:train} tells us that the constellation pair based on the training transmission scheme naturally forms a UFCP.  In this paper, we are interested in the design of non-trival UFCPs each element of which is a complex integer.
To do that, we need to develop a necessary condition which a UFCP must satisfy. 
\begin{proposition}\label{pro:nec-condt-ufcp} Let ${\mathcal X}$ and ${\mathcal Y}$ form a UFCP. If $|{\mathcal Y}|\ge 2$, then, $0\notin{\mathcal X}$.
 ~\hfill\QED
\end{proposition}
\textsc{Proof}: Since $|{\mathcal Y}|\ge 2$, there exist two elements $y_1, y_2\in{\mathcal Y}$ such that $y_1\ne y_2$. Now, suppose that $0\in{\mathcal X}$. Then, we would have 
$0\times y_1=0\times y_2=0$, which contradicts with the assumption that ${\mathcal X}$ and ${\mathcal Y}$ constitute a UFCP. This completes the proof of Proposition~\ref{pro:nec-condt-ufcp}.~\hfill$\Box$

From Proposition~\ref{pro:nec-condt-ufcp} and Definition~\ref{def:UFCP}, we can immediately obtain the following proposition:
\begin{proposition}\label{pro:nece-suf}
For a pair of given constellations ${\mathcal X}$ and ${\mathcal Y}$ with each having finite size and $0\notin{\mathcal X}$, if a new constellation $\mathcal{Z}$
is defined as 
\begin{equation}
\mathcal{Z}=\Big\{z: z=\frac{y}{x}, x\in\mathcal{X}, y\in\mathcal{Y}\Big\},
\nonumber
\end{equation}
then, such a pair of ${\mathcal X}$ and ${\mathcal Y}$ constitutes a UFCP if and only if 
\begin{equation}
|\mathcal{Z}|=|\mathcal{X}|\times |\mathcal{Y}|.
\end{equation}
~\hfill\QED
\end{proposition}
\textsc{Proof}: Define a map $\tau$ from ${\mathcal X}\times {\mathcal Y}$ to ${\mathcal Z}$ as
\begin{eqnarray}
\tau: (x, y)\in{\mathcal X}\times {\mathcal Y}\longrightarrow \frac{y}{x}=z\in{\mathcal Z}.
\end{eqnarray}
It can be verified that $\tau$ is a one-to-one correspondence if and only if $\frac{y}{x}\ne \frac{\widetilde y}{\widetilde x}$, i.e., ${\widetilde x}y\ne x{\widetilde y}$ for $(x, y)\ne (\widetilde x, \widetilde y)$, which is equivalent to saying that a pair of the constellations ${\mathcal X}$ and ${\mathcal Y}$ constitutes a UFCP. On the other hand, since both ${\mathcal X}\times {\mathcal Y}$ and ${\mathcal Z}$ have finite size,  $\tau$ is one-to-one correspondence if and only if $|\mathcal{Z}|=|\mathcal{X}\times \mathcal{Y}|=|\mathcal{X}|\times |\mathcal{Y}|$. This completes the proof of Proposition~\ref{pro:nece-suf}.~\hfill$\Box$

Proposition~\ref{pro:nece-suf} tells us that a UFCP $\mathcal{Y}\sim {\mathcal X}$ can be constructed by factorizing a constellation $\mathcal{Z}$ in such a way that each fraction is unique. For notation simplicity, this kind of construction is specifically denoted by ${\mathcal Z}=\frac{\mathcal Y}{\mathcal X}$.   
In general, when $\mathcal{Z}$ is sizable, it is not easy to find a non-trival unique factorization.
However, for some special constellations, we can utilize this factorization to systematically construct a UFCP.
\begin{definition}\label{rotation-inr}
A constellation is said to be rotation-invariant with respect to $e^{j\theta}$ if
every element in the constellation multipled by $e^{j\theta}$ still belongs to the constellation.
~\hfill\QED
\end{definition}
\begin{example} The square QAM constellation is rotation-invariant with respect to $e^{j\theta}$ with the angle $\theta$ being equal to $\pi/2, \pi$ and $3\pi/2$.~\hfill\QED
\end{example}
In general, we always have the following property, which can be verified directly by the definition and thus, whose poof is omitted. 
\begin{proposition}\label{pro:factor}
Let $\mathcal{Z}$ be rotation-invariant with respect to $e^{j\pi/2}$, $e^{j\pi}, e^{j3\pi/2}$ and $0, r, r j\notin{\mathcal Z}$, where $r$ is real.  
Then, the following statements are true:
\begin{enumerate}
 \item If we let 
\begin{eqnarray}
{\mathcal X}&=&\{1, j\},\nonumber\\
{\mathcal Y}&=&\{z: z=z_{\rm re}+j z_{\rm im} \in{\mathcal Z}, z_{\rm re}, z_{\rm im}>0\}\cup\{z: z=z_{\rm re}+j z_{\rm im} \in{\mathcal Z}, z_{\rm re}, z_{\rm im}<0\}\nonumber,
\end{eqnarray} 
 then, such a pair of ${\mathcal X}$ and ${\mathcal Y}$ constitutes a UFCP and ${\mathcal Z}=\frac{{\mathcal Y}}{{\mathcal X}}$.
 \item If we let  
 \begin{eqnarray}
{\mathcal X}&=&\{1, -1, j, -j\},\nonumber\\
{\mathcal Y}&=&\{z: z=z_{\rm re}+j z_{\rm im} \in{\mathcal Z}, z_{\rm re}, z_{\rm im}>0\}\nonumber,
\end{eqnarray} 
 then, such a pair of ${\mathcal X}$ and ${\mathcal Y}$ forms another UFCP and ${\mathcal Z}=\frac{{\mathcal Y}}{{\mathcal X}}$.
\end{enumerate}
~\hfill\QED
\end{proposition}
The following two examples show us how to obtain the UFCPs by factorizing the 16-QAM constellation by Proposition~\ref{pro:factor}.
\begin{example}\label{exam:qam-factor1} Let ${\mathcal Z}$ be the 16-QAM constellation. By Proposition~\ref{pro:factor}, a UFCP ${\mathcal Y}\sim{\mathcal X}$ is obtained by factorizing ${\mathcal Z}$:   
\begin{eqnarray}
\mathcal{X}&=&\{1,j\},
\nonumber\\
\mathcal{Y}&=&\{3+3j,3+j,1+3j,1+j,-1-j,-3-3j,-3-j,-1-3j\}.
\nonumber
\end{eqnarray}
~\hfill\QED
\end{example}
\begin{example}\label{exam:qam-factor2} Again, applying an idea similar to Proposition~\ref{pro:factor} to the 16-QAM constellation yields another UFCP: 
\begin{eqnarray}
\mathcal{X}&=&\{1,-1,j,-j\},
\nonumber\\
\mathcal{Y}&=&\{3+3j,3+j,1+3j,1+j\}.
\nonumber
\end{eqnarray}
~\hfill\QED
\end{example} 
In addition, it is noticed that for a given UFCP $\mathcal{Y}\sim\mathcal{X}$, when the symbol $x$ is fixed, all the fractions of the form
$z=\frac{y}{x}, y\in{\mathcal Y}$ can be chosen only from a certain subset of $\mathcal{Z}$, which will be employed in the design of the constellation triple ${\mathcal A}, {\mathcal B}_1$ and ${\mathcal B}_2$ in Section~\ref{sec:identi}. Therefore, we particularly give a formal definition as follows:
\begin{definition}
Given a UFCP $\mathcal{Y}\sim\mathcal{X}$  and a fixed $x\in\mathcal{X}$, a set generated from $x$, denoted by ${\mathcal Z}_x$, 
\begin{eqnarray}
{\mathcal Z}_x=\{z:z=\frac{y}{x},y\in\mathcal{Y}\}
\end{eqnarray}
is called a Group-$x$.~\hfill\QED 
\end{definition}
The groups have some interesting properties:
\begin{proposition} \label{pro:decom-group}Let ${\mathcal Z}=\frac{{\mathcal Y}}{{\mathcal X}}$. Then, the following three statements are true:
\begin{enumerate}
 \item \textit{Non-intersection}: For any $x_1, x_2\in{\mathcal X}, x_1\ne x_2$, there is no intersection between Group-$x_1$ and Group-$x_2$, i.e., 
 \begin{eqnarray}
 {\mathcal Z}_{x_1} \cap{\mathcal Z}_{x_2}=\Phi.
  \end{eqnarray}
 \item \textit{Decomposition}: The union of all the groups is equal to the original constellation ${\mathcal Z}$, i.e., 
 \begin{eqnarray}
 \cup_{x\in{\mathcal X}}{\mathcal Z}_{x}={\mathcal Z}.
  \end{eqnarray}  
  \item The number of groups is equal to $|\mathcal{X}|$ and
$|{\mathcal Z}_x|=|\mathcal{Y}|$ for any $x\in{\mathcal X}$.~\hfill\QED  
\end{enumerate} 
\end{proposition}
\textsc{Proof}: Statement 2) is derived directly from the definition of the Group. Here, we only examine Statements~1) and~3). For any $x_1, x_2\in{\mathcal X}$ and $x_1\ne x_2$, suppose that there exists some $z$ belonging to  ${\mathcal Z}_{x_1} \cap{\mathcal Z}_{x_2}$. Then, $z\in {\mathcal Z}_{x_1}$ and $z\in {\mathcal Z}_{x_2}$ and thus, there exist $y_1, y_2\in{\mathcal Y}$ such that $\frac{y_1}{x_1}=\frac{y_2}{x_2}$, which implies that $x_1 y_2=x_2 y_1$. Since ${\mathcal X}$ and ${\mathcal Y}$ form a UFCP, we have $x_1=x_2$ and $y_1=y_2$, which contradicts with the assumption that $x_1\ne x_2$. Hence,   ${\mathcal Z}_{x_1} \cap{\mathcal Z}_{x_2}=\Phi$, i.e., Statement 1) is true.

Using the Statements~1) and~2), we obtain 
\begin{eqnarray}
|{\mathcal Z}|=\sum_{x\in{\mathcal X}}|{\mathcal Z}_{x}|.\label{eq:proof-group}
\end{eqnarray}
By the definition of the group, we have $|{\mathcal Z}_{x}|=|{\mathcal Y}|$. Substituting this into~\eqref{eq:proof-group} yields $|{\mathcal X}||{\mathcal Y}|=|{\mathcal Z}|$. This completes the proof of Statement~3) and thus, of Proposition~\ref{pro:decom-group}.~\hfill$\Box$ 

\subsection{Unique factorizations of the modified cross QAM constellations}
In spite of the fact that Proposition~\ref{pro:nece-suf} tells us that a UFCP $\mathcal{Y}\sim\mathcal{X}$  can be constructed by factorizing a constellation $\mathcal{Z}$, in general, the UFCP ${\mathcal Y}\sim\mathcal{X}$  so derived from the given ${\mathcal Z}$ is not unique. In other words, the same constellation ${\mathcal Z}$ can generate two different UFCPs.  For instance, consider an example below:
\begin{example}Let ${\mathcal Z}$ be the 16-QAM constellation. If we let 
\begin{eqnarray}
\mathcal{X}&=&\{1,-1,j,-j\},
\nonumber\\
\mathcal{Y}&=&\{3+3j,3-j,-1+3j,-1-j\},
\nonumber
\end{eqnarray}
then, it can be verified that such a pair of constellations ${\mathcal X}$ and ${\mathcal Y}$ also forms a UFCP, which is different from the one given in Example~\ref{exam:qam-factor2}.~\hfill\QED
\end{example} 

Now, a natural question is: \textit{for a given constellation ${\mathcal Z}$, which pair of constellations ${\mathcal X}$ and ${\mathcal Y}$ generated by ${\mathcal Z}$ is better?} A general answer to this question is hard to be given. However, in this paper, we are interested in the UFCPs which are derived from the cross $2^K$-ary QAM constellation~\cite{forney-jsac89-1} and every element of which is still a complex integer. This implicitly requires that each element of ${\mathcal X}$ must be of the unit-norm, i.e., $\pm 1$ or $\pm j$. Specifically, we focus on such a UFCP generated from the cross QAM constellation that the minimum distance of ${\mathcal Y}$ is as large as possible. Since the conventional 8-QAM constellation does not satisfy the rotation-invariant property, we need to modify it into a new 8-QAM constellation so as to be rotation-invariant under $e^{j\pi/2}$, $e^{j\pi}$ and $e^{j3\pi/2}$. For discussion self-containment, a formal definition of a $2^K$-ary modified cross QAM constellation is provided here.  
  
\begin{definition}\label{def:qam} A modified $2^K$-ary QAM constellation ${\mathcal Q}$ is defined as follows: 
\begin{enumerate}
\begin{subequations}\label{eq:def-qam}
\item If $K$ is even, ${\mathcal Q}$ is the standard square $2^K$-ary QAM constellation, i.e., 
\begin{eqnarray}
{\mathcal Q}=\Big\{(2m-1)+(2n-1)j: -2^{\frac{K-2}{2}}+1\le m, n\le 2^{\frac{K-2}{2}}\Big\}.\nonumber
 \end{eqnarray}
\item If $K=3$, ${\mathcal Q}$ is a new 8-QAM constellation modified from the conventional 8-ary QAM constellation, i.e.,   
\begin{eqnarray}
{\mathcal Q}=\Big\{1+3j, 1+j, 3-j, 1-j, -1-3j, -1-j, -3+j, -1+j\Big\}.\nonumber
\end{eqnarray}
\item If $K$ is an odd number exceeding 3, ${\mathcal Q}$ is the union of a horizontal rectangular QAM constellation and  a vertical  rectangular QAM constellation, i.e., 
\begin{eqnarray}
{\mathcal Q}=\big\{(2m-1)+(2n-1)j: -3\times 2^{\frac{K-5}{2}}+1\le m \le 3\times 2^{\frac{K-5}{2}}, -2^{\frac{K-3}{2}}+1\le n\le 2^{\frac{K-3}{2}}\big\}\nonumber\\
 \bigcup\big\{(2m-1)+(2n-1)j: -2^{\frac{K-3}{2}}+1\le m\le 2^{\frac{K-3}{2}}, -3\times 2^{\frac{K-5}{2}}+1\le n\le 3\times 2^{\frac{K-5}{2}}\big\}.\nonumber
 \end{eqnarray}
 \end{subequations}
 \end{enumerate}
 ~\hfill\QED
 \end{definition}
 \begin{proposition}\label{pro:ufcp-qam} Let ${\mathcal Z}$ be the given modified $2^K$-cross QAM constellation. Then, subject to ${\mathcal X}\subseteq\{1, -1, j, -j\}$ with a fixed size greater than one, one solution to the following optimization problem:
 \begin{eqnarray}
\{{\mathcal X}_{\rm opt}, {\mathcal Y}_{\rm opt}\}=\arg\max_{\frac{\mathcal Y}{{\mathcal X}}={\mathcal Z}}\min_{y_1\ne y_2\in{\mathcal Y}}| y_1-y_2|
\end{eqnarray}
is given as follows: 
 \begin{enumerate}
  \item If $|{\mathcal X}|=2$, then, 
  \begin{eqnarray}
  {\mathcal X}^{(1)}_{\rm opt}&=&\{1, j\}.\nonumber
      \end{eqnarray}
    \begin{enumerate}  
      \item For $K=3$,
      \begin{eqnarray}
    {\mathcal Y}^{(1)}_{\rm opt}&=&\{1+3j, -1-3j, -1-j, 1+j\}.\label{eq:x1-k3}
      \end{eqnarray}
       \item For $K=5$,
           \begin{eqnarray}
    {\mathcal Y}^{(1)}_{\rm opt}=\{-1+5j, 3+5j, -3+3j, 1+3j, 5+3j, -5+j, -1+j, 3+j, \nonumber\\-3-j, 1-j, 5-j, -5-3j, -1-3j, 3-3j, -3-5j, 1-5j\}.\label{eq:x1-k5}
      \end{eqnarray}
       \item For $K\ge 4$, the optimal ${\mathcal Y}^{(1)}_{\rm opt}$ is determined as follows:  
      \begin{enumerate}
       \item When $K$ is even, 
       \begin{eqnarray}
	       {\mathcal Y}^{(1)}_{\rm
	opt}=\Big\{(2^{\frac{K}{2}}-1-4m)+(2^{\frac{K}{2}}-1-4n) j: ~0\le
	m, n\le 2^{{\frac{K-2}{2}}}-1\Big\} \nonumber\\
	         \bigcup\Big\{(2^{\frac{K}{2}}-3-4m)+(2^{\frac{K}{2}}-3-4n) j:
	~0\le m, n\le 2^{{\frac{K-2}{2}}}-1\Big\}.\label{eq:x1-k-even}
	\end{eqnarray}       
       \item When $K$ is an odd number exceeding 5,
       \begin{small}
	\begin{eqnarray}
	       {\mathcal Y}^{(1)}_{\rm opt}=\Big\{(3\times
	2^{\frac{K-3}{2}}-1-4m)+(2^{\frac{K-1}{2}}-1-4n) j\Big\}_{m=0,
	n=0}^{m=2^{{\frac{K-3}{2}}}-1,\, n=3\times
	2^{{\frac{K-5}{2}}}-1}\nonumber\\
	             \bigcup  \Big\{(2^{\frac{K-1}{2}}-1-4m)+(3\times
	2^{\frac{K-3}{2}}-1-4n) j\Big\}_{m=0, n=0}^{m=3\times
	2^{{\frac{K-5}{2}}}-1,\, n=2^{{\frac{K-3}{2}}}-1}\nonumber\\
	             \bigcup \Big\{(3\times
	2^{\frac{K-3}{2}}-3-4m)+(2^{\frac{K-1}{2}}-3-4n)
	j\Big\}_{m=0, n=0}^{m=2^{{\frac{K-3}{2}}}-1,\, n=3\times
	2^{{\frac{K-5}{2}}}-1}\nonumber\\
	             \bigcup  \Big\{(2^{\frac{K-1}{2}}-3-4m)+(3\times
	2^{\frac{K-3}{2}}-3-4n) j\Big\}_{m=0, n=0}^{m=3\times
	2^{{\frac{K-5}{2}}}-1,\,
	n=2^{{\frac{K-3}{2}}}-1}.\label{eq:x1-k-odd}
	\end{eqnarray}
        \end{small}
        \end{enumerate}
    \end{enumerate}
  \item If $|{\mathcal X}|=4$ and $K\ge 3$, then,   
   \begin{eqnarray}
  {\mathcal X}^{(2)}_{\rm opt}&=&\{1,  -1, j, -j\}.\nonumber
      \end{eqnarray}
      \begin{enumerate}
      \item For $K=3$,
       \begin{eqnarray}
    {\mathcal Y}^{(2)}_{\rm opt}&=&\{1+3j, -1-j\}.\nonumber
      \end{eqnarray}      
      \item For $K=5$,
       \begin{eqnarray}
    {\mathcal Y}^{(2)}_{\rm opt}&=&\{-1+5j, 3+5j, -5+j, -1+j, 3+j, -5-3j, -1-3j, 3-3j\}.\nonumber
      \end{eqnarray}            
      \item For $K\ge 4$, the optimal ${\mathcal Y}^{(2)}_{\rm opt}$ can be determined as follows:     
       \begin{enumerate}
       \item When $K$ is even, 
       \begin{eqnarray}
       {\mathcal Y}^{(2)}_{\rm opt}=\Big\{(4m-2^{\frac{K}{2}}+3)+(2^{\frac{K}{2}}-1-4n) j: ~0\le m, n\le 2^{{\frac{K-2}{2}}}-1 \Big\}.\label{eq:x2-k-even}
        \end{eqnarray}
       \item When $K$ is an odd number exceeding 5,
\begin{eqnarray}
       {\mathcal Y}^{(2)}_{\rm opt}=\Big\{(3\times
2^{\frac{K-3}{2}}-1-4m)+(2^{\frac{K-1}{2}}-1-4n) j\Big\}_{m=0,
n=0}^{m=3\times2^{{\frac{K-5}{2}}}-1,\, n=2^{{\frac{K-3}{2}}}-1}\nonumber\\
             \bigcup  \Big\{(2^{\frac{K-1}{2}}-1-4m)+
(3\times
2^{\frac{K-3}{2}}-1-4n) j\Big\}_{m=0, n=0}^{m=2^{{\frac{K-3}{2}}}-1,\,
n=3\times2^{{\frac{K-5}{2}}}-1}.\label{eq:x2-k-odd}
\end{eqnarray}
           \end{enumerate}           
      \end{enumerate}
      \end{enumerate}  
 ~\hfill\QED     
 \end{proposition}  
\begin{figure}[h]
    \centering
    \subfigure[Diagonal line search over the square QAM constellation for the optimal ${\mathcal Y}_{\rm opt}^{(1)}$: {\color{blue}$\bigcirc$}]{
    \includegraphics[width=6.5cm] {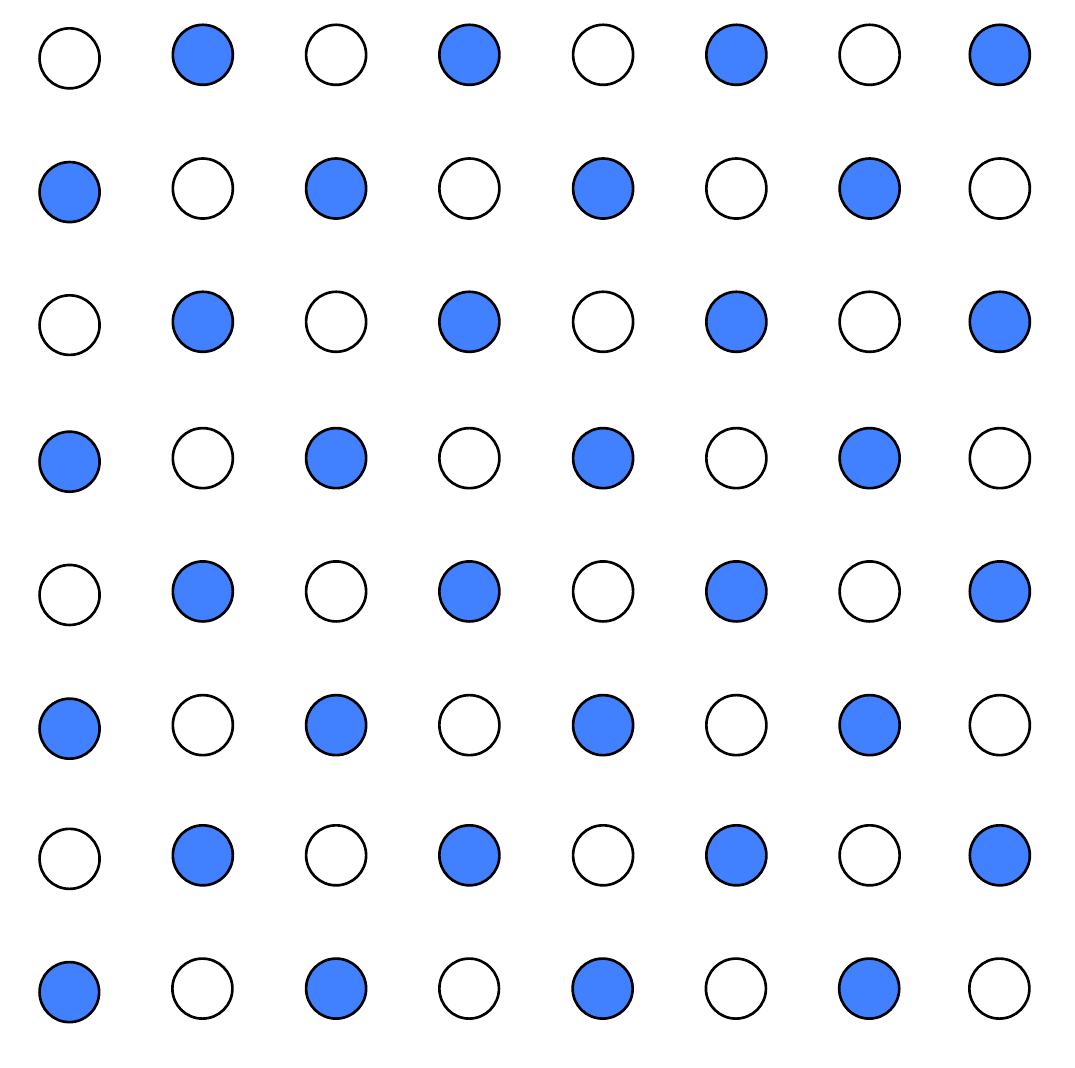}}
    \subfigure[Diagonal search over the cross QAM constellation for the optimal ${\mathcal Y}_{\rm opt}^{(1)}$: {\color{blue}$\bigcirc$}]{
    \includegraphics[width=6.5cm] {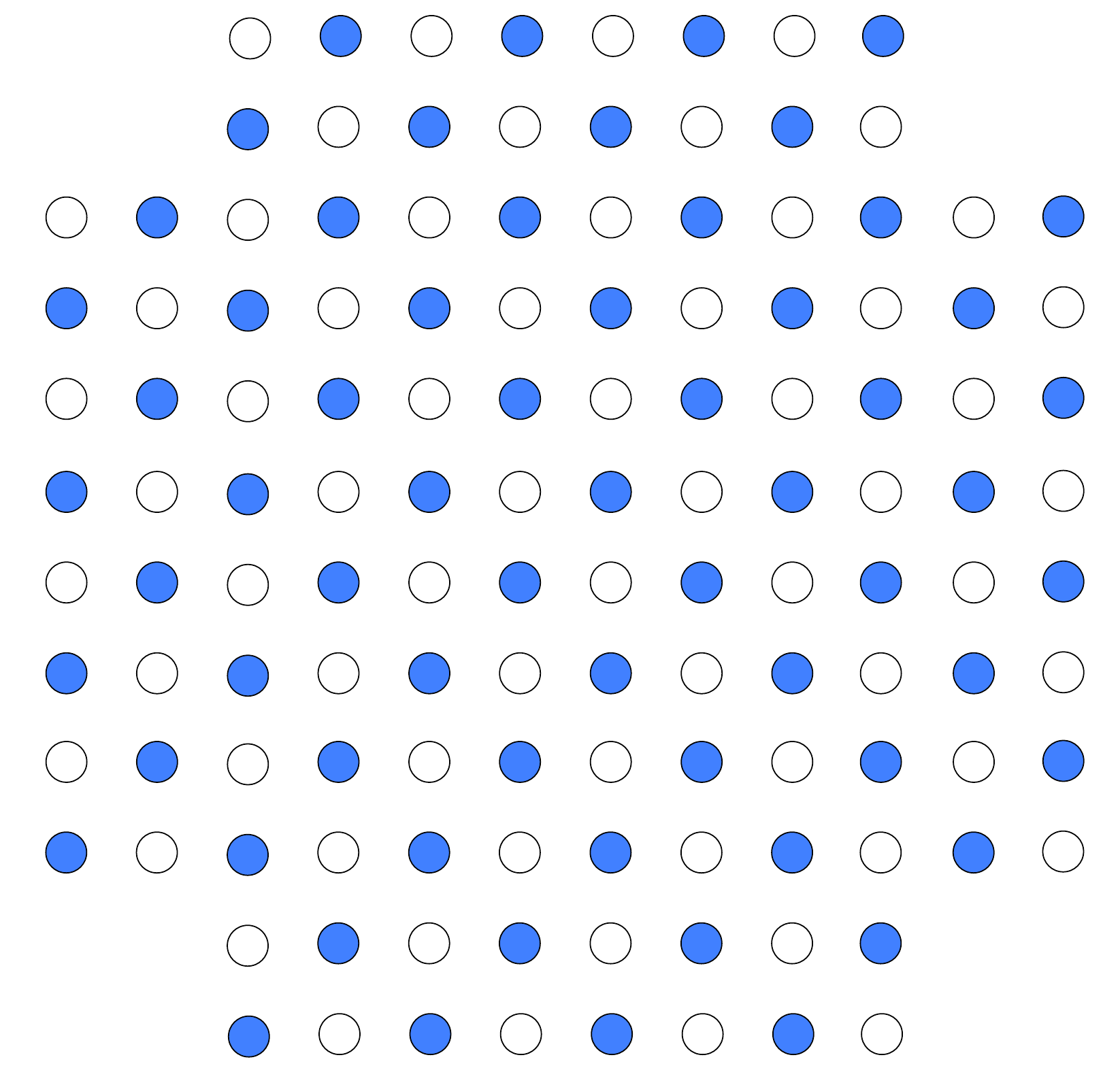}} \\   
    \subfigure[Horizontal and vertical lines search over the square QAM constellation for the optimal ${\mathcal Y}_{\rm opt}^{(2)}$: {\color{green}$\bigcirc$}]{
    \includegraphics[width=6.5cm] {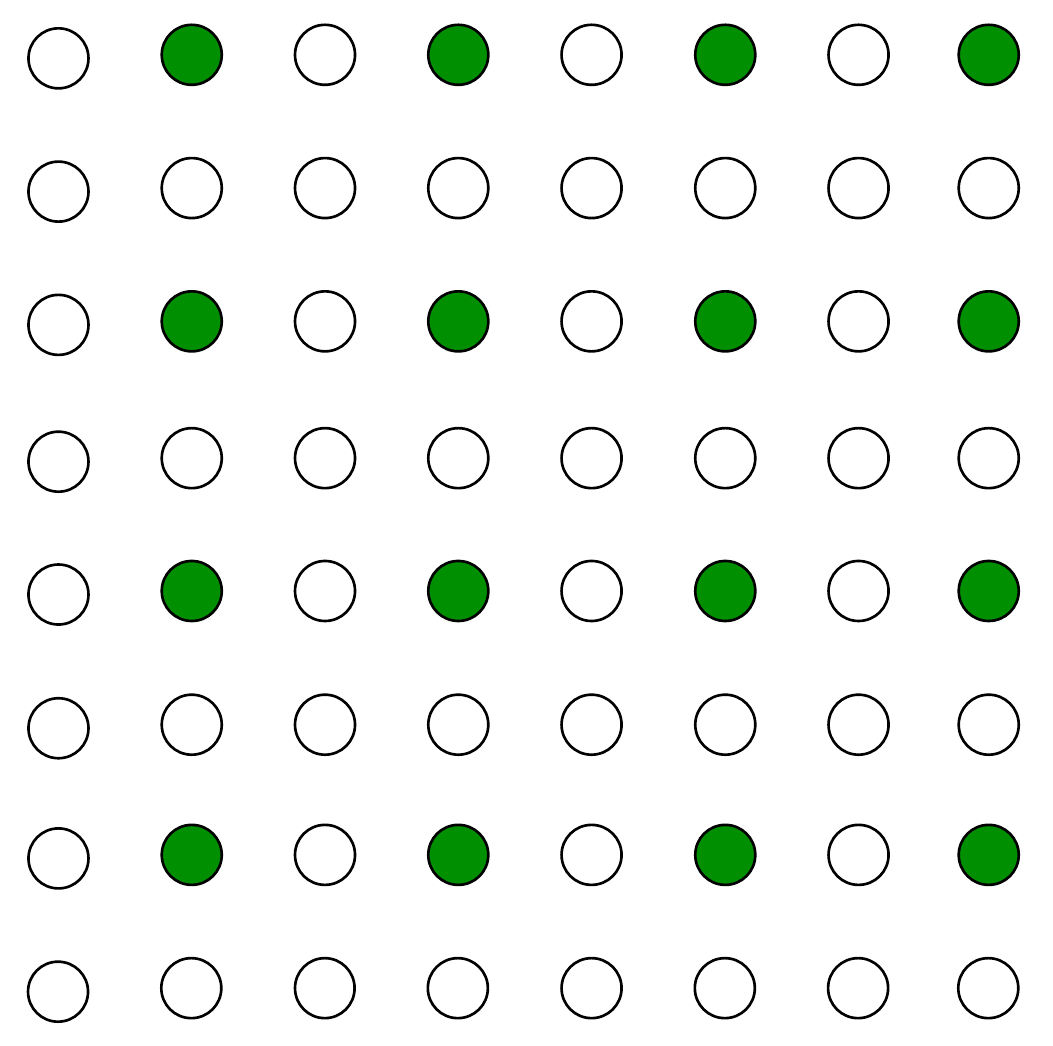}}
    \subfigure[Horizontal and vertical lines search over the cross QAM constellation for the optimal ${\mathcal Y}_{\rm opt}^{(2)}$: {\color{green}$\bigcirc$}]{
    \includegraphics[width=6.5cm] {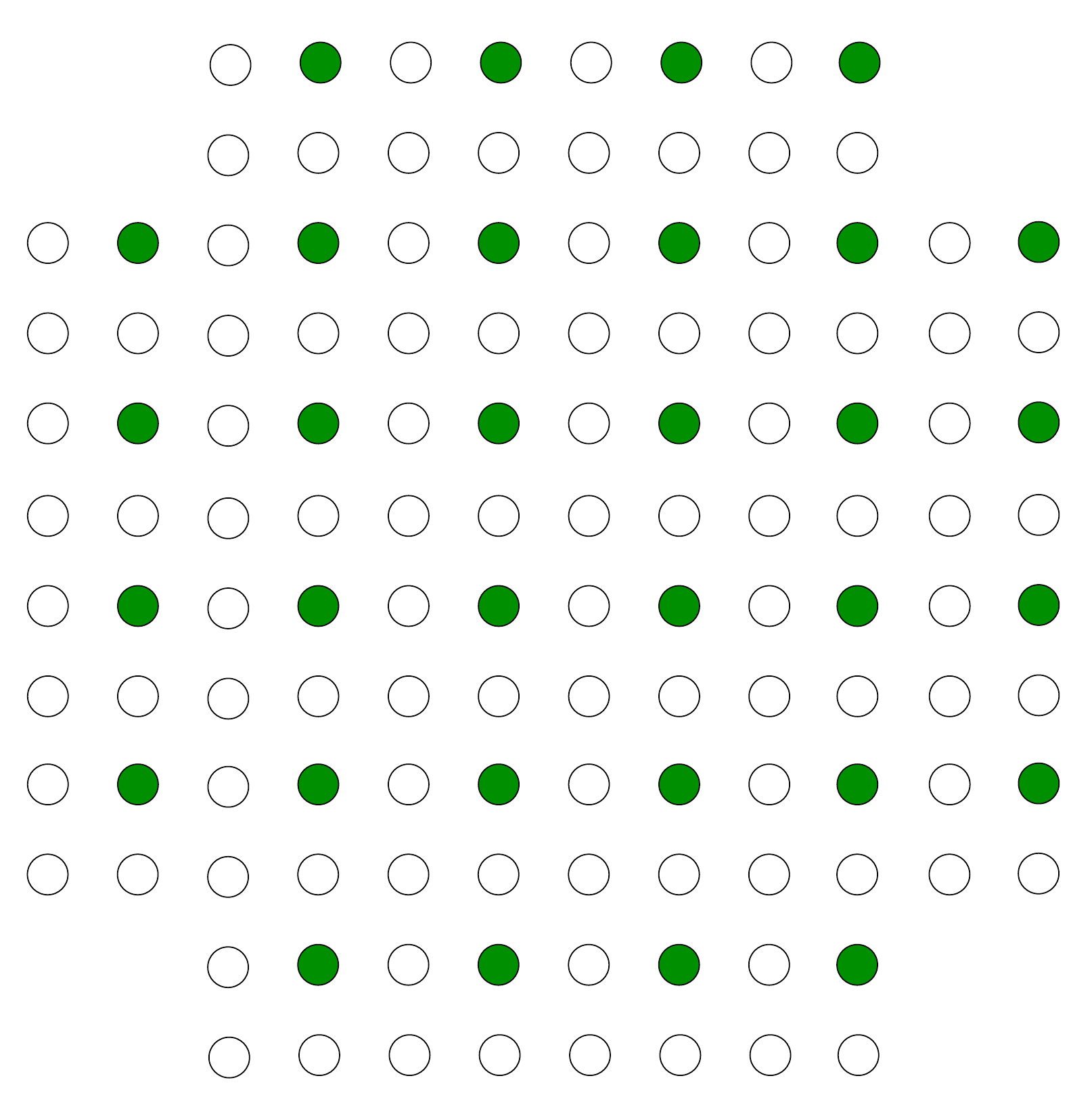}}       
    \caption{The optimal UFCP selections from the square 64QAM and cross 128QAM constellations in Proposition~\ref{pro:ufcp-qam}} \label{fig:ufcp-pick}    
\end{figure}
The proof of Proposition~\ref{pro:ufcp-qam} is given in Appendix~\ref{appendix:pro:ufcp-qam}. In principle,  the optimal constellations ${\mathcal Y}_{\rm opt}^{(1)}$ and ${\mathcal Y}_{\rm opt}^{(2)}$ in Proposition~\ref{pro:ufcp-qam}  can be attained by starting at any corner point in ${\mathcal Z}$ with the largest energy and then, for $|{\mathcal X}|=2$,  
successively selecting all nearest neighbors of the previously already selected points along the diagonal lines 
and for $|{\mathcal X}|=4$,  successively selecting all every other points of the previously already selected points
along the horizontal and vertical lines. 
Fig.~\ref{fig:ufcp-pick} visually demonstrates how the optimal constellations ${\mathcal Y}_{\rm opt}^{(1)}$ and ${\mathcal Y}_{\rm opt}^{(2)}$ in Proposition~\ref{pro:ufcp-qam} have been obtained by starting at the corner point in the first quadrant with the largest energy  
via the diagonal line or the horizontal and vertical lines search over the cross QAM constellations when $K$ is either even or odd. However, we should clearly point out here that the way in obtaining the optimal constellation ${\mathcal Y}_{\rm opt}$ in Proposition~\ref{pro:ufcp-qam} for ${\mathcal X}_{\rm opt}=\{1, j\}$ is  exactly the same as the one in partitioning a large constellation into two small sub-constellations in the trellis coded modulation (TCM) proposed by Urgerboeck in~\cite{ungerboeck87-1, ungerboeck87-2}.  The major difference between these two partition methods is that the partition of a constellation in the TCM is based on a union operation, i.e., the union of the two partitioned sub-constellations is equal to the original constellation, whereas  the partition of a constellation in the UFCP is based on multiplication. As an application of Proposition~\ref{pro:ufcp-qam}, we give the following example.  
 \begin{example} $K=4$. By Proposition~\ref{pro:ufcp-qam}, we have that 
 \begin{enumerate}
  \item if $|{\mathcal X}|=2$, then, 
  \begin{eqnarray}
  {\mathcal X}^{(1)}_{\rm opt}&=&\{1, j\},\nonumber\\
   {\mathcal Y}^{(1)}_{\rm opt}&=&\{3+3j, 1+j, -1+3j, 3-j, -3+j, -1-j, 1-3j, -3-3j \},\nonumber
    \end{eqnarray}  
  \item if $|{\mathcal X}|=4$, then,  
   \begin{eqnarray}
  {\mathcal X}^{(2)}_{\rm opt}&=&\{1,  -1, j, -j\},\nonumber\\
   {\mathcal Y}^{(2)}_{\rm opt}&=&\{3+3j, -1+3j, 3-j, -1-j\}.\nonumber      
   \end{eqnarray}  
  \end{enumerate}
 ~\hfill\QED
 \end{example}

\section{Collaborative Unitary UFCP Space-Time Block Codes }\label{sec:identi} In this section,
we take advantage of UFCPs established in Section~\ref{sec:ufcp} to carefully design the constellation triple ${\mathcal A}, {\mathcal B}_1$ and ${\mathcal B}_2$. We will show that it is the unique factorization of signals that results not only in the unique identification of the channel coefficients and the transmitted signals in the noise-free case, but in full diversity in the noise case as well. 

\subsection{Signal Designs and Unique Identification}\label{subsec:identi}
In order for the code~\eqref{eq:code-b} to enable the unique identification of the channel and the transmitted signals as well as full diversity, the three constellations ${\mathcal A}, {\mathcal B}_1$ and ${\mathcal B}_2$ must work cooperatively.  There may exist many cooperative agreements to make these constellations work together. Different collaborative ways will produce different codes, but in this paper,  we require that ${\mathcal A}, {\mathcal B}_1$ and ${\mathcal B}_2$ cooperate in such a way that ${\mathcal A}$ and ${\mathcal B}_1$, ${\mathcal A}$ and ${\mathcal B}_2$ constitute two pairs of UFCPs. In other words, the same constellation ${\mathcal A}$ collaborates with both the constellations ${\mathcal B}_1$ and ${\mathcal B}_2$. More explanations of why such a collaborative scheme is adopted will be given after Theorem~\ref{th:identi}.  To make the selection of these constellations more clear, we change notation ${\mathcal A}, {\mathcal B}_1$ and ${\mathcal B}_2$ into the respective ${\mathcal X}, {\mathcal Y}_1$ and ${\mathcal Y}_2$. Correspondingly, the code  now takes the following form: 
\begin{eqnarray}\label{eq:blind-stbc}
{\mathcal U}=\left\{{\mathbf U}=\frac{1}{\sqrt{|x|^2+ |y_1|^2+ |y_2|^2}}
\left(%
\begin{array}{cc}
  x & 0\\
  0 & x^*\\
  y_1 & y_2  \\
  - y^*_2  &  y_1^*
\end{array}%
\right):\,\,\,\,x\in{\mathcal X}, y_1\in{\mathcal Y}_1, y_2\in{\mathcal Y}_2\right\},
\end{eqnarray}
where ${\mathcal Y}_1\sim {\mathcal X}$ and ${\mathcal Y}_2\sim {\mathcal X}$ constitute two UFCPs. Such a code ${\mathcal U}$ is called a collaborative unitary UFCP space-time block code. 

Now, we formally state the first main result in this paper.
\begin{theorem}\label{th:identi}
Let ${\mathcal Y}_1\sim{\mathcal X}$,  and ${\mathcal Y}_2\sim{\mathcal X}$ be two UFCPs.  Then, for any given nonzero received signal vector without noise, i.e.,  ${\mathbf r}={\mathbf U}{\mathbf h}$, the code ${\mathcal U}$ designed by~\eqref{eq:blind-stbc} enables the unique identification of the channel coefficients and the transmitted signals. ~\hfill\QED
\end{theorem}
\textsc{Proof}: Using the code structure~\eqref{eq:blind-stbc}, we have 
\begin{subequations}\label{eq:blind}
\begin{eqnarray}
    \mathbf{r}_x&=&\frac{1}{\sqrt{|x|^2+|y_1|^2+ |y_2|^2}}\mathbf{X} \mathbf{h},\\
     \mathbf{r}_y&=&\frac{1}{\sqrt{|x|^2+|y_1|^2+ |y_2|^2}}\mathbf{Y} \mathbf{h},     
\end{eqnarray}
\end{subequations}
where ${\mathbf X}$ and ${\mathbf Y}$ are defined by
\begin{eqnarray}
{\mathbf X}&=&\left(\begin{array}{cc}x & 0 \\0 & x^*\end{array}\right),\nonumber\\
{\mathbf Y}&=&\left(\begin{array}{cc}y_1 & y_2 \\-y_2^* & y_1^*\end{array}\right).\nonumber
\end{eqnarray}
Since $|{\mathcal Y}_i|>1$ for $i=1, 2$, by Proposition~\ref{pro:nec-condt-ufcp} we have $x\ne 0$ and as a result, the matrix ${\mathbf X}$ is invertible.  Eliminating ${\mathbf h}$ from~\eqref{eq:blind} yields 
\begin{eqnarray}\label{eq:re-blind}
{\mathbf r}_y=\mathbf{Y} \mathbf{X}^{-1}{\mathbf r}_x. 
\end{eqnarray}
Notice that $\mathbf{Y} \mathbf{X}^{-1}=\left(
\begin{array}{cc}
\frac{y_1}{x} & -(\frac{y_2}{x})^*\\
\frac{y_2}{x} & (\frac{y_1}{x})^*
\end{array}
\right)$
is the Alamouti codeword matrix. Therefore, equation~\eqref{eq:re-blind} can be rewritten as
\begin{equation}\label{eq:blind2}
     \widetilde{{\mathbf r}}_y= \mathbf{R}_x \mathbf{u},
\end{equation}
where $ \widetilde{{\mathbf r}}_y=(r_y(1), r_y^*(2))^T$, $ \mathbf{u}=(y_1/x, y_2/x)^T$ and $\mathbf{R}_x$ is given by
\begin{equation}
\mathbf{R}_x=\left(
\begin{array}{cc}
 r_x(1) & -r_x(2)\\
r_x^*(2) & r_x^*(1)
\end{array}
\right).\nonumber
\end{equation}
It can be verified that $\mathbf{R}_x$ is unitary up to a
scale. In addition, since ${\mathbf r}=({\mathbf r}_x^T, {\mathbf r}_y^T)^T\ne {\mathbf 0}$ and ${\mathbf X}$ is invertible, we have $\mathbf{h}\neq\mathbf{0}$ and thus, ${\mathbf r}_x\ne {\mathbf 0}$, which is equivalent to the fact
that $\mathbf{R}_x$ is invertible. Therefore, from (\ref{eq:blind2}) we obtain
\begin{equation}\label{eq:blind3}
    \mathbf{u}=\mathbf{R}^{-1}_x\widetilde{\mathbf{r}}_y.
\end{equation}
That is
\begin{subequations}\label{eq:blind-iden}
\begin{eqnarray}
    \frac{y_1}{x}&=&\frac{r_x^*(1)r_y(1)+r_x(2)r_y^*(2)}{|r_x(1)|^2+|r_x(2)|^2},\label{eq:blind-iden-a}\\
    \frac{y_2}{x}&=&\frac{-r_x^*(2)r_y(1)+r_x(1)r_y^*(2)}{|r_x(1)|^2+|r_x(2)|^2}.\label{eq:blind-iden-b}
    \end{eqnarray}
\end{subequations}
Since ${\mathcal X}$ and ${\mathcal Y}_i$  form the two UFCPs and $x\in{\mathcal X}$ and $y_i\in{\mathcal Y}_i$ for $i=1, 2$, once their quotients $y_i/x$ have been determined, then,  $x $ and $y_i$ themselves can be uniquely determined. In other words, there exists a unique triple $x, y_1$ and $y_2$ such that~\eqref{eq:blind-iden} is satisfied. Moreover, after we have determined $x$ and $y_i$, then,  the channel vector $\mathbf{h}$ can be
uniquely determined by $\mathbf{ h}=\sqrt{|x|^2+|y_1|^2+|y_2|^2}\,\mathbf{ X}^{-1}\mathbf{ r}_x=\sqrt{|x|^2+|y_1|^2+|y_2|^2}\,\mathbf{ Y}^{-1}\mathbf{ r}_y$. This completes the proof of Theorem~\ref{th:identi}.~\hfill$\Box$

Some observations on Theorem~\ref{th:identi}  are made as follows:
\begin{enumerate}
 \item Theorem~\ref{th:identi} tells us that using the UFCP code~\eqref{eq:blind-stbc}, the channel
coefficients and the transmitted signals can be uniquely identified by {\em only} processing the four received signals.  Theorem~1 is very desirable in the design of noncoherent constellations, since if a signal design is not able to provide the unique identiÞcation of the channel and the transmitted signals in the noise-free case, then, the reliable estimation of the signal will not be guaranteed, even in high SNR.    
\item From the proof of Theorem~\ref{th:identi}, i.e.,~\eqref{eq:blind-iden},  we can observe that  if either ${\mathcal Y}_1\sim {\mathcal X}$ or ${\mathcal Y}_2\sim {\mathcal X}$, then, the channel coefficients and the transmitted signals can be still uniquely identified. In addition, \eqref{eq:blind-iden-a} divided by \eqref{eq:blind-iden-b} gives us 
  \begin{eqnarray}
   \frac{y_1}{y_2}=\frac{r_x^*(1)r_y(1)+r_x(2)r_y^*(2)}{-r_x^*(2)r_y(1)+r_x(1)r_y^*(2)}\nonumber.
  \end{eqnarray}
Now, it can be seen clearly that if ${\mathcal Y}_1\sim {\mathcal Y}_2$,   then, $y_1, y_2$ and thus, $x$ can be uniquely determined. As a result, the channel coefficients can also be uniquely determined. Therefore, it is worth emphasizing the fact that it is the unique factorization of constellations proposed in Section~\ref{sec:ufcp} that enables the channel coefficients and the transmitter signals to be uniquely identified. In a word, the unique identification of both the channel coefficients and the transmitted signals requires three constellations collaboration.   
  \item The aforementioned Observations 2) tells us that there are several ways in which the three constellations ${\mathcal A}, {\mathcal B}_1$ and ${\mathcal B}_2$ can collaborate so that both the channel coefficients and the transmitted signals are able to be uniquely identified. However, in this paper we adopt ${\mathcal Y}_1\sim {\mathcal X}$ and ${\mathcal Y}_2\sim {\mathcal X}$, since the same constellation ${\mathcal X}$ collaborating with the other two constellations ${\mathcal Y}_1$ and ${\mathcal Y}_2$ not only enables the unique identification and thus, full diversity (see Theorem~\ref{th:ufcp-fd}), but also provides an opportunity for the two transmitter antennas to accumulate their minimum Euclidean distances such that the coding gain is increased. See more details in Section~\ref{sec:gain}.     
   \end{enumerate}

\subsection{Full Diversity}\label{subsec:fd}In order to analyze the full diversity of the UFCP code~\eqref{eq:blind-stbc}, let us first consider a general space-time block coded noncoherent MIMO system with $M$ transmitter antennas, $N$ receiver antennas and flat fading channels as follows:
 \begin{eqnarray}
{\mathbf\Upsilon}={\mathbf
 S}{\mathbf H}+{\boldsymbol \Xi},
\end{eqnarray}
where ${\mathbf\Upsilon}$ denotes a $T\times N$ received signal
matrix, ${\mathbf H}$ denotes an $M\times N$ channel matrix, ${\mathbf
S}$ is a $T\times M$ codeword matrix and ${\boldsymbol \Xi}$ denotes a $T\times N$ noise matrix. We assume that the elements ${\xi_{tn}}$ of ${\boldsymbol \Xi}$ are samples of independent circularly-symmetric
zero-mean complex Gaussian random variables with variance $\sigma^2$. Under these assumptions,  the probability density function of the received signal matrix ${\mathbf\Upsilon}$ conditioned on the transmitted signal matrix ${\mathbf S}$ is the Gaussian distribution, i.e.,  
\begin{eqnarray*}
\frac{1}{\pi^{TN}\det({\mathbf S}{\mathbf S}^H+\sigma^2{\mathbf I})}\times \exp\left(-\frac{{\rm Tr}\big({\mathbf\Upsilon}^H({\mathbf S}{\mathbf S}^H+\sigma^2{\mathbf I})^{-1}{\mathbf \Upsilon}\big)}{\sigma^2}\right), 
\end{eqnarray*}
and thus, its likelihood is given by 
\begin{eqnarray*}
-\frac{{\rm Tr}\big({\mathbf\Upsilon}^H({\mathbf S}{\mathbf S}^H+\sigma^2{\mathbf I})^{-1}{\mathbf\Upsilon}\big)}{\sigma^2}-\ln\det({\mathbf S}{\mathbf S}^H+\sigma^2{\mathbf I})-TN\ln\pi. 
\end{eqnarray*}
Then, the maximum likelihood (ML) receiver for the noncoherent MIMO system is equivalent to solving the following optimization problem: $\hat{{\mathbf S}}=\arg\min_{{\mathbf
S}} \{\Delta_s-{\rm Tr}\left({\mathbf\Upsilon}^H{\mathbf \Theta}_s{\mathbf
\Upsilon}\right)\}$,
where ${\mathbf \Theta}_s=\frac{1}{{\sigma}^2}{\mathbf
S}\left({\sigma^2}{\mathbf I}+{\mathbf S}^H{\mathbf
S}\right)^{-1}{\mathbf S}^H$ and $
\Delta_s=\ln\det\left({\sigma^2}{\mathbf I}+{\mathbf S}^H{\mathbf
S}\right)$.
To avoid estimating the variance of noise, consider the conditional probability density function of the received signal matrix ${\mathbf\Upsilon}$ given the channel matrix ${\mathbf H}$ and  the transmitted signal matrix ${\mathbf S}$, i.e., 
\begin{eqnarray*}
\frac{1}{\pi^{TN}\sigma^{2TN}}\times \exp\left(-\frac{\|{\mathbf\Upsilon}-{\mathbf S}{\mathbf H}\|^2_{F}}{\sigma^2}\right),
\end{eqnarray*}
and thus, its likelihood is given by 
\begin{eqnarray*}
-\frac{\|{\mathbf\Upsilon}-{\mathbf S}{\mathbf H}\|^2_{F}}{\sigma^2}-TN\ln\pi-2TN\ln\sigma.
\end{eqnarray*} 
The \textit{generalized likelihood ratio test} (GLRT) receiver for the joint estimation of ${\mathbf H}$ and ${\mathbf S}$ is to maximize the likelihood, which is essentially equivalent to solving the following
nonlinear least square error optimization problem~\cite{golub73, narayan98, brehler01}: 
 \begin{eqnarray}\label{sphere-grlt}
\{\hat{\mathbf H}, \hat{\mathbf S}\}&=&\arg\min_{{\mathbf H}, {\mathbf S}}\|{\mathbf\Upsilon}-{\mathbf S}{\mathbf H}\|^2_F.
\end{eqnarray} 
Its solution can be obtained by first estimating the transmitted signal matrix ${\mathbf S}$ as 
\begin{eqnarray}\label{glrt}
\hat{{\mathbf S}}=\arg\max_{{\mathbf S}} {\rm
Tr}\left({\mathbf\Upsilon}^H{\mathbf S}\left({\mathbf S}^H{\mathbf
S}\right)^{-1}{\mathbf S}^H{\mathbf\Upsilon}\right),
\end{eqnarray}
and then, estimating the channel matrix ${\mathbf H}$ as $\hat{\mathbf H}=\big(\hat{\mathbf S}^H\hat{\mathbf S}\big)^{-1}\hat{\mathbf S}^H{\mathbf \Upsilon}$. Particularly for any unitary code, the ML receiver and the GLRT receiver for the optimal estimation of the transmitted signal matrix are equivalent, i.e., $\hat{{\mathbf S}}=\arg\max_{{\mathbf S}} {\rm
Tr}\left({\mathbf\Upsilon}^H{\mathbf S}{\mathbf S}^H{\mathbf\Upsilon}\right)$.
In addition, Brehler and Varanasi~\cite{brehler01} analyzed the asymptotic performance of the GLRT detector for the noncoherent MIMO system and proved the following lemma.
\begin{lemma}\label{lem:brehler}Let a $2M\times 2M$ matrix ${\mathbf R}_{s\hat{s}}$ be defined as ${\mathbf R}_{s\hat{s}}=\big({\mathbf S}, \hat{\mathbf S}\big)^H\big({\mathbf S}, \hat{\mathbf S}\big)$.
If each matrix ${\mathbf R}_{s\hat{s}}$ has full rank for all pairs of distinct
codewords ${\mathbf S}$ and $\hat{\mathbf S}$, then, the
resulting space-time block code provides full diversity for the GLRT
receiver, and moreover, the pair-wise error probability ${\rm P_{GLRT}}({\mathbf S}\rightarrow\hat{\mathbf S})$ of transmitting
${\mathbf S}$ and deciding in favor of $\hat{\mathbf S}\ne {\mathbf
S}$ has the following asymptotic formula: 
\begin{eqnarray}
{\rm P_{GLRT}}({\mathbf S}\rightarrow\hat{\mathbf S})
=\frac{\left(%
\begin{array}{c}
  2MN-1 \\
  MN \\
\end{array}%
\right)\det^N(\hat{\mathbf S}^H\hat{\mathbf S})}{\det^N({\mathbf R}_{s\hat{s}})}\times{\rm SNR}^{-MN} +o\big({\rm SNR}^{-MN}\big).\nonumber
\end{eqnarray}
 ~\hfill\QED
\end{lemma}
Lemma~\ref{lem:brehler} tells us that the full rank of the matrices ${\mathbf R}_{s\hat{s}}$ for all the distinct codewords ${\mathbf S}$ and $\hat{\mathbf S}$ assures full diversity. It is not difficult to prove that a necessary condition for ${\mathbf R}_{s\hat{s}}$ to have full rank is $T\ge 2M$. Hence, in this paper, we consider the case of the shortest coherent time slots when it is possible for the UFCP code design to enables unique identification of both the channel coefficients and the transmitted signals as well as full diversity. Now, we are in a position to state the second main result in this paper.
\begin{theorem}\label{th:ufcp-fd}Let ${\mathcal Y}_1\sim{\mathcal X}$,  and ${\mathcal Y}_2\sim{\mathcal X}$ constitute two UFCPs.  Then, the code ${\mathcal U}$ designed by~\eqref{eq:blind-stbc} enables full diversity for the noncoherent ML receiver. Furthermore, the pair-wise error probability ${\rm P_{ML}}({\mathbf U}\rightarrow\hat{\mathbf U})$ of transmitting
${\mathbf U}$ and deciding in favor of $\hat{\mathbf U}\ne {\mathbf
U}$ has the following asymptotic formula: 
\begin{eqnarray}
{\rm P_{ML}}({\mathbf U}\rightarrow\hat{\mathbf U})=\frac{3}{\det({\mathbf R}_{u\hat{u}})}\times{\rm SNR}^{-2}+o\big({\rm SNR}^{-2}\big),\nonumber
\end{eqnarray}
where ${\mathbf R}_{u\hat{u}}=\big({\mathbf U}, \hat{\mathbf U}\big)^H\big({\mathbf U}, \hat{\mathbf U}\big)$.~\hfill\QED
\end{theorem}
\textsc{Proof}: We first note that since the UFCP code~\eqref{eq:blind-stbc} is unitary, we have $\det({\mathbf U}^H{\mathbf U})=\det(\hat{\mathbf U}^H\hat{\mathbf U})=1$. By Lemma~\ref{lem:brehler}, we only need to prove that ${\mathbf R}_{u\hat{u}}$ is invertible for any pair of distinct ${\mathbf U}$ and $\hat{\mathbf U}$. Since $\big({\mathbf U}, \hat{\mathbf U}\big)$ is a square matrix and ${\mathbf R}_{u\hat{u}}=\big({\mathbf U}, \hat{\mathbf U}\big)^H\big({\mathbf U}, \hat{\mathbf U}\big)$,  proving that the matrix ${\mathbf R}_{u\hat{u}}$ is invertible is equivalent to proving that the matrix $\big({\mathbf U}, \hat{\mathbf U}\big)$ is invertible. Notice that 
\begin{eqnarray}
\big({\mathbf U}, \hat{\mathbf U}\big)=\frac{1}{\sqrt{|x|^2+|y_1|^2+|y_2|^2}}\left(\begin{array}{cccc}
x & 0 & \hat{x} & 0 \\
0 & x^* & 0 & \hat{x}^* \\
y_1 & y_2 & \hat{y}_1 & \hat{y}_2 \\
-y_2^* & y_1^* & -\hat{y}_2^* & \hat{y}_1^*
\end{array}\right).
\end{eqnarray}
By some algebraic manipulations, we can obtain that the determinant of $\big({\mathbf U}, \hat{\mathbf U}\big)$ is given by $\det\big({\mathbf U}, \hat{\mathbf U}\big)=\big(\left |\frac{y_1}{x}-\frac{\hat{y}_1}{\hat{x}}\right |^2+\left |\frac{y_2}{x}-\frac{\hat{y}_2}{\hat{x}}\right |^2\big)/\big(|x|^2+|y_1|^2+|y_2|^2\big)^2$.   
Since ${\mathcal X}$ and ${\mathcal Y}_i$ for $i=1, 2$ forms two UFCPs for any ${\mathbf U}\ne \hat{\mathbf U}$, i.e., $(x, y_1, y_2)\ne (\hat{x}, \hat{y}_1, \hat{y}_2)$, we have $\det\big({\mathbf U}, \hat{\mathbf U}\big)\ne 0$. In addition, the asymptotic formula can be immediately obtained by utilizing Lemma~\ref{lem:brehler} again with $M=2, N=1$ and ${\mathbf S}={\mathbf U}$. This completes the proof of Theorem~\ref{th:ufcp-fd}.~\hfill$\Box$  

From the proof of Theorem~\ref{th:ufcp-fd} we can observe that it is the unique factorization of our designed constellations that enables the matrix ${\mathbf R}_{y\hat{y}}$ to have full rank. In addition, we can also observe that the condition in Comment~2) on Theorem~\ref{th:identi} is still a sufficient condition for the matrix ${\mathbf R}_{y\hat{y}}$ to have full rank. Therefore, noncoherent full diversity also requires the three constellations collaboration.  
\section{Optimal Designs of Unitary UFCP Stace-Time Block Codes}\label{sec:gain}
Our main task in this section is to efficiently and effectively optimize the coding gain for the unitary UFCP space-time block codes generated from the energy-efficient cross QAM constellations. 
\subsection{Problem Formulation} Theorems~\ref{th:identi} and~\ref{th:ufcp-fd} together tell us that the unitary UFCP code designed by~\eqref{eq:blind-stbc} enables the unique identification of the channel coefficients and the transmitted signals as well as full diversity for the noncoherent ML receiver. Therefore, the code design partially gives a solution to Problem~\ref{prob:start}. In order to further optimize its error performance, we can see from the asymptotic formula of the pair-wise error probability in Theorem~\ref{th:ufcp-fd} that when SNR is large, the error performance is dominated by       
the term $\det\big((\mathbf{ U},\hat{\mathbf{ U}})^H(\mathbf{ U},\hat{\mathbf{ U}})\big)$. 
Hence, following the way similar to coherent MIMO communications~\cite{tarokh98}, we define the coding gain for the unitary code as
\begin{equation}\label{eq:def-gain}
G({\mathcal X}, {\mathcal Y}_1, {\mathcal Y}_2)=\min_{{\mathbf U}\ne\hat{\mathbf U}, {\mathbf U}, \hat{\mathbf U}\in{\mathcal U}}\sqrt{\det\big((\mathbf{ U},\hat{\mathbf{ U}})^H(\mathbf{ U},\hat{\mathbf{ U}})\big)}.
\end{equation}
Theoretically speaking, we should maximize the coding gain $G({\mathcal X}, {\mathcal Y}_1, {\mathcal Y}_2)$ directly  among all two UFCPs ${\mathcal Y}_1\sim {\mathcal X}$ and ${\mathcal Y}_2\sim {\mathcal X}$. However, this optimization problem, in general,  is too difficult to be solved, since the optimal design of constellations is generally extremely difficult to be reformulated into a tractable optimization problem, even for an additive white Gaussian noise channel~\cite{forney-it88-1, forney-it88-2, forney-jsac89-1, forney-jsac89-2, conway-book98, forney-it98, gallager-book08}. To make the problem tractable, in this paper, we restrict ourselves to using the energy-efficient cross QAM constellations to generate the two UFCPs ${\mathcal Y}_1\sim{\mathcal X}$ and ${\mathcal Y}_2\sim{\mathcal X}$ with ${\mathcal X}\subseteq \{-1, -j, 1, j\}$.  Specifically, let ${\mathcal Z}_1$ and ${\mathcal Z}_2$ be two given $2^p$-ary and $2^q$-ary cross QAM constellations, respectively.  Then, the three constellations ${\mathcal X}, {\mathcal Y}_1$ and ${\mathcal Y}_2$ are selected in such a way that ${\mathcal Z}_1=\frac{{\mathcal Y}_1}{\mathcal X}$ and ${\mathcal Z}_2=\frac{{\mathcal Y}_2}{\mathcal X}$. In addition, it is not difficult to verify that if ${\mathcal Z}_1=\frac{{\mathcal Y}_1}{\mathcal X}$ and ${\mathcal Z}_2=\frac{{\mathcal Y}_2}{\mathcal X}$, then, $\alpha{\mathcal Z}_1=\frac{\alpha{\mathcal Y}_1}{\mathcal X}$ and $\alpha{\mathcal Z}_2=\frac{\alpha{\mathcal Y}_2}{\mathcal X}$ for any positive $\alpha$. Therefore, a family of UFCP codes resulting from the cross QAM constellations and an energy scale $\alpha$  is characterized by
\begin{eqnarray}\label{eq:ufcp-stbc}
{\mathcal U}_\alpha({\mathcal X}, {\mathcal Y}_1, {\mathcal Y}_2)\!=\!\!\left\{\!\!{\mathbf U}_\alpha\!\!=\!\!\frac{1}{\sqrt{|x|^2+\alpha^2 |y_1|^2+\alpha^2 |y_2|^2}}\!\!
\left(\!\!\!%
\begin{array}{cc}
  x & 0\\
  0 & x^*\\
 \alpha y_1 &\alpha y_2  \\
  -\alpha y^*_2  & \alpha y_1^*
\end{array}%
\!\!\!\right):\,x\in{\mathcal X}, y_1\in{\mathcal Y}_1, y_2\in{\mathcal Y}_2\!\!\right\}.
\end{eqnarray}
By employing the code structure~\eqref{eq:ufcp-stbc} and performing some algebraic manipulations, the expression~\eqref{eq:def-gain} can be further simplified into
\begin{eqnarray}\label{eq:gain}
G_{\alpha}({\mathcal X}, {\mathcal Y}_1, {\mathcal Y}_2)
\!\!&=&\!\!\!\!\!\min_{(x, y_1, y_2)\ne (\hat{x}, \hat{y}_1, \hat{y}_2), x, \hat{x}\in{\mathcal X}, y_1, \hat{y}_1\in{\mathcal Y}_1, y_2, \hat{y}_2\in{\mathcal Y}_2}\frac{\alpha^2\Big(|\frac{y_1}{x}-\frac{\hat{y}_1}{\hat{x}}|^2
+|\frac{y_2}{x}-\frac{\hat{y}_2}{\hat{x}}|^2\Big)}
{(1+\alpha^2(|\frac{y_1}{x}|^2+|\frac{y_2}{x}|^2))(1+\alpha^2(|\frac{\hat{y}_1}{\hat{x}}|^2+|\frac{\hat{y}_2}{\hat{x}}|^2))}\nonumber\\
\!\!&=&\!\!\min_{(x, y_1, y_2)\ne (\hat{x}, \hat{y}_1, \hat{y}_2), x, \hat{x}\in{\mathcal X}, y_1, \hat{y}_1\in{\mathcal Y}_1, y_2, \hat{y}_2\in{\mathcal Y}_2} G_{\alpha}(x, {\mathbf y}; \hat{x}, \hat{\mathbf y}|{\mathcal X}, {\mathcal Y}_1, {\mathcal Y}_2),
\end{eqnarray}
where notation $G_{\alpha}(x, {\mathbf y}; \hat{x}, \hat{\mathbf y}|{\mathcal X}, {\mathcal Y}_1, {\mathcal Y}_2)$ is defined as
\begin{eqnarray}\label{def:obj}
G_{\alpha}(x, {\mathbf y}; \hat{x}, \hat{\mathbf y}|{\mathcal X}, {\mathcal Y}_1, {\mathcal Y}_2)=\frac{\alpha^2\Big(|\frac{y_1}{x}-\frac{\hat{y}_1}{\hat{x}}|^2
+|\frac{y_2}{x}-\frac{\hat{y}_2}{\hat{x}}|^2\Big)}
{(1+\alpha^2(|\frac{y_1}{x}|^2+|\frac{y_2}{x}|^2))(1+\alpha^2(|\frac{\hat{y}_1}{\hat{x}}|^2+|\frac{\hat{y}_2}{\hat{x}}|^2))},
\end{eqnarray}
which is called a \textit{coding gain function}. Our design problem is now formally stated as follows:
\begin{problem}\label{prob:design} Let $|{\mathcal U}_\alpha({\mathcal X}, {\mathcal Y}_1, {\mathcal Y}_2)|=2^r$ ($r\ge 4$) be fixed. Find an energy scale $\alpha$, three nonnegative integers $\delta, p$ and $q$ satisfying a total transmission bits constraint: $p+q-\delta=r$, and the unique factorizations of a pair of the $2^p$-ary and $2^q$-ary cross QAM constellations ${\mathcal Z}_1$ and ${\mathcal Z}_2$, i.e., ${\mathcal Z}_1={\mathcal Y}_1/{\mathcal X}$ and  
${\mathcal Z}_2={\mathcal Y}_2/{\mathcal X}$ with ${\mathcal X}\subseteq \{-1, -j, 1, j\}$ and $|{\mathcal X}|=2^{\delta}$, such that the coding gain $G_\alpha({\mathcal X}, {\mathcal Y}_1, {\mathcal Y}_2)$ is maximized, i.e.,
\begin{eqnarray}\label{eq:prob}
\big\{\widetilde\alpha, \widetilde\delta, \widetilde p, \widetilde q, \widetilde{\mathcal X}, \widetilde{\mathcal Y}_1, \widetilde{\mathcal Y}_2\big\}=\arg \max_{p+q-\delta=r}\max_{{\mathcal Z}_1={\mathcal Y}_1/{\mathcal X}, {\mathcal Z}_2={\mathcal Y}_2/{\mathcal X}} \max_{\alpha}G_{\alpha}({\mathcal X}, {\mathcal Y}_1, {\mathcal Y}_2).\nonumber
\end{eqnarray}
~\hfill\QED   
\end{problem}

\subsection{The Solution to Problem~\ref{prob:design}}\label{optimal}
In order to solve each individual optimization problem in Problem~\ref{prob:design}, we first introduce some notation for discussion simplicity. Recall that ${\mathcal Q}$ denotes the modified $2^K$-ary cross QAM defined in Definition~\ref{def:qam}. Let $P$ denote one of its corner points with the largest energy $E$, and let $P_1$ and $P_2$ denote the two nearest neighbors of this corner point, with the respective energies $E_{\rm s} $ and $E_{\rm t}$, where $E_{\rm s}\ge E_{\rm t}$. For given positive integers $u, v$ and $w$ satisfying $u+v=w$ with $u\ge v$, let ${\mathcal Q}_1$ and ${\mathcal Q}_2$ denote the respective modified  $2^u$-ary and $2^v$-ary cross QAM constellations; $Z_i$ denotes one of the corner point in ${\mathcal Q}_i$ with the largest energy $E_i$; Its two nearest neighbors are denoted by $Z_{i1}$ and $Z_{i2}$, respectively, with energies being $E_{i1}$ and $E_{i2}$, where $E_{i1}\ge E_{i2}$. Specifically, for a given positive integer $w\ge 4$, 
integers $\widetilde u$ and $\widetilde v$ are defined as follows:
  \begin{eqnarray}\label{eq:def-pq}
\Bigg\{
  \begin{array}{ll}
 {\widetilde u}={\widetilde v}=\frac{w}{2}   &~\mathrm{if~}w~{\rm is~even},\\
\widetilde u=\frac{w+1}{2}, \widetilde v=\frac{w-1}{2}  &~ \mathrm{if~} w~{\rm is~odd}. 
  \end{array} 
\end{eqnarray}   
Correspondingly, all notations  $\widetilde{\mathcal Q}_i, {\widetilde E}_{i1}$ and ${\widetilde E}_{i2}$ 
are defined in the same way as ${\mathcal Q}_i, E_{i1}$ and $E_{i2}$. As will be seen shortly in the following discussions on the solution to Problem~\ref{prob:design}, the integers $u, v$ and $w$ are to be replaced by $p, q$ and $r-\delta$, respectively, and thus, ${\mathcal Q}_i$ is to be regarded as ${\mathcal Z}_i$. Hence, the corresponding positive integers $\widetilde{p}$ and $\widetilde {q}$ and energy notation ${\widetilde E}_{i}^{(\delta)}, {\widetilde E}_{i1}^{(\delta)}$ and ${\widetilde E}_{i2}^{(\delta)}$ are used directly without the need of redefinitions.  
Some properties regrading these energies are collected as the following Lemma.
\begin{lemma}\label{lem:energy} For the modified $2^K$-ary cross QAM ${\mathcal Q}$, 
the following statements are true: 
\begin{enumerate}
\item  If $K=3$, then, each corner point has only one nearest neighbor, $E=10$ and $E_{\rm s}=2$.
\item If $K$ is even, then, 
$E=2 (2^{\frac{K}{2}}-1)^2$ and $E_{\rm s}=E_{\rm t}=(2^{\frac{K}{2}}-1)^2+(2^{\frac{K}{2}}-3)^2$.
\item If $K$ is odd and greater than $3$, then, $E=(2^{\frac{K-1}{2}}-1)^2+(3\times 2^{\frac{K-3}{2}}-1)^2$,\\$E_{\rm s}=(2^{\frac{K-1}{2}}-3)^2+(3\times 2^{\frac{K-3}{2}}-1)^2$ and $E_{\rm t}=(2^{\frac{K-1}{2}}-1)^2+(3\times 2^{\frac{K-3}{2}}-3)^2$.~\hfill\QED
\end{enumerate}
\end{lemma}
Lemma~\ref{lem:energy} can be verified directly by the calculation and thus, its proof is omitted.  Now, applying Lemma~\ref{lem:energy} to a pair of QAM constellations ${\mathcal Q}_1$ and ${\mathcal Q}_2$ yields Lemma~\ref{lem:energy-ineq}:  
\begin{lemma}\label{lem:energy-ineq} Let ${\mathcal Q}_1$ and ${\mathcal Q}_2$ denote the modified $2^u$-ary and $2^v$-ary QAM constellations, respectively, with $u\ge v\ge 2$. Then, 
\begin{subequations} 
\begin{eqnarray}
E_2+E_{11}&\le& E_1+E_{21},\label{eq:energy-ineq1}\\
E_2+E_{12}&\le& E_1+E_{22}\label{eq:energy-ineq2}.
\end{eqnarray}
\end{subequations}
~\hfill\QED
\end{lemma}
The proof of Lemma~\ref{lem:energy-ineq} is provided in Appdendix~\ref{appendix:lem:energy-ineq}.
\begin{lemma}\label{lem:energy-min} 
Let ${\mathcal Q}_1$ and ${\mathcal Q}_2$ denote the modified $2^u$-ary and $2^v$-ary QAM constellations, respectively, with $u\ge v\ge 2$. Then, the following four inequalities hold: 
\begin{subequations}\label{eq:energy-min}
\begin{eqnarray}
E_1+E_2&\ge& {\widetilde{E}_1+\widetilde{E}_2},\label{eq:energy-min1}\\
\sqrt{E_1+E_{21}}+\sqrt{E_1+E_{22}}&\ge& \sqrt{\widetilde{E}_1+\widetilde{E}_{21}}+\sqrt{\widetilde{E}_1+\widetilde{E}_{22}},\label{eq:energy-min2}\\
\sqrt{E_1+E_{2}}+\sqrt{E_1+E_{21}}&\ge& \sqrt{\widetilde{E}_1+\widetilde{E}_2}+\sqrt{\widetilde{E}_1+\widetilde{E}_{21}},\label{eq:energy-min3}\\
\sqrt{E_1+E_{21}}+\sqrt{E_2+E_{11}}&\ge& \sqrt{\widetilde{E}_1+\widetilde{E}_{21}}+\sqrt{\widetilde{E}_2+\widetilde{E}_{11}}.\label{eq:energy-min4}
\end{eqnarray}
\end{subequations}
~\hfill\QED
\end{lemma}
The proof of Lemma~\ref{lem:energy-min} is postponed to Appendix~\ref{appendix:lem:energy-min}. We also need the following lemma, whose proof is given in Appendix~\ref{appendix:lem:energy-ineq-delta2}.   
\begin{lemma}\label{lem:energy-ineq-delta2}
Let $\mathcal{\widetilde{Q}}_1$ and $\mathcal{ \widetilde{Q}}_2$ represent the modified $2^{\widetilde{u}}$-ary and $2^{\widetilde{v}}$-ary QAM constellations with $\widetilde{u}$ and $\widetilde{v}$ defined in~\eqref{eq:def-pq}. Then,
\begin{subequations}
\begin{eqnarray}\label{eq:energy-ineq-delta2}
\widetilde{E}_1+\widetilde{E}_{22} &\le& \widetilde{E}_2+\widetilde{E}_{11},
~~\mathrm{when}~w~\mathrm{is~even},\\
\widetilde{E}_1+\widetilde{E}_{22} &\ge& \widetilde{E}_2+\widetilde{E}_{11},
~~\mathrm{when}~w~\mathrm{is~odd}.
\end{eqnarray}
\end{subequations}
~\hfill\QED
\end{lemma}
Now, it is time to start solving Problem~\ref{prob:design}. Let us consider the following three cases of all the possible values of $\delta$. 
\subsubsection{$\delta=0$}\label{subsec:delta0}
In this case, the constellation ${\mathcal X}$ includes only one element, i.e., ${\mathcal X}=\{x\}$, where $x=\pm 1, \pm j$. Since ${\mathbf r}={\mathbf U}{\mathbf h}+{\boldsymbol\xi}$ and matrix ${\rm diag}(x^*{\mathbf I}_2, {\mathbf I}_2)$ is unitary, we have  ${\rm diag}(x^*{\mathbf I}_2, {\mathbf I}_2){\mathbf r}={\rm diag}(x^*{\mathbf I}_2, {\mathbf I}_2){\mathbf U}{\mathbf h}+{\rm diag}(x^*{\mathbf I}_2, {\mathbf I}_2){\boldsymbol\xi}$, with ${\rm diag}(x^*{\mathbf I}_2, {\mathbf I}_2){\boldsymbol\xi}$ having the same statistical property as ${\boldsymbol\xi}$ and 
\begin{eqnarray}
{\rm diag}(x^*{\mathbf I}_2, {\mathbf I}_2){\mathbf U}_\alpha=\frac{1}{\sqrt{1+\alpha^2(|y_1|^2+|y_2|^2)}}\left(\begin{array}{cccc}
1 & 0 \\
0 & 1 \\
\alpha y_1 & \alpha y_2 \\
-\alpha y_2^* & \alpha y_1^* 
\end{array}\right).
\end{eqnarray}
Therefore, we only need to consider the case where $x=1$. In this situation, ${\mathcal Z}_1={\mathcal Y}_1, {\mathcal Z}_2={\mathcal Y}_2$ and $G_{\alpha}(x, {\mathbf y}; \hat{x}, \hat{\mathbf y}|{\mathcal X}, {\mathcal Y}_1, {\mathcal Y}_2)$ is reduced to  
\begin{eqnarray}\label{eq:obj-train}
G_{\alpha}(1, {\mathbf y}; 1, \hat{\mathbf y}|{\mathcal X}, {\mathcal Y}_1, {\mathcal Y}_2)=\frac{\alpha^2 (|y_1-\hat{y}_1|^2
+|y_2-\hat{y}_2|^2)}
{(1+\alpha^2(|y_1|^2+|y_2|^2))(1+\alpha^2(|\hat{y}_1|^2+|\hat{y}_2|^2))}.
\end{eqnarray}
Let us first consider a special case where $|{\mathcal Z}_2|=4$. In this case, $|y_2|=|\hat{y}_2|=2$ and hence,
\begin{eqnarray}
G_{\alpha}(1, {\mathbf y};1, \hat{\mathbf y}|{\mathcal X}, {\mathcal Y}_1, {\mathcal Y}_2)=\frac{\alpha^2(|y_1-\hat{y}_1|^2
+|y_2-\hat{y}_2|^2)}
{(1+\alpha^2(|y_1|^2+2))(1+\alpha^2(|\hat{y}_1|^2+2))}.
\end{eqnarray}
It is very interesting to observe that for any fixed $\alpha, {\mathcal X}$ and ${\mathcal Y}_i$, the objection function $G_{\alpha}(x, {\mathbf y}; \hat{x}, \hat{\mathbf y})|{\mathcal X}, {\mathcal Y}_1, {\mathcal Y}_2)$ with respect  to the variables  $x, \hat{x}, {\mathbf y}$ and $\hat{\mathbf y}$ is minimized when its numerator achieves the minimum and simultaneously, its denominator achieves the maximum,  both optimums being achieved when  $y_2$ is the nearest neighbor of $\hat{y}_2$ and $y_1=\hat{y}_1=Z_1$ is the corner point with the largest energy $E_1$ in ${\mathcal Z}_1$. Therefore, the minimum in this case is given by 
\begin{eqnarray}\label{eq:spe}
G_{\alpha}({\mathcal X}, {\mathcal Y}_1, {\mathcal Y}_2)=\min_{{\mathbf y}\ne \hat{\mathbf y}}G_{\alpha}(1, {\mathbf y}; 1, \hat{\mathbf y}|{\mathcal X}, {\mathcal Y}_1, {\mathcal Y}_2)=4\alpha/(1+\alpha(2+E_1))^2. 
\end{eqnarray}
Now, we consider a general case where $|{\mathcal Z}_2|\ge 8$. In this case, notice that
\begin{eqnarray}\label{eq:min}
G_{\alpha}({\mathcal X}, {\mathcal Y}_1, {\mathcal Y}_2)=\min\{\min_{(|y_1|, |y_2|)= (|\hat{y}_1|, |\hat{y}_2|), {\rm either}\,|y_1|=\sqrt{2}\, {\rm or}\, |y_2|=\sqrt{2}} G_{\alpha}(x, {\mathbf y}; \hat{x}, \hat{\mathbf y}|{\mathcal X}, {\mathcal Y}_1, {\mathcal Y}_2), \nonumber\\
\min_{(|y_1|, |y_2|)= (|\hat{y}_1|, |\hat{y}_2|), |y_1|>\sqrt{2}, |y_2|>\sqrt{2}} G_{\alpha}(x, {\mathbf y}; \hat{x}, \hat{\mathbf y}|{\mathcal X}, {\mathcal Y}_1, {\mathcal Y}_2), \nonumber\\
\min_{(|y_1, |y_2|)\ne (|\hat{y}_1|, |\hat{y}_2|)} G_{\alpha}(x, {\mathbf y}; \hat{x}, \hat{\mathbf y}|{\mathcal X}, {\mathcal Y}_1, {\mathcal Y}_2)\}.
\end{eqnarray}
Realizing the following facts is key to obtaining the minimum of the objection function~\eqref{eq:obj-train} with respect to variables $x, \hat{x}, {\mathbf y}$ and $\hat{\mathbf y}$:
\begin{enumerate}
 \item [(a)]Under the conditions that ${\mathbf y}\ne\hat{\mathbf y}, (|y_1|, |y_2|)= (|\hat{y}_1|, |\hat{y}_2|)$ and either $|y_1|=\sqrt{2}$ or $|y_2|=\sqrt{2}$, the numerator of the objection function~\eqref{eq:obj-train} is lower-bounded by $4\alpha^2$, i.e., 
 \begin{subequations}
 \begin{eqnarray}\label{eq:lbound-train}
 \alpha^2(|y_1-\hat{y}_1|^2
+|y_2-\hat{y}_2|^2)\ge 4\alpha^2,
 \end{eqnarray}
where the equality holds when either $y_2$ is the nearest neighbor of $\hat{y}_2$, i.e., $|y_2-\hat{y}_2|=2$, and $y_1=\hat{y}_1$ or $y_1$ is the nearest neighbor of $\hat{y}_1$ and $y_2=\hat{y}_2$.  The denominator of the objection function~\eqref{eq:obj-train} is upper-bounded by  
 \begin{eqnarray}\label{eq:ubound-train}
(1+\alpha^2(|y_1|^2+|y_2|^2))(1+\alpha^2(|\hat{y}_1|^2+|\hat{y}_2|^2))\le  (1+\alpha^2(E_1+2))^2,
 \end{eqnarray}
 \end{subequations}
where the equality holds when both $y_1$ and $\hat{y}_1$ are the corner points in ${\mathcal Z}_1$ with each having the largest energy $E_1$. Therefore, both equalities in~\eqref{eq:lbound-train} and~\eqref{eq:ubound-train} hold simultaneously when both $y_1$ and $\hat{y}_1$ are the corner points in ${\mathcal Z}_1$ with each having the largest energy $E_1$ and $y_2$ is the nearest neighbor of $\hat{y}_2$.
 \item [(b)] Under the conditions that $(|y_1|, |y_2|)= (|\hat{y}_1|, |\hat{y}_2|), |y_1|>\sqrt{2}, |y_2|>\sqrt{2}$ and ${\mathbf y}\ne\hat{\mathbf y}$, the numerator of the objection function~\eqref{eq:obj-train} is lower-bounded by  
 \begin{subequations}
 \begin{eqnarray}\label{eq:condt2-lbound-train}
 \alpha^2(|y_1-\hat{y}_1|^2
+|y_2-\hat{y}_2|^2)\ge \Bigg\{
  \begin{array}{ll}
  8\alpha^2 &~ \mathrm{if~}|y_1|> E_{11}~{\rm and}~|y_2|> E_{21},\\  
  4\alpha^2 &~ \mathrm{if~either}~|y_1|\le E_{11}~{\rm or}~|y_2|\le E_{21}.
    \end{array}
 \end{eqnarray}
The denominator of the objection function~\eqref{eq:obj-train} is upper-bounded by  
 \begin{eqnarray}\label{eq:condt2-ubound-train}
(1+\alpha^2(|y_1|^2+|y_2|^2))(1+\alpha^2(|\hat{y}_1|^2+|\hat{y}_2|^2))\nonumber\\
\le \Bigg\{
  \begin{array}{ll}
   (1+\alpha^2(E_1+E_2))^2\quad\mathrm{if~}|y_1|> E_{11}~{\rm and}~|y_2|> E_{21},\\  
  (1+\alpha^2(E_1+E_{21}))^2\quad\mathrm{if~either}~|y_1|\le E_{11}~{\rm or}~|y_2|\le E_{21}.  
  \end{array}
 \end{eqnarray}
 \end{subequations}
\item [(c)] If $|{\mathcal Z}_1|\ge |{\mathcal Z}_2|$ and $(|y_1|, |y_2|)\ne (|\hat{y}_1|, |\hat{y}_2|)$, then, $E_1\ge E_2$. Note that the numerator of the objection function~\eqref{eq:obj-train} is lower-bounded by  
 \begin{subequations}
 \begin{eqnarray}\label{eq:condt3-lbound-train}
 \alpha^2(|y_1-\hat{y}_1|^2
+|y_2-\hat{y}_2|^2)\ge 4\alpha^2,
 \end{eqnarray}
where the equality holds when either $y_2$ is the nearest neighbor of $\hat{y}_2$ and $y_1=\hat{y}_1$ or $y_1$ is the nearest neighbor of $\hat{y}_1$ and $y_2=\hat{y}_2$. In addition, the denominator of the objection function~\eqref{eq:obj-train} is upper-bounded by  
 \begin{eqnarray}\label{eq:condt3-ubound-train}
(1+\alpha^2(|y_1|^2+|y_2|^2))(1+\alpha^2(|\hat{y}_1|^2+|\hat{y}_2|^2))\le  (1+\alpha^2(E_1+E_{2}))(1+\alpha^2(E_1+E_{21})),\nonumber
\\
\end{eqnarray}
 \end{subequations}
 where the equality holds when  both $y_1$ and $\hat{y}_1$ are the corner points in ${\mathcal Z}_1$ with each having the largest energy $E_1$, while one of $y_2$ and $\hat{y}_2$  is the corner point in ${\mathcal Z}_2$ with the largest energy $E_2$, and the other is of the second largest energy $E_{21}$. In other words, the upper bound $(1+\alpha^2(E_1+E_2))(1+\alpha^2(E_1+E_{21}))$ is the second largest maximum of the denominator of the objection function~\eqref{eq:obj-train}. Very interestingly, under this condition, both equalities in~\eqref{eq:condt3-lbound-train} and~\eqref{eq:condt3-ubound-train} are able to be achieved at the same time when  $y_1=\hat{y}_1$ is the corner point having the largest energy $E_1$  in ${\mathcal Z}_1$, $y_2$ is the nearest neighbor of $\hat{y}_2$ and of the largest energy $E_2$  and $\hat{y}_2$ has the second largest energy $E_{21}$ in ${\mathcal Z}_2$.
 \end{enumerate}
The above three observations reveal that
\begin{subequations}
\begin{eqnarray}\label{eq:min1}
\min_{(|y_1|, |y_2|)= (|\hat{y}_1|, |\hat{y}_2|), |y_1|=\sqrt{2}\, {\rm or}\, |y_2|=\sqrt{2}} G_{\alpha}(1, {\mathbf y}; 1, \hat{\mathbf y}|{\mathcal X}, {\mathcal Y}_1, {\mathcal Y}_2)\nonumber\\
=\frac{4\alpha^2}{(1+\alpha^2(E_1+2))^2},
\end{eqnarray}
\begin{eqnarray}\label{eq:min2}
\min_{(|y_1|, |y_2|)= (|\hat{y}_1|, |\hat{y}_2|), |y_1|>\sqrt{2}, |y_2|>\sqrt{2}} G_{\alpha}(1, {\mathbf y}; 1, \hat{\mathbf y}|{\mathcal X}, {\mathcal Y}_1, {\mathcal Y}_2)\nonumber\\
\ge\frac{4\alpha^2}{(1+\alpha^2(E_1+E_{21}))^2},
\end{eqnarray}
\begin{eqnarray}\label{eq:min3}
\min_{(|y_1|, |y_2|)\ne (|\hat{y}_1|, |\hat{y}_2|)} G_{\alpha}(1, {\mathbf y}; 1, \hat{\mathbf y}|{\mathcal X}, {\mathcal Y}_1, {\mathcal Y}_2)\nonumber\\
=\frac{4\alpha^2}{(1+\alpha^2(E_1+E_2))(1+\alpha^2(E_1+E_{21}))}.
\end{eqnarray}
\end{subequations}
When $|{\mathcal Z}_2|>4$, $E_1>E_{21}\ge 2$ and as a result, comparing~\eqref{eq:min1} with~\eqref{eq:min3} gives us  
 \begin{eqnarray}
 \frac{4\alpha^2}{(1+\alpha^2(E_1+2))^2}> \frac{4\alpha^2}{(1+\alpha^2(E_1+E_2))(1+\alpha^2(E_1+E_{21}))}\nonumber
  \end{eqnarray}   
and thus, we have 
\begin{subequations}\label{eq:sub-min}
\begin{eqnarray}
\min_{(|y_1|, |y_2|)= (|\hat{y}_1|, |\hat{y}_2|), |y_1|=\sqrt{2}\, {\rm or}\, |y_2|=\sqrt{2}} G_{\alpha}(x, {\mathbf y}; \hat{x}, \hat{\mathbf y}|{\mathcal X}, {\mathcal Y}_1, {\mathcal Y}_2)\nonumber\\
>\min_{(|y_1|, |y_2|)\ne (|\hat{y}_1|, |\hat{y}_2|)} G_{\alpha}(x, {\mathbf y}; \hat{x}, \hat{\mathbf y}|{\mathcal X}, {\mathcal Y}_1, {\mathcal Y}_2).
\end{eqnarray}
In addition, since $E_2>E_{21}$, we can obtain from~\eqref{eq:min2} and~\eqref{eq:min3} that
\begin{eqnarray}
\min_{(|y_1|, |y_2|)= (|\hat{y}_1|, |\hat{y}_2|), |y_1|>\sqrt{2}, |y_2|>\sqrt{2}} G_{\alpha}(x, {\mathbf y}; \hat{x}, \hat{\mathbf y}|{\mathcal X}, {\mathcal Y}_1, {\mathcal Y}_2)\nonumber\\
>\min_{(|y_1|, |y_2|)\ne (|\hat{y}_1|, |\hat{y}_2|)} G_{\alpha}(x, {\mathbf y}; \hat{x}, \hat{\mathbf y}|{\mathcal X}, {\mathcal Y}_1, {\mathcal Y}_2).
\end{eqnarray}
\end{subequations}
Combining~\eqref{eq:min} with~\eqref{eq:sub-min} yields 
\begin{eqnarray}\label{eq:gen}
G_{\alpha}({\mathcal X}, {\mathcal Y}_1, {\mathcal Y}_2)=\frac{4\alpha^2}{(1+\alpha^2(E_1+E_2))(1+\alpha^2(E_1+E_{21}))}.
\end{eqnarray}
Since if $|{\mathcal Z}_2|=4$, $E_2=E_{21}$, the conclusion~\eqref{eq:gen} includes~\eqref{eq:spe} as a special case. Therefore, no matter whether or not $|{\mathcal Z}_2|>4$, we always have~\eqref{eq:gen}.    
Notice that~\eqref{eq:gen} can be rewritten as 
\begin{subequations}
\begin{eqnarray}
G_{\alpha}({\mathcal X}, {\mathcal Y}_1, {\mathcal Y}_2)&=&\frac{4}{(\alpha^{-1}+\alpha(E_1+E_2))(\alpha^{-1}+\alpha(E_1+E_{21}))}\nonumber\\
&=&\frac{4}{\alpha^{-2}+\alpha^2(E_1+E_2)(E_1+E_{21})+(2E_1+E_2+E_{21})}\label{eq:a-mean}\\
&\le&\frac{4}{2\sqrt{(E_1+E_2)(E_1+E_{21})}+(2E_1+E_2+E_{21})}\label{eq:g-mean}\\
&=&\frac{4}{\big(\sqrt{E_1+E_2}+\sqrt{E_1+E_{21}}\big)^2}\nonumber,
\end{eqnarray}
\end{subequations}
where we have used the arithmetic mean and the geometric mean inequality from~\eqref{eq:a-mean} to~\eqref{eq:g-mean}: $a+b\geq2\sqrt{ab}$, and the equality holds when $\alpha=1/\sqrt[4]{(E_1+E_2)(E_1+E_{21})}$. Thus, we have 
\begin{eqnarray}\label{eq:gain-delta0}
\max_{\alpha}G_{\alpha}({\mathcal X}, {\mathcal Y}_1, {\mathcal Y}_2)&=&\frac{4}{\big(\sqrt{E_1+E_2}+\sqrt{E_1+E_{21}}\big)^2}\nonumber\\
&\le &\frac{4}{\big(\sqrt{\widetilde{E}^{(0)}_1+\widetilde{E}^{(0)}_2}+\sqrt{\widetilde{E}^{(0)}_1+\widetilde{E}^{(0)}_{21}}\big)^2},
\end{eqnarray}
where we have used Lemma~\ref{lem:energy-min} with $w=r, u=p, v=q$ and ${\mathcal Q}_i={\mathcal Z}_i$ and the inequality in~\eqref{eq:gain-delta0} holds when $p=\widetilde p$ and $q=\widetilde q$. All the above discussions can be summarized as Property~\ref{pro:delta0}.  
\begin{property}\label{pro:delta0}
When $p\ge q$ and $\delta=0$, the optimal solution to Problem~\ref{prob:design} is given as follows:
\begin{enumerate}
 \item [(1)] If $r$ is even, then, $\widetilde{p}=\widetilde{q}=r/2$.
 \item [(2)]If $r$ is odd, then, $\widetilde{p}=(r+1)/2, \widetilde{q}=(r-1)/2$.
\end{enumerate}
Once the optimal $\widetilde{p}$ and $\widetilde{q}$ have been determined, then, $\widetilde{\mathcal X}^{(0)}=\{1\}$, $\widetilde{\mathcal Y}_1^{(0)}$ is the $2^{\widetilde p}$-ary QAM constellation $\widetilde{\mathcal Z}_1$, $\widetilde{\mathcal Y}_2^{(0)}$ is the $2^{\widetilde q}$-ary QAM constellation  $\widetilde{\mathcal Z}_2$ and the optimal energy scale $\widetilde{\alpha}$ is determined by 
\begin{eqnarray}
\widetilde{\alpha}=\frac{1}{\sqrt[4]{(\widetilde{E}^{(0)}_1+\widetilde{E}^{(0)}_2)(\widetilde{E}^{(0)}_1+\widetilde{E}^{(0)}_{21})}}.
\end{eqnarray}
Moreover, the optimal coding gain is given by
\begin{equation}
G(\widetilde{\mathcal X}^{(0)}, \widetilde{\mathcal Y}_1^{(0)},\widetilde{\mathcal Y}_2^{(0)})=\frac{4}{\Big(\sqrt{\widetilde{E}^{(0)}_1+\widetilde{E}^{(0)}_2}+\sqrt{\widetilde{E}^{(0)}_1+\widetilde{E}^{(0)}_{21}}\,\Big)^2}.
\end{equation}
~\hfill\QED
\end{property}
In fact, Property~\ref{pro:delta0} gives us an optimal design of the unitary training Alamouti code based on the cross QAM constellations.  
\subsubsection{$\delta=1$}\label{subsec:delta1} In this case, there are in total six possibilities in choosing the constellation ${\mathcal X}$, i.e., ${\mathcal X}=\{1, -1\},{\mathcal X}=\{1, j\}, {\mathcal X}=\{1, -j\},{\mathcal X}=\{-1, j\},{\mathcal X}=\{-1, -j\}, {\mathcal X}=\{j, -j\}$. In order to attain the optimal solution, we  
take the following two steps. 

{\bf Step 1}: Determine $G_{\alpha}({\mathcal X}^{(1)}, {\mathcal Y}^{(1)}_1, {\mathcal Y}^{(1)}_2)=\min_{(x, {\mathbf y}^T)\ne (\hat{x}, \hat{\mathbf y}^T)} G_{\alpha}(x, {\mathbf y}; \hat{x}, \hat{\mathbf y}|{\mathcal X}^{(1)}, {\mathcal Y}^{(1)}_1, {\mathcal Y}^{(1)}_2)$, where constellations ${\mathcal X}^{(1)}$ and ${\mathcal Y}^{(1)}_i$ are derived from Proposition~\ref{pro:ufcp-qam} with ${\mathcal Z}={\mathcal Z}_i$, i.e., ${\mathcal X}^{(1)}=\{1, j\}$ and ${\mathcal Y}^{(1)}_i={\mathcal Y}_{i, \rm opt}$. To do that, we first represent $G_{\alpha}({\mathcal X}^{(1)}, {\mathcal Y}^{(1)}_1, {\mathcal Y}^{(1)}_2)$ as
\begin{eqnarray}\label{eq:opt-step1}
G_{\alpha}({\mathcal X}^{(1)}, {\mathcal Y}^{(1)}_1, {\mathcal Y}^{(1)}_2)=\min\Big\{\min_{(x, {\mathbf y}^T)\ne (\hat{x}, \hat{\mathbf y}^T), x=\hat{x}} G_{\alpha}(x, {\mathbf y}; \hat{x}, \hat{\mathbf y}|{\mathcal X}^{(1)}, {\mathcal Y}^{(1)}_1, {\mathcal Y}^{(1)}_2), \nonumber\\
\min_{(x, {\mathbf y}^T)\ne (\hat{x}, \hat{\mathbf y}^T), x\ne\hat{x}} G_{\alpha}(x, {\mathbf y}; \hat{x}, \hat{\mathbf y}|{\mathcal X}^{(1)}, {\mathcal Y}^{(1)}_1, {\mathcal Y}^{(1)}_2)\Big\},
\end{eqnarray}
which leads us to individually considering the following two optimization problems.
\begin{enumerate}
 \item [(a)] $x=\hat{x}$. In this case, the objection function in~\eqref{eq:gain} is simplified into 
 \begin{eqnarray}
  G_{\alpha}(x, {\mathbf y}; \hat{x}, \hat{\mathbf y}|{\mathcal X}^{(1)}, {\mathcal Y}^{(1)}_1, {\mathcal Y}^{(1)}_2)=\frac{\alpha^2(|y_1-\hat{y}_1|^2
+|y_2-\hat{y}_2|^2)}
{(1+\alpha^2(|y_1|^2+|y_2|^2))(1+\alpha^2(|\hat{y}_1|^2+|\hat{y}_2|^2))},
  \end{eqnarray}
  since $|x|=|\hat{x}|=1$. Following the discussion very similar to the case when $\delta=0$, we can obtain 
  \begin{eqnarray}\label{eq:opt-step11}
&&  \min_{(x, {\mathbf y}^T)\ne (\hat{x}, \hat{\mathbf y}^T), x=\hat{x}}G_{\alpha}(x, {\mathbf y}; \hat{x}, \hat{\mathbf y}|{\mathcal X}^{(1)}, {\mathcal Y}^{(1)}_1, {\mathcal Y}^{(1)}_2)\nonumber\\
&&=\left\{
  \begin{array}{ll}
   \frac{8\alpha^2}{(1+\alpha^2({E}_1+{E}_2))^2}&~ \mathrm{if~}q=5~{\rm or~} q=2~{\rm or}~q=4~{\rm and}~p=5, \\  
    \frac{8\alpha^2}{(1+\alpha^2({E}_1+{E}_2))(1+\alpha^2({E}_1+{E}_{21}))}&~ \mathrm{if~}q=3,\\     
    \frac{8\alpha^2}{(1+\alpha^2({E}_1+{E}_{21}))^2}&~\mathrm{if~}q~{\rm is~even~and~greater~than~2}~{\rm and}~p\ne 5,\\
  \frac{8\alpha^2}{(1+\alpha^2({E}_1+{E}_{21}))(1+\alpha^2({E}_1+{E}_{22}))}&~ \mathrm{if~} q~{\rm is~an~odd~integer~exceeding~5}. 
  \end{array}\right. 
\end{eqnarray}  
 \item [(b)] $x\ne\hat{x}$.  Under this condition, we further split the feasible domain of the optimization problem: $\min_{(x, {\mathbf y}^T)\ne (\hat{x}, \hat{\mathbf y}^T), x\ne\hat{x}} G_{\alpha}(x, {\mathbf y}; \hat{x}, \hat{\mathbf y}|{\mathcal X}^{(1)}, {\mathcal Y}^{(1)}_1, {\mathcal Y}^{(1)}_2)$, into four disjoint sub-domians as follows: 
 \begin{eqnarray}\label{eq:domain-split}
 {\mathcal D}_{11}=\{(x, {\mathbf y}^T, \hat{x}, \hat{\mathbf y}^T): x\ne\hat{x}, (|y_1|, |\hat{y}_1|)=(E_1, E_1), (|y_2|, |\hat{y}_2|)=(E_2, E_2)\},\nonumber\\
 {\mathcal D}_{12}=\{(x, {\mathbf y}^T, \hat{x}, \hat{\mathbf y}^T): x\ne\hat{x}, (|y_1|, |\hat{y}_1|)=(E_1, E_1), (|y_2|, |\hat{y}_2|)\ne(E_2, E_2)\},\nonumber\\
 {\mathcal D}_{13}=\{(x, {\mathbf y}^T, \hat{x}, \hat{\mathbf y}^T): x\ne\hat{x}, (|y_1|, |\hat{y}_1|)\ne(E_1, E_1), (|y_2|, |\hat{y}_2|)=(E_2, E_2)\},\nonumber\\  
  {\mathcal D}_{14}=\{(x, {\mathbf y}^T, \hat{x}, \hat{\mathbf y}^T): x\ne\hat{x}, (|y_1|, |\hat{y}_1|)\ne(E_1, E_1), (|y_2|, |\hat{y}_2|)\ne(E_2, E_2)\}.\nonumber 
  \end{eqnarray}
 Therefore, we have
\begin{eqnarray}\label{eq:opt-step12-min}
\min_{(x, {\mathbf y}^T)\ne (\hat{x}, \hat{\mathbf y}^T), x\ne\hat{x}} G_{\alpha}(x, {\mathbf y}; \hat{x}, \hat{\mathbf y}|{\mathcal X}^{(1)}, {\mathcal Y}^{(1)}_1, {\mathcal Y}^{(1)}_2)\nonumber\\
=\min_{1\le k\le 4}\min_{(x, {\mathbf y}^T, \hat{x}, \hat{\mathbf y}^T)\in{\mathcal D}_{1k}} G_{\alpha}(x, {\mathbf y}; \hat{x}, \hat{\mathbf y}|{\mathcal X}^{(1)}, {\mathcal Y}^{(1)}_1, {\mathcal Y}^{(1)}_2).
\end{eqnarray} 
Now, let us first consider each inner minimization problem.
\begin{enumerate}
 \item When $(x, {\mathbf y}^T, \hat{x}, \hat{\mathbf y}^T)\in{\mathcal D}_{11}\cup{\mathcal D}_{12}$ and $|{\mathcal Z}_1|\ge 8$, $|\frac{y_1}{x}-\frac{\hat{y}_1}{\hat{x}}|\ge\sqrt{20}$ and as a result, the numerator of $G_{\alpha}(x, {\mathbf y}; \hat{x}, \hat{\mathbf y}|{\mathcal X}^{(1)}, {\mathcal Y}^{(1)}_1, {\mathcal Y}^{(1)}_2)$ is lower bounded by   
 \begin{eqnarray}\label{eq:lb-delta1}
 \Big|\frac{y_1}{x}-\frac{\hat{y}_1}{\hat{x}}\Big|^2+\Big|\frac{y_2}{x}-\frac{\hat{y}_2}{\hat{x}}\Big|^2\ge 24\alpha^2.
 \end{eqnarray}
Under the same condition, the denominator of $G_{\alpha}(x, {\mathbf y}; \hat{x}, \hat{\mathbf y}|{\mathcal X}^{(1)}, {\mathcal Y}^{(1)}_1, {\mathcal Y}^{(1)}_2)$ is upper bounded by  
 \begin{eqnarray}\label{eq:ub-delta1}
\Big(1+\big|\frac{y_1}{x}\big|^2+\big|\frac{y_2}{x}\big|^2\Big)\Big(1+\big|\frac{\hat{y}_1}{\hat{x}}\big|^2+\big|\frac{\hat{y}_2}{\hat{x}}\big|^2\Big)\le (1+\alpha^2(E_1+E_2))^2.
 \end{eqnarray}
Combining~\eqref{eq:lb-delta1} with~\eqref{eq:ub-delta1} results in
\begin{eqnarray}\label{eq:a}
\min_{(x, {\mathbf y}^T, \hat{x}, \hat{\mathbf y}^T)\in{\mathcal D}_{11}\cup{\mathcal D}_{12}} G_{\alpha}(x, {\mathbf y}; \hat{x}, \hat{\mathbf y}|{\mathcal X}^{(1)}, {\mathcal Y}^{(1)}_1, {\mathcal Y}^{(1)}_2)
\ge\frac{24\alpha^2}{(1+\alpha^2(E_1+E_2))^2}. 
\end{eqnarray}
 \item When $(x, {\mathbf y}^T, \hat{x}, \hat{\mathbf y}^T)\in{\mathcal D}_{14}$, we first notice that the numerator of the objective is lower bounded by 
 \begin{eqnarray}\label{eq:lb-b}
 \Big|\frac{y_1}{x}-\frac{\hat{y}_1}{\hat{x}}\Big|^2+\Big|\frac{y_2}{x}-\frac{\hat{y}_2}{\hat{x}}\Big|^2\ge 8\alpha^2,
 \end{eqnarray}
since $\big|\frac{y_1}{x}-\frac{\hat{y}_1}{\hat{x}}\big|^2\ge 4$ and $\big|\frac{y_2}{x}-\frac{\hat{y}_2}{\hat{x}}\big|^2\ge 4$. In addition, the denominator of the objective is upper bounded by
\begin{eqnarray}\label{eq:ub-b}
 \Big(1+\big|\frac{y_1}{x}\big|^2+\big|\frac{y_2}{x}\big|^2\Big)\Big(1+\big|\frac{\hat{y}_1}{\hat{x}}\big|^2+\big|\frac{\hat{y}_2}{\hat{x}}\big|^2\Big)\nonumber\\
 \le (1+\alpha^2(E_1+E_{21}))(1+\alpha^2(E_{11}+E_2)).
 \end{eqnarray}
Now, let us argue that both equalities can be achieved simultaneously. Since $x, \hat{x}\in{\mathcal X}=\{1, j\}$, without loss of generality, we can always assume that $x=1$ and $\hat{x}=j$. Recall the definition and properties of the Group-$x$ which we have discussed in Section~\ref{sec:ufcp}. If we use ${\mathcal Z}_{i, 1}$ and ${\mathcal Z}_{i, j}$ to denote two groups generated by ${\mathcal Y}_i\sim{\mathcal X}$ for $i=1, 2$, then, we have ${\mathcal Z}_{i, 1}\cup{\mathcal Z}_{i, j}={\mathcal Z}_i$ and ${\mathcal Z}_{i, 1}\cap{\mathcal Z}_{i, j}=\Phi$. Also, recall that notation $Z_i$ denotes one of the corner points in ${\mathcal Z}_i$ with the largest energy $E_i$,  and notation $Z_{i1}$ and $Z_{i2}$ denote its two nearest neighbors in ${\mathcal Z}_i$, i.e.,  $|Z_i-Z_{i1}|=|Z_i-Z_{i2}|=2$, with the respective energies $E_{i1}$ and $E_{i2}$. 
Then, $Z_{i1}\in{\mathcal Z}_{i,j}$ for $i=1, 2$, since ${\mathcal Z}_{i, 1}={\mathcal Y}_i$ and $d_{\min}({\mathcal Y}_i)>2$. Hence, if we set $y_i=Z_{i}$ and $\hat{y}_i=j Z_{i1}$, then, both the equalities in~\eqref{eq:lb-b} and~\eqref{eq:ub-b} hold at the same time and thus,
\begin{eqnarray}\label{eq:b}
\min_{(x, {\mathbf y}^T, \hat{x}, \hat{\mathbf y}^T)\in{\mathcal D}_{14}} G_{\alpha}(x, {\mathbf y}; \hat{x}, \hat{\mathbf y}|{\mathcal X}^{(1)}, {\mathcal Y}^{(1)}_1, {\mathcal Y}^{(1)}_2)\nonumber\\
=\frac{8\alpha^2}{(1+\alpha^2(E_1+E_{21}))(1+\alpha^2(E_{11}+E_2))}. 
\end{eqnarray}
\item  When $(x, {\mathbf y}^T, \hat{x}, \hat{\mathbf y}^T)\in{\mathcal D}_{13}$, we need to consider two possibilities: $|{\mathcal Z}_2|=4$ and $|{\mathcal Z}_2|\ge 8$. If $|{\mathcal Z}_2|\ge 8$, then, following the discussion similar to Situation a), we obtain 
\begin{eqnarray}\label{eq:lb-c}
\min_{(x, {\mathbf y}^T, \hat{x}, \hat{\mathbf y}^T)\in{\mathcal D}_{13}} G_{\alpha}(x, {\mathbf y}; \hat{x}, \hat{\mathbf y}|{\mathcal X}^{(1)}, {\mathcal Y}^{(1)}_1, {\mathcal Y}^{(1)}_2)
\ge\frac{16\alpha^2}{(1+\alpha^2(E_1+E_2))^2}. 
\end{eqnarray}
If $|{\mathcal Z}_2|=4$, then, $|y_2|=|\hat{y}_2|=2$ and $|y_2/x-\hat{y}_2/\hat{x}|=2$. Following the same discussion as Situation b) and choosing $y_1= Z_{1}, \hat{y}_1=j Z_{11}$, we have 
\begin{eqnarray}\label{eq:min-c}
\min_{(x, {\mathbf y}^T, \hat{x}, \hat{\mathbf y}^T)\in{\mathcal D}_{13}} G_{\alpha}(x, {\mathbf y}; \hat{x}, \hat{\mathbf y}|{\mathcal X}^{(1)}, {\mathcal Y}^{(1)}_1, {\mathcal Y}^{(1)}_2)\nonumber\\
=\frac{8\alpha^2}{(1+\alpha^2(E_1+2))(1+\alpha^2(E_{11}+2))}. 
\end{eqnarray}
 \end{enumerate}
 If we make the convention that $E_{2}=E_{21}=2$ when ${\mathcal Z}_2$ is the 4-QAM constellation, then, equation~\eqref{eq:b} includes~\eqref{eq:min-c} as a special case. To compare~\eqref{eq:a} with~\eqref{eq:b}, we need to prove that the following inequality
 \begin{eqnarray}\label{eq:energy-ship}
  3(1+\alpha^2(E_1+E_{21}))(1+\alpha^2(E_{11}+E_2))\ge (1+\alpha(E_1+E_2))^2 
 \end{eqnarray}
 is true for any positive $\alpha$. To do that, we first establish an energy inequality:
 \begin{eqnarray}\label{eq:energy-ineq}
  3E_1 E_{11}+3E_1 E_2+3E_{11} E_{21}+3E_2 E_{21}\ge (E_1+E_2)^2.
  \end{eqnarray} 
To show this,  let us discuss the following four possibilities.
 \begin{enumerate}
  \item $|{\mathcal Z}_1|=|{\mathcal Z}_2|=8$. In this special case, it can be verified by calculation that $3E_1 E_{11}+3E_1 E_2+3E_{11} E_{21}+3E_2 E_{21}=432$ and $(E_1+E_2)^2=400$.  
 Thus, the inequality~\eqref{eq:energy-ineq} is true in this particular situation. 
  \item $|{\mathcal Z}_2|=4$. Since $3E_{11}+4\ge E_1$, we have 
 \begin{subequations}
 \begin{eqnarray}
 && 3E_1 E_{11}+3E_1 E_2+3E_{11} E_{21}+3E_2 E_{21}\nonumber \\
 && =E_1(3E_{11}+4)+ 2(3E_{11}+4-E_1)+4E_1+4\nonumber\\
 && \ge E_1^2+2E_1 E_2+4=(E_1+E_2)^2.\label{eq:energy-4}
  \end{eqnarray}
 \item $|{\mathcal Z}_1|>|{\mathcal Z}_2|=8$, In this case, note that $2E_{11}\ge E_1, E_1>E_{11}>E_2$. Hence, we attain  
\begin{eqnarray}
 &&3E_1 E_{11}+3E_1 E_2+3E_{11} E_{21}+3E_2 E_{21}\nonumber\\
 &&>2E_1 E_{11}+2E_1 E_2+E_1E_{11}\nonumber\\
  &&=E_1^2+2E_1 E_2+E_2^2=(E_1+E_2)^2.\label{eq:energy-8}   
 \end{eqnarray}
\item $|{\mathcal Z}_1|\ge |{\mathcal Z}_2|>8$. Then, $2E_{11}\ge E_1$ and $2 E_{21}>E_2$ and as a result, 
\begin{eqnarray}
&& 3E_1 E_{11}+3E_1 E_2+3E_{11} E_{21}+3E_2 E_{21}\nonumber\\
&&>2E_1 E_{11}+2E_1 E_2+2E_2E_{21}\nonumber\\
 &&>E_1^2+2E_1 E_2+E_2^2=(E_1+E_2)^2.\label{eq:energy>8}  
 \end{eqnarray}
 \end{subequations}
 \end{enumerate}
 Combining~\eqref{eq:energy-4},~\eqref{eq:energy-8} with~\eqref{eq:energy>8} gives us 
 \begin{eqnarray}\label{eq:energy-ship-4}
&&3(1+\alpha^2(E_1+E_{21}))(1+\alpha^2(E_{11}+E_2))\nonumber\\
 &&=3+3(E_1+E_2+E_{11}+E_{21})\alpha^2+(3E_1 E_{11}+3E_1 E_2+3E_{11} E_{21}+3E_2 E_{21}))\alpha^4\nonumber\\
&&\ge 1+2(E_1+E_2)\alpha^2+(E_1+E_2)^2\alpha^4\nonumber\\
&&=(1+\alpha(E_1+E_2))^2.
 \end{eqnarray}  
Hence, the inequality~\eqref{eq:energy-ship} is true. 
Now, comparing~\eqref{eq:a} with~\eqref{eq:b} and using~\eqref{eq:energy-ship} lead to 
 \begin{eqnarray}\label{eq:com-ab}
 \frac{\min_{(x, {\mathbf y}^T, \hat{x}, \hat{\mathbf y}^T)\in{\mathcal D}_{11}\cup{\mathcal D}_{12}} G_{\alpha}(x, {\mathbf y}; \hat{x}, \hat{\mathbf y}|{\mathcal X}^{(1)}, {\mathcal Y}^{(1)}_1, {\mathcal Y}^{(1)}_2)}{\min_{(x, {\mathbf y}^T, \hat{x}, \hat{\mathbf y}^T)\in{\mathcal D}_{14}} G_{\alpha}(x, {\mathbf y}; \hat{x}, \hat{\mathbf y}|{\mathcal X}^{(1)}, {\mathcal Y}^{(1)}_1, {\mathcal Y}^{(1)}_2)}\nonumber\\
 \ge\frac{2(1+\alpha^2(E_1+E_{21}))(1+\alpha^2(E_{11}+E_2)}{(1+\alpha^2(E_1+E_2))^2} \ge 1
 \end{eqnarray}
 if $|{\mathcal Z}_2|\ge 8$, which is equivalent to the fact that
 \begin{eqnarray}\label{eq:com-11-12-14}
 \min_{(x, {\mathbf y}^T, \hat{x}, \hat{\mathbf y}^T)\in{\mathcal D}_{11}\cup{\mathcal D}_{12}} G_{\alpha}(x, {\mathbf y}; \hat{x}, \hat{\mathbf y}|{\mathcal X}^{(1)}, {\mathcal Y}^{(1)}_1, {\mathcal Y}^{(1)}_2)\nonumber\\
 \ge\min_{(x, {\mathbf y}^T, \hat{x}, \hat{\mathbf y}^T)\in{\mathcal D}_{14}} G_{\alpha}(x, {\mathbf y}; \hat{x}, \hat{\mathbf y}|{\mathcal X}^{(1)}, {\mathcal Y}^{(1)}_1, {\mathcal Y}^{(1)}_2).  
 \end{eqnarray} 
Using the same argument, we can derive that 
\begin{eqnarray}\label{eq:com-13-14}
\min_{(x, {\mathbf y}^T, \hat{x}, \hat{\mathbf y}^T)\in{\mathcal D}_{14}} G_{\alpha}(x, {\mathbf y}; \hat{x}, \hat{\mathbf y}|{\mathcal X}^{(1)}, {\mathcal Y}^{(1)}_1, {\mathcal Y}^{(1)}_2)\nonumber\\
\le\min_{(x, {\mathbf y}^T, \hat{x}, \hat{\mathbf y}^T)\in{\mathcal D}_{13}} G_{\alpha}(x, {\mathbf y}; \hat{x}, \hat{\mathbf y}|{\mathcal X}^{(1)}, {\mathcal Y}^{(1)}_1, {\mathcal Y}^{(1)}_2).
\end{eqnarray}  
 Combining~\eqref{eq:opt-step12-min} with \eqref{eq:com-11-12-14} and~\eqref{eq:com-13-14} altogether tells us that 
 \begin{eqnarray}\label{eq:opt-step12}
  \min_{(x, {\mathbf y}^T, \hat{x}, \hat{\mathbf y}^T), x\ne\hat{x}}G_{\alpha}(x, {\mathbf y}; \hat{x}, \hat{\mathbf y}|{\mathcal X}^{(1)}, {\mathcal Y}^{(1)}_1, {\mathcal Y}^{(1)}_2)\nonumber\\
=\frac{8\alpha^2}{(1+\alpha^2({E}_1+{E}_{21}))(1+\alpha^2({E}_{11}+{E}_2))}.
\end{eqnarray}       
 \end{enumerate}
Substituting~\eqref{eq:opt-step11} and~\eqref{eq:opt-step12} into~\eqref{eq:opt-step1} yields 
\begin{eqnarray}\label{eq:opt-step1-bound}
G_{\alpha}({\mathcal X}^{(1)}, {\mathcal Y}^{(1)}_1, {\mathcal Y}^{(1)}_2)
=\left\{
  \begin{array}{ll}
   \frac{8\alpha^2}{(1+\alpha^2({E}_1+{E}_2))^2}&~ \mathrm{if~}p\ge q=5~{\rm or}~p=5, q=4,\\  
    \frac{8\alpha^2}{(1+\alpha^2({E}_1+{E}_2))(1+\alpha^2({E}_1+{E}_{21}))}&~ \mathrm{if~}p\ge q=3,\\     
  \frac{8\alpha^2}{(1+\alpha^2({E}_1+{E}_{21}))(1+\alpha^2({E}_1+{E}_{22}))}&~ \mathrm{if~} p> q, q\ne 3, 4, 5,\\
     \frac{8\alpha^2}{(1+\alpha^2({E}_1+{E}_{21}))(1+\alpha^2({E}_{11}+{E}_2))}&~ \mathrm{if~} p=q\ne 3, 5.\\  
  \end{array}\right. 
  \end{eqnarray}
 
{\bf Step 2}: Establish the achievable upper bound of the coding gain for any fixed UFCP, i.e., 
\begin{eqnarray}\label{eq:ufcp-bound}
G_{\alpha}({\mathcal X}, {\mathcal Y}_1, {\mathcal Y}_2)\le G_{\alpha}({\mathcal X}^{(1)}, {\mathcal Y}^{(1)}_1, {\mathcal Y}^{(1)}_2) 
\end{eqnarray}
for any positive $\alpha$. As we have mentioned before, when $\delta=1$, there are totally six candidates regarding the constellation ${\mathcal X}$, i.e., ${\mathcal X}=\{1, -1\},{\mathcal X}=\{1, j\}, {\mathcal X}=\{1, -j\},{\mathcal X}=\{-1, j\},{\mathcal X}=\{-1, -j\}, {\mathcal X}=\{j, -j\}$. Here we only consider the case where ${\mathcal X}=\{1, -1\}$, since the discussion for the other cases are very similar.   Now, we examine the following possibilities:

{\bf Case 1}: $q=3$. In this case, since ${\mathcal Z}_2$ includes the 4-QAM constellation as a subset, there are two points $y_{20}$ and $\hat{y}_{20}$ in ${\mathcal Y}_2$ such that $|y_{20}-\hat{y}_{20}|=2$ and $|y_{20}|=|\hat{y}_{20}|=2$ and thus, 
\begin{subequations}\label{eq:case-delta1}
 \begin{eqnarray}
 G_{\alpha}({\mathcal X}, {\mathcal Y}_1, {\mathcal Y}_2)&=& \min_{(x, {\mathbf y}^T)\ne (\hat{x}, \hat{\mathbf y}^T)}G_{\alpha}(x, {\mathbf y}; \hat{x}, \hat{\mathbf y}|{\mathcal X}, {\mathcal Y}_1, {\mathcal Y}_2)\nonumber\\
 &\le&G_{\alpha}((1, Z_1, y_{20}; 1, Z_1, \hat{y}_{20})|{\mathcal X}, {\mathcal Y}_1, {\mathcal Y}_2)\nonumber\\
 &=&\frac{4\alpha^2}{(1+\alpha^2({E}_1+2))(1+\alpha^2({E}_1+2))}. 
  \end{eqnarray}
   
{\bf Case 2}: $q=5$. There are totally 8 corner points in ${\mathcal Z}_2$ with the largest energy and ${\mathcal Y}_2$ includes  four of them. If two of the four points, say $y_{20}$ and $\hat{y}_{20}$, are in the same quadrant, i.e.,   
$|y_{20}-\hat{y}_{20}|=2\sqrt{2}$, then, we have 
\begin{eqnarray}
 G_{\alpha}({\mathcal X}, {\mathcal Y}_1, {\mathcal Y}_2)&=& \min_{(x, {\mathbf y}^T)\ne (\hat{x}, \hat{\mathbf y}^T)}G_{\alpha}(x, {\mathbf y}; \hat{x}, \hat{\mathbf y}|{\mathcal X}, {\mathcal Y}_1, {\mathcal Y}_2)\nonumber\\
 &\le&G_{\alpha}(1, Z_1, y_{20}; 1, Z_1, \hat{y}_{20}|{\mathcal X}, {\mathcal Y}_1, {\mathcal Y}_2)\nonumber\\
 &=&\frac{8\alpha^2}{(1+\alpha^2({E}_1+E_2))^2}. 
  \end{eqnarray}
  
{\bf Case 3}: $p>q$ and $q\ne 3, 5$.
\begin{enumerate}
 \item [(a)] $Z_{21}\in {\mathcal Y}_2$. Then, we have 
 \begin{eqnarray}
 G_{\alpha}({\mathcal X}, {\mathcal Y}_1, {\mathcal Y}_2)&=& \min_{(x, {\mathbf y}^T)\ne (\hat{x}, \hat{\mathbf y}^T)}G_{\alpha}(x, {\mathbf y}; \hat{x}, \hat{\mathbf y}|{\mathcal X}, {\mathcal Y}_1, {\mathcal Y}_2)\nonumber\\
 &\le&G_{\alpha}((1, Z_1, Z_2; 1, Z_1, Z_{21})|{\mathcal X}, {\mathcal Y}_1, {\mathcal Y}_2)\nonumber\\
 &=&\frac{4\alpha^2}{(1+\alpha^2({E}_1+{E}_{21}))(1+\alpha^2({E}_1+{E}_2))}. 
  \end{eqnarray}
 \item [(b)] $Z_{22}\in{\mathcal Y}_2$. Similarly, we can obtain  
 \begin{eqnarray}
 G_{\alpha}({\mathcal X}, {\mathcal Y}_1, {\mathcal Y}_2)&=& \min_{(x, {\mathbf y}^T)\ne (\hat{x}, \hat{\mathbf y}^T)}G_{\alpha}(x, {\mathbf y}; \hat{x}, \hat{\mathbf y}|{\mathcal X}, {\mathcal Y}_1, {\mathcal Y}_2)\nonumber\\
 &\le&G_{\alpha}(1, Z_1, Z_2; 1, Z_1, Z_{22})|{\mathcal X}, {\mathcal Y}_1, {\mathcal Y}_2)\nonumber\\
 &=&\frac{4\alpha^2}{(1+\alpha^2({E}_1+{E}_{22}))(1+\alpha^2({E}_1+{E}_2))}. 
\end{eqnarray} 
\item [(c)] $Z_{21}, Z_{22}\notin{\mathcal Y}_2$. Then, $-Z_{21}, -Z_{22}\in{\mathcal Y}_2$ and as a consequence, we arrive at 
 \begin{eqnarray}
 G_{\alpha}({\mathcal X}, {\mathcal Y}_1, {\mathcal Y}_2)&=& \min_{(x, {\mathbf y}^T)\ne (\hat{x}, \hat{\mathbf y}^T)}G_{\alpha}(x, {\mathbf y}; \hat{x}, \hat{\mathbf y}|{\mathcal X}, {\mathcal Y}_1, {\mathcal Y}_2)\nonumber\\
 &\le&G_{\alpha}((1, Z_1, -Z_{21}; 1, Z_1, -Z_{22})|{\mathcal X}, {\mathcal Y}_1, {\mathcal Y}_2)\nonumber\\
 &=&\frac{8\alpha^2}{(1+\alpha^2({E}_1+{E}_{21}))(1+\alpha^2({E}_1+{E}_{22}))}. 
  \end{eqnarray}
 \end{enumerate}
 
 {\bf Case 4}: $p=q$ and $q\ne 3, 5$. We consider the following possibilities:
 \begin{enumerate}
 \item [(a)] If either $Z_{11}\in {\mathcal Y}_1$ or $Z_{21}\in {\mathcal Y}_2$, then, following Case~3-(a), we can have 
 \begin{eqnarray}
 G_{\alpha}({\mathcal X}, {\mathcal Y}_1, {\mathcal Y}_2)\le \frac{4\alpha^2}{(1+\alpha^2({E}_1+{E}_{21}))(1+\alpha^2({E}_1+{E}_2))},
  \end{eqnarray}
  where $E_1=E_2$ and $E_{11}=E_{21}$. 
 \item [(b)] If either $Z_{12}\in {\mathcal Y}_1$ or $Z_{22}\in {\mathcal Y}_2$, then, similar to Case~3-(b), we can obtain  
 \begin{eqnarray}
 G_{\alpha}({\mathcal X}, {\mathcal Y}_1, {\mathcal Y}_2)\le \frac{4\alpha^2}{(1+\alpha^2({E}_1+{E}_{22}))(1+\alpha^2({E}_1+{E}_2))},
\end{eqnarray} 
 where $E_1=E_2$ and $E_{12}=E_{22}$. 
 \item [(c)] $Z_{11}, Z_{12}\notin{\mathcal Y}_1$ and $Z_{21}, Z_{22}\notin{\mathcal Y}_2$. Then, $-Z_{11}, -Z_{12}\in{\mathcal Y}_1, -Z_{21}, -Z_{22}\in{\mathcal Y}_2$ and as a result, we attain
 \begin{eqnarray}
 G_{\alpha}({\mathcal X}, {\mathcal Y}_1, {\mathcal Y}_2)&=& \min_{(x, {\mathbf y}^T)\ne (\hat{x}, \hat{\mathbf y}^T)}G_{\alpha}(x, {\mathbf y}; \hat{x}, \hat{\mathbf y}|{\mathcal X}, {\mathcal Y}_1, {\mathcal Y}_2)\nonumber\\
 &\le&G_{\alpha}((1, Z_{1}, -Z_{21}; -1, -Z_{11}, Z_2)|{\mathcal X}, {\mathcal Y}_1, {\mathcal Y}_2)\nonumber\\
 &=&\frac{8\alpha^2}{(1+\alpha^2({E}_1+{E}_{21}))(1+\alpha^2({E}_1+{E}_{11}))}. 
  \end{eqnarray}
 \end{enumerate}
 \end{subequations}
 Comparing~\eqref{eq:case-delta1} with~\eqref{eq:opt-step1} gives~\eqref{eq:ufcp-bound} as required. Now, using the geometrical and arithmetical mean inequality and then,  applying Lemma~\ref{lem:energy-min} to~\eqref{eq:ufcp-bound} and~\eqref{eq:opt-step1-bound} result in
 \begin{eqnarray}
G_{\alpha}({\mathcal X}, {\mathcal Y}_1, {\mathcal Y}_2)\le G(\widetilde{\mathcal X}^{(1)}, \widetilde{\mathcal Y}^{(1)}_1, \widetilde{\mathcal Y}^{(1)}_2), 
\end{eqnarray} 
where 
 \begin{eqnarray}
G(\widetilde{\mathcal X}^{(1)}, \widetilde{\mathcal Y}^{(1)}_1, \widetilde{\mathcal Y}^{(1)}_2) =\left\{
  \begin{array}{ll}
   \frac{2}{\widetilde E^{(1)}_1+\widetilde E^{(1)}_2}&~ \mathrm{if~}p\ge q=5~{\rm or}~p=5, q=4,\\  
    \frac{8}{\big(\sqrt{\widetilde E^{(1)}_1+\widetilde E^{(1)}_2}+\sqrt{\widetilde E^{(1)}_1+\widetilde E^{(1)}_{21}}\big)^2}&~ \mathrm{if~}p\ge q=3,\\     
  \frac{8}{\big(\sqrt{\widetilde E^{(1)}_1+\widetilde E^{(1)}_{21}}+\sqrt{\widetilde E^{(1)}_1+\widetilde E^{(1)}_{22}}\big)^2}&~ \mathrm{if~} p> q, q\ne 3, 4, 5,\\
     \frac{8}{\big(\sqrt{\widetilde E^{(1)}_1+\widetilde E^{(1)}_{21}}+\sqrt{\widetilde E^{(1)}_{11}+\widetilde E^{(1)}_2}\big)^2}&~ \mathrm{if~} p=q\ne 3, 5.\\  
  \end{array}\right. 
  \end{eqnarray}
 All the above discussions can be concluded as Property~\ref{pro:delta1}:  
\begin{property}\label{pro:delta1}
When $\delta=1$, one of the optimal solution to Problem~\ref{prob:design} is that $\widetilde{\mathcal X}^{(1)}=\{1, j\}$ and that $\widetilde{p}, \widetilde{q}$ and $\widetilde{\mathcal Y}^{(1)}_i$ are determined as follows:
\begin{enumerate}
 \item [(1)] If $r$ is even, then, $\widetilde{p}=(r+2)/2, \widetilde{q}=r/2$, $\widetilde{\mathcal Y}^{(1)}_i=\widetilde{\mathcal Y}_{i, \rm opt}$ and 
 \begin{eqnarray}
\widetilde{\alpha}=\left\{
  \begin{array}{ll}
   \frac{1}{\sqrt{\widetilde{E}^{(1)}_1+\widetilde{E}^{(1)}_2}}&~ \mathrm{if~}r=10~{\rm or}~r=8,\\  
    \frac{1}{\sqrt[4]{(\widetilde{E}^{(1)}_1+\widetilde{E}^{(1)}_2))(\widetilde{E}^{(1)}_1+\widetilde{E}^{(1)}_{21})}}&~ \mathrm{if~}r=6,\\     
  \frac{1}{\sqrt[4]{(\widetilde{E}^{(1)}_1+\widetilde{E}^{(1)}_{21}))(\widetilde{E}^{(1)}_1+\widetilde{E}^{(1)}_{22})}}&~ \mathrm{if~}r\ne 6, 8, 10.
  \end{array}\right.  
  \end{eqnarray}
 Furthermore, the optimal coding gain is given by
 \begin{eqnarray}
 G_{\widetilde{\alpha}}(\widetilde{\mathcal X}, \widetilde{\mathcal Y}_1, \widetilde{\mathcal Y}_2)
=\left\{
  \begin{array}{ll}
   \frac{2}{\widetilde{E}^{(1)}_1+\widetilde{E}^{(1)}_2}&~ \mathrm{if~}r=10~{\rm or}~r=8,\\  
    \frac{8}{(\sqrt{\widetilde{E}^{(1)}_1+\widetilde{E}^{(1)}_2}+\sqrt{\widetilde{E}^{(1)}_1+\widetilde{E}^{(1)}_{21}})^2}&~ \mathrm{if~}r=6,\\     
  \frac{8}{(\sqrt{\widetilde{E}^{(1)}_1+\widetilde{E}^{(1)}_{21}}+\sqrt{\widetilde{E}^{(1)}_1+\widetilde{E}^{(1)}_{22}})^2}&~ \mathrm{if~} r\ne 6, 8, 10.\\
  \end{array}\right. 
  \end{eqnarray}   
 \item [(2)] If $r$ is odd, then, $\widetilde{p}=\widetilde{q}=(r+1)/2$, $\widetilde{\mathcal Y}_i^{(1)}=\widetilde{\mathcal Y}_{i, \rm opt}$ and 
 \begin{eqnarray}
\widetilde{\alpha}=\left\{
  \begin{array}{ll}
   \frac{1}{\sqrt{\widetilde{E}^{(1)}_1+\widetilde{E}^{(1)}_2}}&~ \mathrm{if~}r=9,\\  
    \frac{1}{\sqrt[4]{(\widetilde{E}^{(1)}_1+\widetilde{E}^{(1)}_2))(\widetilde{E}^{(1)}_1+\widetilde{E}^{(1)}_{21})}}&~ \mathrm{if~}r=5,\\     
  \frac{1}{\sqrt[4]{(\widetilde{E}^{(1)}_1+\widetilde{E}^{(1)}_{21}))(\widetilde{E}^{(1)}_{11}+\widetilde{E}^{(1)}_2)}}&~ \mathrm{if~}r\ne 5, 9.
  \end{array}\right.  
  \end{eqnarray}
 Furthermore, the optimal coding gain is given by
 \begin{eqnarray}
 G_{\widetilde{\alpha}}(\widetilde{\mathcal X}, \widetilde{\mathcal Y}_1, \widetilde{\mathcal Y}_2)
=\left\{
  \begin{array}{ll}
   \frac{2}{\widetilde{E}^{(1)}_1+\widetilde{E}^{(1)}_2}&~ \mathrm{if~}r=9,\\  
    \frac{8}{(\sqrt{\widetilde{E}^{(1)}_1+\widetilde{E}^{(1)}_2}+\sqrt{\widetilde{E}^{(1)}_1+\widetilde{E}^{(1)}_{21}})^2}&~ \mathrm{if~}r=5,\\     
  \frac{8}{(\sqrt{\widetilde{E}^{(1)}_1+\widetilde{E}^{(1)}_{21}}+\sqrt{\widetilde{E}^{(1)}_{11}+\widetilde{E}^{(1)}_2})^2}&~ \mathrm{if~} r\ne 5, 9.\\
  \end{array}\right. 
  \end{eqnarray}   
 \end{enumerate}
~\hfill\QED
\end{property} 
Actually, Property~\ref{pro:delta1} also suggests us that for $\delta=1$, the optimal UFCP designed by Proposition~\ref{pro:ufcp-qam} is still optimal in the sense of maximizing the coding gain.   

\subsubsection{$\delta=2$}\label{subsec:delta2} In this case, ${\mathcal X}=\{1, -1, j, -j\}$. 
 \begin{subequations}\label{eq:case2-delta2} We examine the following possibilities:
 \begin{enumerate}
 \item [(a)] If either $Z_{11}\in {\mathcal Y}_1$ or $Z_{21}\in {\mathcal Y}_2$, then,  we can have either  
\begin{eqnarray}
 G_{\alpha}({\mathcal X}, {\mathcal Y}_1, {\mathcal Y}_2)&=&\min_{(x, {\mathbf y}^T)\ne (\hat{x}, \hat{\mathbf y}^T)}G_{\alpha}(x, {\mathbf y}; \hat{x}, \hat{\mathbf y}|{\mathcal X}, {\mathcal Y}_1, {\mathcal Y}_2)\nonumber\\
 &\le& G_{\alpha}(1, Z_1, Z_2; 1, Z_{11}, Z_2|{\mathcal X}, {\mathcal Y}_1, {\mathcal Y}_2)\nonumber\\
&=&\frac{4\alpha^2}{(1+\alpha^2({E}_1+{E}_2))(1+\alpha^2({E}_{11}+{E}_2))},
  \end{eqnarray}  or 
  \begin{eqnarray}
 G_{\alpha}({\mathcal X}, {\mathcal Y}_1, {\mathcal Y}_2)&\le& G_{\alpha}(1, Z_1, Z_{21}; 1, Z_1, Z_2|{\mathcal X}, {\mathcal Y}_1, {\mathcal Y}_2),\nonumber\\
&=&\frac{4\alpha^2}{(1+\alpha^2({E}_1+{E}_{21}))(1+\alpha^2({E}_1+{E}_2))}.
  \end{eqnarray}
 \item [(b)] If either $Z_{12}\in {\mathcal Y}_1$ or $Z_{22}\in {\mathcal Y}_2$, then, similar to situation (a), we can obtain  
 \begin{eqnarray}
 G_{\alpha}({\mathcal X}, {\mathcal Y}_1, {\mathcal Y}_2)\le \max\{\frac{4\alpha^2}{(1+\alpha^2({E}_1+{E}_2))(1+\alpha^2({E}_{12}+{E}_2))}, \nonumber\\
 \frac{4\alpha^2}{(1+\alpha^2({E}_1+{E}_{22}))(1+\alpha^2({E}_1+{E}_2))}\}\nonumber\\
 =\frac{4\alpha^2}{(1+\alpha^2({E}_1+{E}_2))(1+\alpha^2({E}_{12}+{E}_2))},
 \end{eqnarray} 
 since ${E}_{12}+{E}_2\le {E}_1+{E}_{22}$ according to Lemma~\ref{lem:energy-ineq}.
 \item [(c)] If $Z_{11}, Z_{12}\notin{\mathcal Y}_1$ and $Z_{21}, Z_{22}\notin{\mathcal Y}_2$, then, $Z_{11}, Z_{12}\in{\mathcal Z}_{1, -1}\cup {\mathcal Z}_{1, j}\cup {\mathcal Z}_{1, -j}$ and $Z_{21}, Z_{22}\in {\mathcal Z}_{2, -1}\cup {\mathcal Z}_{2, j}\cup {\mathcal Z}_{2, -j}$. In this case, one of the following two possibilities must occur:
 \begin{enumerate}
  \item If either $Z_{11}, Z_{12}\in {\mathcal Z}_{1, x_1}$ or $Z_{21}, Z_{22}\in {\mathcal Z}_{2, x_2}$, then, we have either 
   \begin{eqnarray}
 &&G_{\alpha}({\mathcal X}, {\mathcal Y}_1, {\mathcal Y}_2)=\min_{(x, {\mathbf y}^T)\ne (\hat{x}, \hat{\mathbf y}^T)}G_{\alpha}(x, {\mathbf y}; \hat{x}, \hat{\mathbf y}|{\mathcal X}, {\mathcal Y}_1, {\mathcal Y}_2)\nonumber\\
 &&\le G_{\alpha}(1, Z_{11}, Z_2; 1, Z_{12}, Z_2|{\mathcal X}, {\mathcal Y}_1, {\mathcal Y}_2)\nonumber\\
&&=\frac{8\alpha^2}{(1+\alpha^2({E}_{11}+{E}_2))(1+\alpha^2({E}_{12}+{E}_2))},
  \end{eqnarray}
or   
  \begin{eqnarray}
 G_{\alpha}({\mathcal X}, {\mathcal Y}_1, {\mathcal Y}_2)&\le& G_{\alpha}(1, Z_1, Z_{21}; 1, Z_1, Z_{22}|{\mathcal X}, {\mathcal Y}_1, {\mathcal Y}_2), \nonumber\\
&=&\frac{8\alpha^2}{(1+\alpha^2({E}_1+{E}_{21}))(1+\alpha^2({E}_1+{E}_{22}))}.
  \end{eqnarray}    
  \item If neither $Z_{11}, Z_{12}$ nor $Z_{21}, Z_{22}$ belong to the same group, then, by the pigeonhole principle, there exists an $x_0\in{\mathcal X}$ such that  one of the following four statements must be true:
  \begin{enumerate}
  \item $Z_{11}\in {\mathcal Z}_{1, x_0}$ and $Z_{21}\in {\mathcal Z}_{2, x_0}$. As a result, we attain
 \begin{eqnarray}
 G_{\alpha}({\mathcal X}, {\mathcal Y}_1, {\mathcal Y}_2)&=& \min_{(x, {\mathbf y}^T)\ne (\hat{x}, \hat{\mathbf y}^T)}G_{\alpha}(x, {\mathbf y}; \hat{x}, \hat{\mathbf y}|{\mathcal X}, {\mathcal Y}_1, {\mathcal Y}_2)\nonumber\\
 &\le&G_{\alpha}((1, Z_1, Z_{21}; x_0, Z_{11}, Z_2)|{\mathcal X}, {\mathcal Y}_1, {\mathcal Y}_2)\nonumber\\
 &=&\frac{8\alpha^2}{(1+\alpha^2({E}_1+{E}_{21}))(1+\alpha^2({E}_2+{E}_{11}))}. 
  \end{eqnarray}
   \item $Z_{12}\in {\mathcal Z}_{1, x_0}$ and $Z_{21}\in {\mathcal Z}_{2, x_0}$. Similarly, we can arrive at
 \begin{eqnarray}
 G_{\alpha}({\mathcal X}, {\mathcal Y}_1, {\mathcal Y}_2)
 &\le&G_{\alpha}((1, Z_1, Z_{21}; x_0, Z_{12}, Z_2)|{\mathcal X}, {\mathcal Y}_1, {\mathcal Y}_2)\nonumber\\
 &=&\frac{8\alpha^2}{(1+\alpha^2({E}_1+{E}_{21}))(1+\alpha^2({E}_2+{E}_{12}))}. 
  \end{eqnarray}
   \item $Z_{11}\in {\mathcal Z}_{1, x_0}$ and $Z_{22}\in {\mathcal Z}_{2, x_0}$. Then, 
 \begin{eqnarray}
 G_{\alpha}({\mathcal X}, {\mathcal Y}_1, {\mathcal Y}_2)
 &\le&G_{\alpha}((1, Z_1, Z_{22}; x_0, Z_{11}, Z_2)|{\mathcal X}, {\mathcal Y}_1, {\mathcal Y}_2)\nonumber\\
 &=&\frac{8\alpha^2}{(1+\alpha^2({E}_1+{E}_{22}))(1+\alpha^2({E}_2+{E}_{11}))}. 
  \end{eqnarray}
   \item $Z_{12}\in {\mathcal Z}_{1, x_0}$ and $Z_{22}\in {\mathcal Z}_{2, x_0}$. Also, we can have
 \begin{eqnarray}
 G_{\alpha}({\mathcal X}, {\mathcal Y}_1, {\mathcal Y}_2)
 &\le&G_{\alpha}((1, Z_1, Z_{22}; x_0, Z_{12}, Z_2)|{\mathcal X}, {\mathcal Y}_1, {\mathcal Y}_2)\nonumber\\
 &=&\frac{8\alpha^2}{(1+\alpha^2({E}_1+{E}_{22}))(1+\alpha^2({E}_2+{E}_{12}))}. 
  \end{eqnarray}
  \end{enumerate} 
  \end{enumerate} 
 \end{enumerate}
 \end{subequations}
Now,  comparing~\eqref{eq:case2-delta2} results in a common upper bound on $ G_{\alpha}({\mathcal X}, {\mathcal Y}_1, {\mathcal Y}_2)$ in Case 2 as follows:
 \begin{eqnarray}
 G_{\alpha}({\mathcal X}, {\mathcal Y}_1, {\mathcal Y}_2)
 \le\max\Big\{\frac{8\alpha^2}{(1+\alpha^2({E}_{11}+{E}_2))(1+\alpha^2({E}_{12}+{E}_2))}, \nonumber\\
 \frac{8\alpha^2}{(1+\alpha^2({E}_1+{E}_{21}))(1+\alpha^2({E}_1+{E}_{22}))}, \nonumber\\
 \frac{8\alpha^2}{(1+\alpha^2({E}_1+{E}_{22}))(1+\alpha^2({E}_2+{E}_{12}))}\Big\}\nonumber\\
 =\max\Big\{\frac{8\alpha^2}{(1+\alpha^2({E}_{11}+{E}_2))(1+\alpha^2({E}_{12}+{E}_2))}, \nonumber\\
 \frac{8\alpha^2}{(1+\alpha^2({E}_1+{E}_{22}))(1+\alpha^2({E}_2+{E}_{12}))}\Big\},
 \end{eqnarray}  
since $E_2+E_{12}\le E_1+E_{21}$, $E_{12}+E_2\le E_1+E_{22}$ and $E_2+E_{11}\le E_1+E_{21}$ by Lemma~\ref{lem:energy-ineq}. Following the same trick as the case when $\delta=1$, using the geometrical and arithmetical mean inequality first and then, Lemma~\ref{lem:energy-min} arrive at the fact that 
\begin{eqnarray}
\frac{8\alpha^2}{(1+\alpha^2({E}_{11}+{E}_2))(1+\alpha^2({E}_{12}+{E}_{2}))}&\le&\frac{8}{\Big(\sqrt{\widetilde{E}_1^{(2)}+\widetilde{E}_{22}^{(2)}}+\sqrt{\widetilde{E}_2^{(2)}+\widetilde{E}_{12}^{(2)}}\Big)^2},\nonumber\\
 \frac{8\alpha^2}{(1+\alpha^2({E}_1+{E}_{22}))(1+\alpha^2({E}_2+{E}_{12}))}&\le&\frac{8}{\Big(\sqrt{\widetilde{E}_{11}^{(2)}+\widetilde{E}_2^{(2)}}+\sqrt{\widetilde{E}_{12}^{(2)}+\widetilde{E}_2^{(2)}}\Big)^2}.\nonumber
 \end{eqnarray}
Therefore, it follows from this that
\begin{eqnarray}
 G({\mathcal X}, {\mathcal Y}_1, {\mathcal Y}_2)=\max_{\alpha} G_{\alpha}({\mathcal X}, {\mathcal Y}_1, {\mathcal Y}_2)\nonumber\\
 \le \max_{\alpha} \max\Big\{ \frac{8\alpha^2}{(1+\alpha^2({E}_{11}+{E}_2))(1+\alpha^2({E}_{12}+{E}_{2}))}, \nonumber\\
 \frac{8\alpha^2}{(1+\alpha^2({E}_1+{E}_{22}))(1+\alpha^2({E}_2+{E}_{12}))}\Big\}\nonumber\\
 \le \max\Big\{\frac{8}{\Big(\sqrt{\widetilde{E}_1^{(2)}+\widetilde{E}_{22}^{(2)}}+\sqrt{\widetilde{E}_2^{(2)}+\widetilde{E}_{12}^{(2)}}\Big)^2},\nonumber\\   \frac{8}{\Big(\sqrt{\widetilde{E}_{11}^{(2)}+\widetilde{E}_2^{(2)}}+\sqrt{\widetilde{E}_{12}^{(2)}+\widetilde{E}_2^{(2)}}\Big)^2}\Big\}\nonumber\\
 =\left\{
  \begin{array}{ll}  
  \frac{8}{\Big(\sqrt{\widetilde{E}_1^{(2)}+\widetilde{E}_{22}^{(2)}}+\sqrt{\widetilde{E}_2^{(2)}+\widetilde{E}_{12}^{(2)}}\Big)^2},~~{\rm if}~r ~\mathrm{is~even},\\  
  \frac{8}{\Big(\sqrt{\widetilde{E}_{11}^{(2)}+\widetilde{E}_2^{(2)}}+\sqrt{\widetilde{E}_{12}^{(2)}+\widetilde{E}_2^{(2)}}\Big)^2},~~{\rm if}~r ~\mathrm{is~odd},  \end{array}\right.\nonumber
  \end{eqnarray}
where we have used Lemma~\ref{lem:energy-ineq-delta2} in the last step. All the above discussions can be conclude as the following property: 
\begin{property}\label{pro:delta2} For given $r$ and ${\mathcal X}=\{1, -1, j, -j\}$,  we have
\begin{eqnarray}
 G({\mathcal X}, {\mathcal Y}_1, {\mathcal Y}_2)\le\left\{
  \begin{array}{ll}  
  \frac{8}{\Big(\sqrt{\widetilde{E}_1^{(2)}+\widetilde{E}_{22}^{(2)}}+\sqrt{\widetilde{E}_2^{(2)}+\widetilde{E}_{12}^{(2)}}\Big)^2},~~{\rm if}~r ~\mathrm{is~even},\\  
  \frac{8}{\Big(\sqrt{\widetilde{E}_{11}^{(2)}+\widetilde{E}_2^{(2)}}+\sqrt{\widetilde{E}_{12}^{(2)}+\widetilde{E}_2^{(2)}}\Big)^2},~~{\rm if}~r ~\mathrm{is~odd}.  \end{array}\right.
  \end{eqnarray}
~\hfill\QED
\end{property}
Properties~\ref{pro:delta0}, ~\ref{pro:delta1} and~~\ref{pro:delta2} lead us to giving the following theorem as one of the optimal solutions to Problem~\ref{prob:design}.
\begin{table}
\centering
\begin{tabular}{|c|c|c|c|c|}
  \hline
  Bit Rate &
$G_{\widetilde{\alpha}}(\mathcal{\widetilde{X}},\mathcal{\widetilde{Y}}_1,\mathcal{\widetilde{Y}}_2)$
& $|\mathcal{\widetilde{X}}|$ & $\mathcal{\widetilde{Z}}_1,\mathcal{\widetilde{Z}}_2$ & $\widetilde{\alpha}$\\\hline
  1 & 0.250 & 1 & 4-QAM, 4-QAM & 0.5 \\\hline
  1.25 & 0.127 & 2 & 8-QAM, 8-QAM & 0.254 \\\hline
  1.5 & 0.0839 & 2 & 8-QAM, 16-QAM & 0.206\\\hline
  1.75 & 0.0614 & 2 & 16-QAM, 16-QAM & 0.189\\\hline
  2 & 0.0385 & 2 & 16-QAM, 32-QAM & 0.137 \\\hline
  2.25 & 0.0294 & 2 & 32-QAM, 32-QAM & 0.121\\\hline
  2.5 & 0.0156 & 1 & 32-QAM, 32-QAM & 0.125 \\\hline
  2.75 & 0.0116 & 2 & 64-QAM, 64-QAM & 0.0762\\\hline
  3 & 0.00820 & 2 & 64-QAM, 128-QAM & 0.0640 \\\hline
  3.25 & 0.00633 & 2 & 128-QAM, 128-QAM & 0.0563\\
  \hline
\end{tabular}
\caption{Maximum coding gains for different
transmission bit rates using optimal designs\label{tab:gain}}
\end{table}
\begin{theorem}\label{th:solu}
One of the optimal solutions to Problem~\ref{prob:design} is given as follows:
\begin{enumerate}
 \item [(1)] If $r=4$, then, $\widetilde{\delta}=0, \widetilde{p}=\widetilde{q}=2, \widetilde{\mathcal X}=\{1\}, \widetilde{\mathcal Y}_1=\widetilde{\mathcal Y}_2={\mathcal Z}_1={\mathcal Z}_2$ is the 4-QAM constellation and $\widetilde{\alpha}=1/\sqrt{2}$. Moreover,  the optimal coding gain is $ G_{\widetilde{\alpha}}(\widetilde{\mathcal X}, \widetilde{\mathcal Y}_1, \widetilde{\mathcal Y}_2)=1/4$.
 \item [(2)] If $r=10$, then, $\widetilde{\delta}=0, \widetilde{p}=\widetilde{q}=5, \widetilde{\mathcal X}=\{1\}, \widetilde{\mathcal Y}_1=\widetilde{\mathcal Y}_2={\mathcal Z}_1={\mathcal Z}_2$ is the 32-QAM constellation and $\widetilde{\alpha}=\frac{1}{2\sqrt[4]{255}}$. Moreover,  the optimal coding gain is $ G_{\widetilde{\alpha}}(\widetilde{\mathcal X}, \widetilde{\mathcal Y}_1, \widetilde{\mathcal Y}_2)=\frac{1}{(\sqrt{15}+\sqrt{17})^2}$.
 \item [(3)] If $r$ is even, then, $\widetilde{p}=(r+2)/2, \widetilde{q}=r/2$, $\widetilde{\mathcal Y}_i=\widetilde{\mathcal Y}_{i, \rm opt}$ and 
 \begin{eqnarray}
\widetilde{\alpha}=\left\{
  \begin{array}{ll}
    \frac{1}{\sqrt[4]{(\widetilde{E}_1+\widetilde{E}_2))(\widetilde{E}_1+\widetilde{E}_{21})}}&~ \mathrm{if~}r=6,\\     
  \frac{1}{\sqrt[4]{(\widetilde{E}_1+\widetilde{E}_{21}))(\widetilde{E}_1+\widetilde{E}_{22})}}&~ \mathrm{if~}r\ne 6, 10.
  \end{array}\right.  
  \end{eqnarray}
 Furthermore, the optimal coding gain is given by
 \begin{eqnarray}
 G_{\widetilde{\alpha}}(\widetilde{\mathcal X}, \widetilde{\mathcal Y}_1, \widetilde{\mathcal Y}_2)
=\Bigg\{
  \begin{array}{ll}
    \frac{8}{(\sqrt{\widetilde{E}_1+\widetilde{E}_2}+\sqrt{\widetilde{E}_1+\widetilde{E}_{21}})^2}&~ \mathrm{if~}r=6,\\     
  \frac{8}{(\sqrt{\widetilde{E}_1+\widetilde{E}_{21}}+\sqrt{\widetilde{E}_1+\widetilde{E}_{22}})^2}&~ \mathrm{if~} r\ne 6, 10.\\
  \end{array} 
  \end{eqnarray}    
  \item [(4)] If $r$ is an odd integer exceeding 4, then, $\delta=1, \widetilde{p}=\widetilde{q}=(r+1)/2$, $\widetilde{\mathcal Y}_i=\widetilde{\mathcal Y}_{i, \rm opt}$ and 
 \begin{eqnarray}
\widetilde{\alpha}=\left\{
  \begin{array}{ll}
   \frac{1}{\sqrt{\widetilde{E}_{1}^{(1)}+\widetilde{E}_{2}^{(1)}}}&~ \mathrm{if~}r=9,\\  
    \frac{1}{\sqrt[4]{(\widetilde{E}_{1}^{(1)}+\widetilde{E}_{2}^{(1)}))(\widetilde{E}_{1}^{(1)}+\widetilde{E}_{21}^{(1)})}}&~ \mathrm{if~}r=5,\\     
  \frac{1}{\sqrt[4]{(\widetilde{E}_{1}^{(1)}+\widetilde{E}_{21}^{(1)})(\widetilde{E}_{11}^{(1)}+\widetilde{E}_{2}^{(1)})}}&~ \mathrm{if~}r\ne 5, 9.
  \end{array}\right.  
  \end{eqnarray}
 Furthermore, the optimal coding gain is given by
 \begin{eqnarray}
 G_{\widetilde{\alpha}}(\widetilde{\mathcal X}, \widetilde{\mathcal Y}_1, \widetilde{\mathcal Y}_2)
=\left\{
  \begin{array}{ll}
   \frac{2}{\widetilde{E}_{1}^{(1)}+\widetilde{E}_{2}^{(1)}}&~ \mathrm{if~}r=9,\\  
    \frac{8}{\Big(\sqrt{\widetilde{E}_{1}^{(1)}+\widetilde{E}_{2}^{(1)}}+\sqrt{\widetilde{E}_{1}^{(1)}+\widetilde{E}_{21}^{(1)}}\Big)^2}&~ \mathrm{if~}r=5,\\     
  \frac{8}{\Big(\sqrt{\widetilde{E}_{1}^{(1)}+\widetilde{E}_{21}^{(1)}}+\sqrt{\widetilde{E}_{11}^{(1)}+\widetilde{E}_{2}^{(1)}}\Big)^2}&~ \mathrm{if~} r\ne 5, 9.\\
  \end{array}\right. 
  \end{eqnarray}   
 \end{enumerate}
~\hfill\QED
\end{theorem}
\begin{figure}[h]
\centering
\includegraphics[width=10cm]{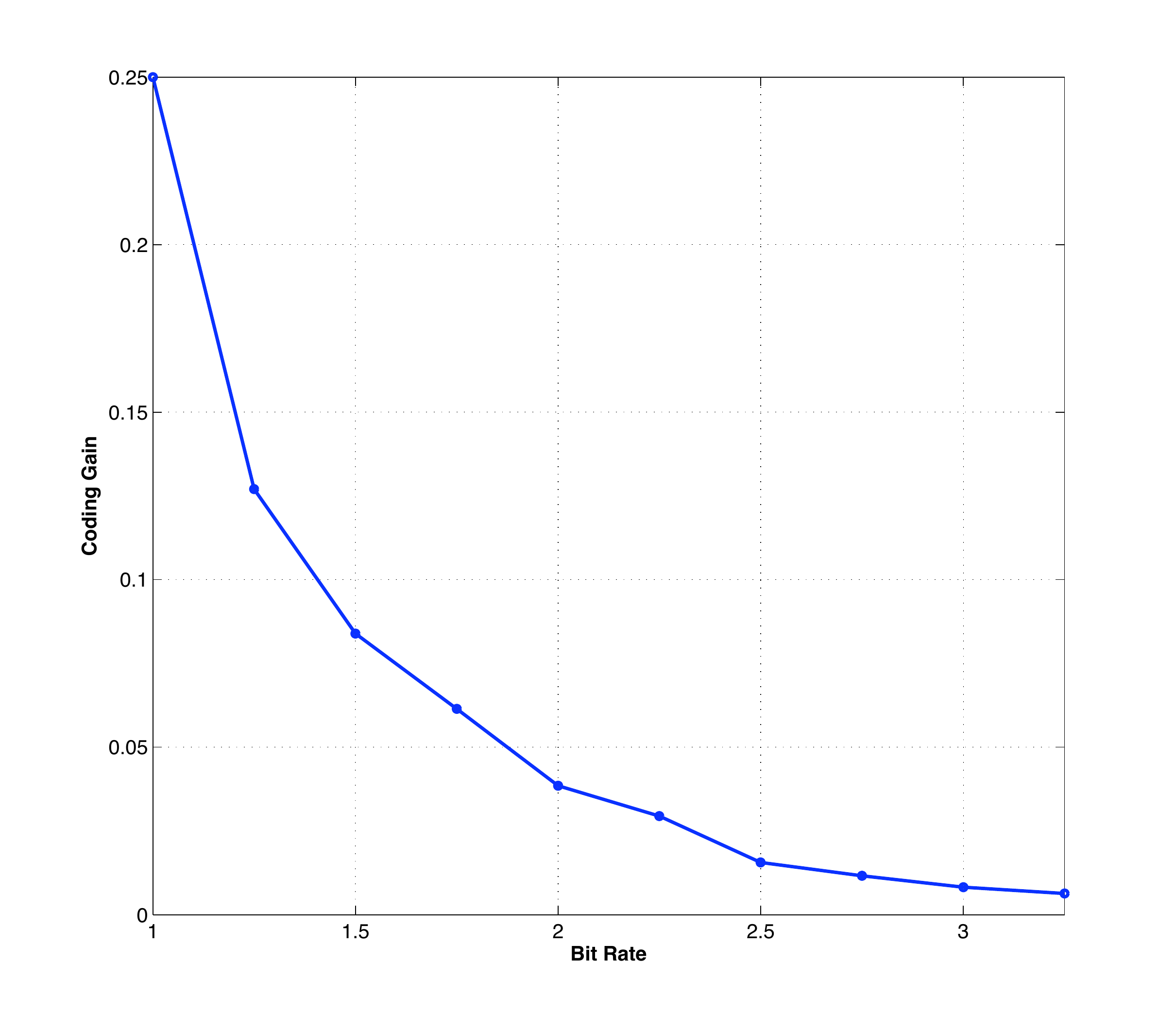}
\caption{The optimal coding gains versus transmission bit rates\label{fig:gain}}
\end{figure}
The proof of Theorem~\ref{th:solu} is given in Appendix~\ref{appendix:th:solu}. The maximum coding gains using the optimal UFCPs determined by Theorem~\ref{th:solu} are listed in Table~\ref{tab:gain} for various transmission bit rates, which is also shown in Fig.~\ref{fig:gain}. Some observations on Theorem~\ref{th:solu} are made as follows:  
\begin{enumerate}
 \item [(1)]Theorem~\ref{th:solu} tells us that the training scheme based on the Alamouti code using the 4-QAM and 32-QAM constellations is optimal when either one bit or 2.5 bits per channel use is transmitted. 
 \item [(2)] In spite of the fact that from Proposition~\ref{pro:ufcp-qam} we know that increasing the number of the groups is increasing the minimum Euclidian distance of the constellation ${\mathcal Y}$, the accumulated minimum Euclidian distance along the two transmitter antennas between two distinct groups is always equal to 8. In addition, increasing the number of groups is also increasing the size of the constellations and thus, increasing the energies of the three corner points. As a result, the UFCP code using four groups cannot enable the optimal coding gain.       
  \end{enumerate}

\begin{figure}[h]
    \centering
    \subfigure[$R_b=1.25$ bits per channel use]{
    \includegraphics[width=7.9cm] {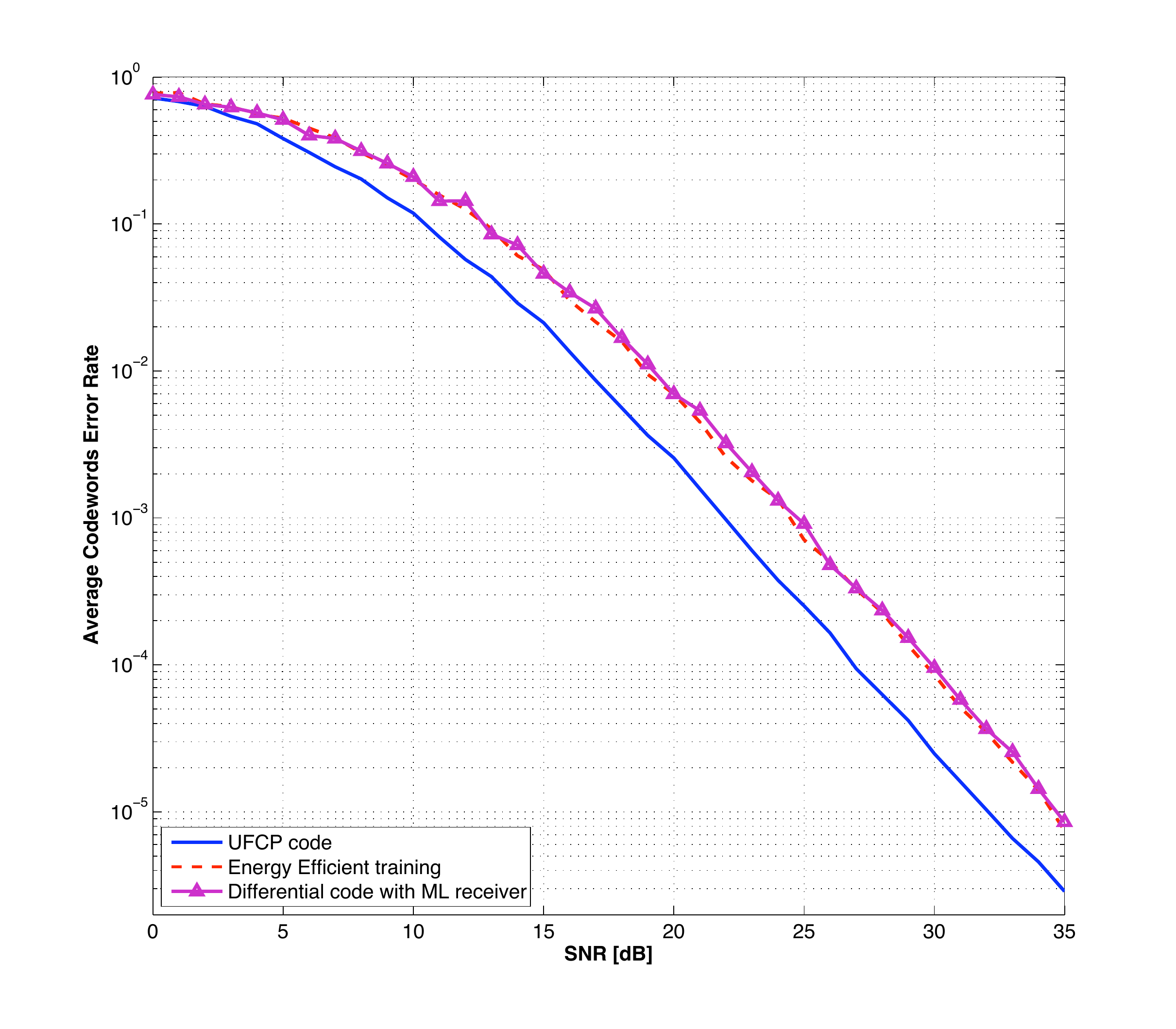}}
    \subfigure[$R_b=1.5$ bits per channel use]{
    \includegraphics[width=7.9cm] {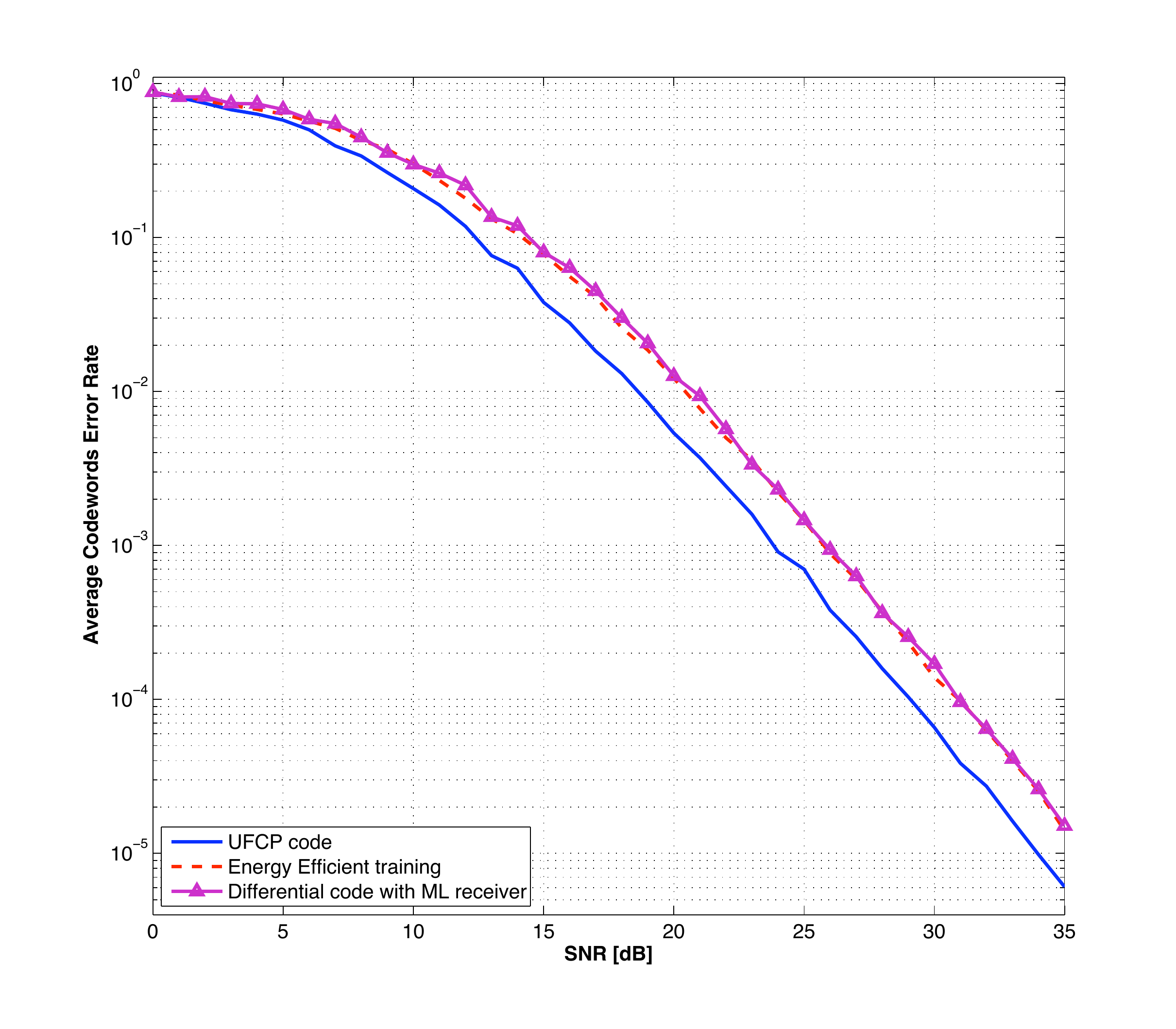}} \\   
    \subfigure[$R_b=1.75$ bits per channel use]{
    \includegraphics[width=7.9cm] {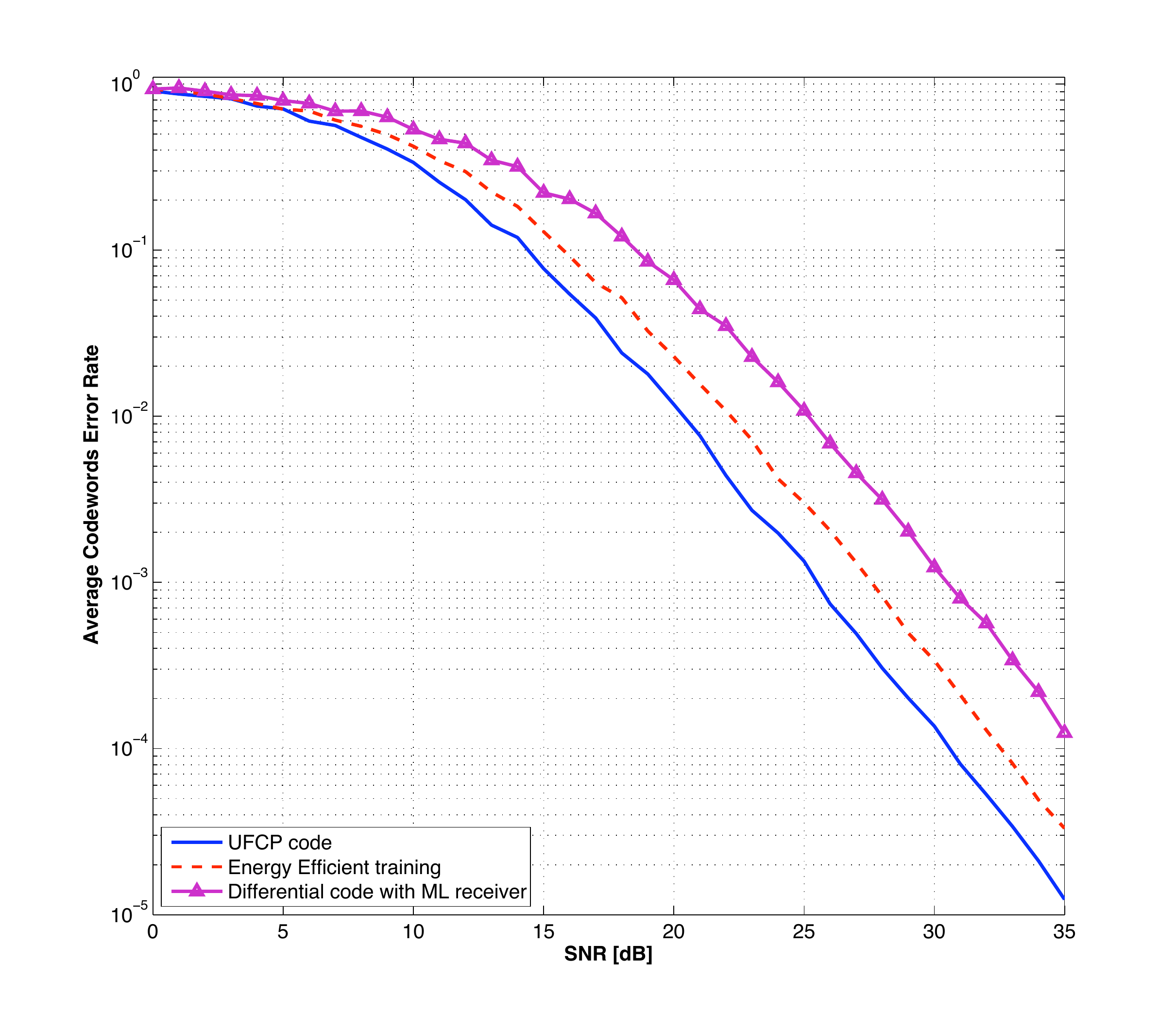}}
    \subfigure[$R_b=2.25$ bits per channel use]{
    \includegraphics[width=7.9cm] {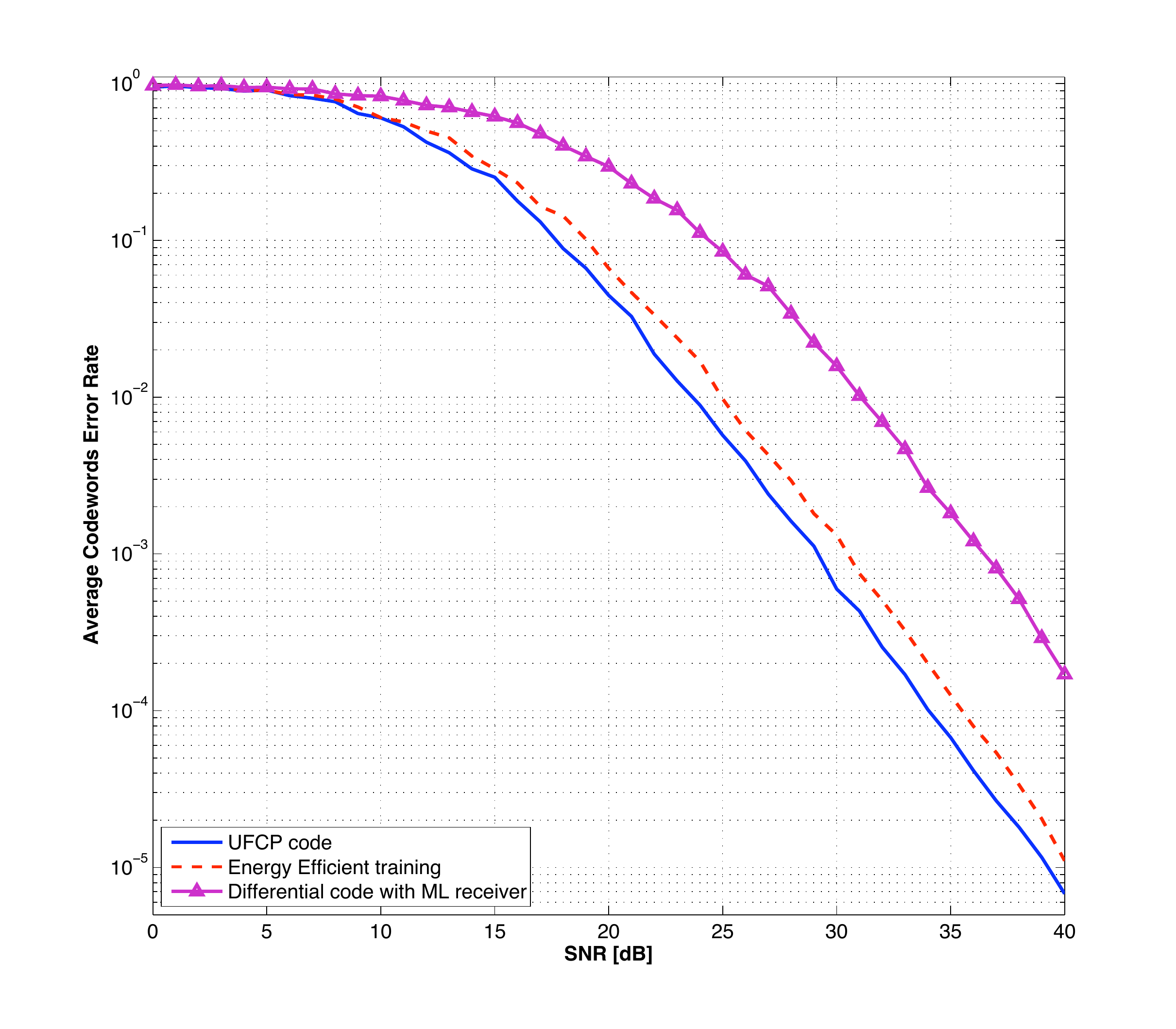}}       
    \caption{Performance comparison of unitary UFCP codes with currently available noncoherent codes } \label{fig:simu}    
\end{figure}

\section{Simulations}\label{sec:simu}
In this section, we carry out computer simulations and compare the error performance of the unitary UFCP code design proposed in this paper with those of  other schemes in the literatures which can
be used in a small noncoherent MISO system having two transmitter antennas and a single receiver antenna, where channel state information is completely unknown at both the transmitter and the receiver and the coherence time is $T=2M=4$. All the schemes that we would like to  
compare here are described as follows:

(a)  \textit{Differential unitary code based on Alamouti coding scheme and PSK constellations}. This design with the fast closed-form ML decoder was proposed in~\cite{ganesan02, liang05} and two unitary codeword matrices are ${\mathbf U}_1={\mathbf I}_2$ and 
\begin{eqnarray}
{\mathbf U}_2=\frac{1}{\sqrt{2}}\times \left(\begin{array}{cc} s_1 & s_2 \\-s_2^* & s_1^*\end{array}\right),
\end{eqnarray}
 where $s_1$ and $s_2$ are randomly, independently and equally likely chosen from the $2^{r_1}$-ary and $2^{r_2}$-ary phase shift keying (PSK) constellations, respectively, with the two integers $a$ and $b$ determined as follows:
 \begin{eqnarray}\label{eq:bit}
 \Bigg\{
  \begin{array}{ll}
   r_1=r_2=\frac{r}{2}&~ \mathrm{if~}r~{\rm is~even},\\  
   r_1=\frac{r+1}{2}, r_2=\frac{r-1}{2}&~ \mathrm{if~}r~{\rm is~odd}.
  \end{array}  
 \end{eqnarray}
 For the necessity of performance comparison and decoding with the GLRT receiver,  these two unitary matrices are normalized and then,  stacked into one codeword matrix, which is denoted by ${\mathbf S}_a$, 
 \begin{eqnarray}\label{eq:dif-code}
 {\mathbf S}_a=\frac{1}{\sqrt{2}}\times \left(\begin{array}{c}{\mathbf U}_1 \\{\mathbf U}_2\end{array}\right),
 \end{eqnarray}
where the normalization constant assures ${\rm E}\big[{\rm tr}\big({\mathbf S}_a^H{\mathbf S}_a\big)\big]=2$. 
 
(b) \textit{SNR-efficient training Alamouti code}. This SNR-efficient training scheme using the Alamouti code was presented in~\cite{dayal04}. The codeword matrices are characterized by 
\begin{eqnarray}
{\mathbf S}_b=\frac{1}{\sqrt{E_b}}\times\left(\begin{array}{cc}\sqrt{E_b/2} & 0 \\0 & \sqrt{E_b/2} \\s_1 & s_2 \\-s_2^* & s_1^*\end{array}\right),
\end{eqnarray}
 where $s_1$ and $s_2$ are randomly and equally likely chosen from either the $2^{r_1}$-ary and $2^{r_2}$-ary PSK constellations or cross QAM constellations, respectively, with the determination of the two integers $r_1$ and $r_2$ being the same as~\eqref{eq:bit}. 
The energy constant $E_b$ is normalized in such a way that  ${\rm E}\big[{\rm tr}\big({\mathbf S}_b^H{\mathbf S}_b\big)\big]=2$. Here, the optimal average energy distribution over the training phase and communication phase is attained by maximizing the training efficiency~\cite{hassibi02, zheng02, dayal04}.     

(c) \textit{Optimal unitary UFCP code}. The code design is proposed in this paper and the codeword matrix is of the form: 
\begin{eqnarray}
{\mathbf S}_b=\frac{1}{\sqrt{1+{\widetilde\alpha}^2 |y_1|^2+{\widetilde\alpha}^2 |y_2|^2}}\times\left(\begin{array}{cc}x & 0 \\0 & x \\ {\widetilde\alpha}y_1 & {\widetilde\alpha} y_2 \\-{\widetilde\alpha} y_2^* & {\widetilde\alpha} y_1^*\end{array}\right),\qquad x\in\widetilde{\mathcal X}, y_1\in\widetilde{\mathcal Y}_1, y_2\in\widetilde{\mathcal Y}_2,
\end{eqnarray}
where the optimal energy scale $\widetilde{\alpha}$ and three constellations $\widetilde{\mathcal X}, \widetilde{\mathcal Y}_1$ and $\widetilde{\mathcal Y}_2$ are determined according to Theorem~\ref{th:solu}. 

It can be seen that the above three transmission schemes have the same spectrum efficiency, i.e., each transmission rate is $R_b=r/4$ bits per channel use. To make all error performance comparisons fair, we decode all the codes using the GLRT detector, i.e., 
\begin{eqnarray}
\hat{\mathbf S}&=&\arg\max_{{\mathbf S}\in {\mathcal S}}{\rm
Tr}\left({\mathbf\Upsilon}^H{\mathbf S}\left({\mathbf S}^H{\mathbf
S}\right)^{-1}{\mathbf S}^H{\mathbf\Upsilon}\right).\nonumber
\end{eqnarray}  
All the average codeword error rates against SNR are shown Fig.~\ref{fig:simu}. It is observed  that the optimal unitary UFCP code designed in this paper performs the best error performance among all the three coding schemes. 
\section{Conclusion And Discussions}\label{sec:conclu}
In this paper, we have considered a wireless communication system having two transmitter antennas and a single receiver antenna, in which the channel coefficients are assumed to be unknown at either the transmitter or the receiver, but remain constant
for the first $4$ time slots, after which they change to new independent values that are fixed for the next $4$ time slots, and
so on.  For such a system, we have developed a novel concept called the uniquely factorable constellation pair for the systematic design of full diversity unitary space-time block code. 
By simply normalizing the two Alamouti codes and carefully selecting three constellations, a full diversity unitary code design with a symbol rate $3/4$ has been attained. It has been shown that it is the unique factorization of constellation pairs that guarantees that the unique identification of both the channel coefficients and the transmitted signals in the noise-free case as well as full diversity in the noise case. In other  words, both the unique identification and full diversity require that the constellation pair must be designed in such a cooperative way that factorization in the product sense is unique-able. It is for this reason that we have named the code proposed in this paper as the \textit{UFCP code}. In addition, to  further enhance error performance, the optimal unitary UFCP code enabling the maximum coding gain has been designed from a pair of energy-efficient cross QAM constellations  subject to a bit rate constraint. After a careful examination of the fractional coding gain  function,  
in this paper we have taken two major steps maximizing the coding gain:  
\begin{enumerate}
 \item The energy scale has been carefully designed  to compress the first three largest energy points of the QAM constellations in the denominator of the objective; 
 \item The two UFCPs have been designed so carefully that the one constellation collaborates with the other two constellations through the two transmitter antennas maximizing the minimum of the numerator and at the same time, avoiding the corner points with the largest energy as many as possible achieving the minimum.
 \end{enumerate}
  In other words, the optimal coding gain has been obtained by constellations collaboration and energy compression. It is for this reason that we have also called the optimal UFCP code designed in this paper as the \textit{energy-efficient collaborative UFCP code}. Computer simulations have demonstrated that error performance of the optimal unitary UFCP code presented in this paper outperforms those of the differential code and the SNR-efficient training code, which, to the best knowledge of the authors, is the best code in current literatures for the system.    
 
 As we have seen, the concept of the UFCP plays an important role in the systematic design of energy-efficient full diversity unitary space-time block codes for the small MIMO system having the two transmitter antennas and a single receiver antenna. However, the constructions and properties on the UFCP and the related transmission scheme which have been reported in this paper are just initiative. Some significant issues still remain unsolved:
\begin{enumerate}
 \item The construction of the optimal UFCPs for the design of the unitary space-time block code has been derived from the cross QAM constellations. How about the hexagonal constellations? since the hexagonal constellations carved from the Eisenstein integer ring are supposed to be more energy-efficient than the QAM constellations carved from the Gaussian integer ring~\cite{conway-book98}. Generally, which constellation is optimal to generate a unitary UFCP space-time block code with the optimal coding gain? 
 \item Instead of a pair of coprime PSK constellations, whether is the UFCP constructed in this paper used to systematically design full diversity noncoherent space-time block codes for a general MIMO system by following the way similar to~\cite{jkz-it09}?        
 \item The coding scheme which has been adopted in this paper is the Alamouti scheme. In spite of the fact that the Alamouti code is optimal in many senses for such coherent system, it is not optimal anymore for such noncoherent system, since it was proved that unitary codes are optimal for general noncoherent MIMO communications, whereas the Alamouti code resulting from the QAM constellation is not unitary in general. Only when the constellation is the PSK, the resulting Alamouti code is unitary.  However, the PSK constellation is not as energy-efficient as the QAM constellation.  In addition, although this paper has proposed a simple method for the design of the unitary code jusy by normalizing the two Alamouti codes,   a deep insight into the fractional coding gain function exposures the drawback of the Alamouti scheme in the nocoherent case, i.e., Too large energies are contributed to the denominator. Hence, a question is: is it possible to find another coding scheme that has the same minimum of the numerator but a smaller maximum of the denominator as the Alamouti scheme?       
\end{enumerate}
This paper has just casted a brick so that the jade may be attracted.


\appendix
\section{Appendix}
\subsection{Proof of Proposition~\ref{pro:ufcp-qam}}\label{appendix:pro:ufcp-qam}
We consider the following two cases: 

{\bf Case 1}: $|{\mathcal X}|=2$, i.e., there are only two elements in ${\mathcal X}$. Let ${\mathcal X}=\{x_1, x_2\}$, where $x_1, x_2\in\{1, -1, -j, j\}$. By Proposition~\ref{pro:ufcp-qam}, we have ${\mathcal Z}={\mathcal Z}_{x_1}\cup{\mathcal Z}_{x_2}$ with ${\mathcal Z}_{x_1}\cap{\mathcal Z}_{x_2}=\Phi$. In fact, ${\mathcal Z}_{x_1}=x_1^*{\mathcal Y}=x_2^*{\mathcal Y}$ and thus, $d_{\min}({\mathcal Z}_{x_1})=d_{\min}({\mathcal Z}_{x_2})=d_{\min}({\mathcal Y})$. Let $P$ be one of the corner point in ${\mathcal Z}$ with the largest energy. Without loss of generality, we can always assume $P\in {\mathcal Z}_{x_1}$. The following three possibilities need to be considered seperately. 
\begin{enumerate}
 \item $K=3$. In this case, let $P_1$ is the first nearest neighbor of $P$, i.e., $|P-P_1|=2$, and $P_2$ is the second nearest neighbor of $P$, i.e, $|P-P_2|=2\sqrt{2}$. If $P_1\in{\mathcal Z}_{x_1}$, then, $d_{\min}({\mathcal Y})=d_{\min}({\mathcal Z}_{x_1})\le |P-P_1|\le 2$; If $P_2\in{\mathcal Z}_{x_1}$, then, $d_{\min}({\mathcal Y})=d_{\min}({\mathcal Z}_{x_1})\le |P-P_2|\le 2\sqrt{2}$; Otherwise, both $P_1$ and $P_2$ must lie in ${\mathcal Z}_{x_2}$ and as a result, $d_{\min}({\mathcal Y})=d_{\min}({\mathcal Z}_{x_2})\le |P_1-P_2|=2\sqrt{2}$. At any rate, the minimum distance of ${\mathcal Y}$ is upper-bounded by
\begin{eqnarray}
d_{\min}({\mathcal Y})\le 2\sqrt{2}.
\end{eqnarray}
On the other hand, it can be verified directly by calculation that the constellation given by~\eqref{eq:x1-k3} satisfies three conditions: (a) $|{\mathcal Y}_{\rm opt}^{(1)}|=4$; (b) $d_{\min}({\mathcal Y}_{\rm opt}^{(1)})=2\sqrt{2}$; (c) ${\mathcal Y}_{\rm opt}^{(1)}$ and ${\mathcal X}_{\rm opt}^{(1)}$ indeed form a UFCP. Therefore, ${\mathcal Y}_{\rm opt}^{(1)}$ is optimal.   
 \item $K=5$. In this case, $P$ has the two nearest neighbors, which are denoted by $P_1$ and $P_2$, i.e., $|P-P_1|=|P-P_2|=2$. If one of $P_1$ and $P_2$, say, $P_1$, belongs to  the Group-$x_1$, ${\mathcal Z}_{x_1}$, then,  $d_{\min}({\mathcal Y})=d_{\min}({\mathcal Z}_{x_1})\le |P-P_1|=2$. Otherwise, both $P_1$ and $P_2$ must lie in ${\mathcal Z}_{x_2}$ and as a result, $d_{\min}({\mathcal Y})=d_{\min}({\mathcal Z}_{x_2})\le |P_1-P_2|=2\sqrt{2}$. Hence, the minimum distance of ${\mathcal Y}$ is always upper-bounded by 
  \begin{eqnarray}
d_{\min}({\mathcal Y})\le 2\sqrt{2}.
\end{eqnarray}
Now, following the argument similar to the Possibility~1) of $K=3$, we can say that the constellation ${\mathcal Y}_{\rm opt}^{(1)}$ given by~\eqref{eq:x1-k5} is indeed optimal.  
\item $K\ge 4$ is even. Similar to the possibility of $K=5$, we can prove $d_{\min}({\mathcal Y})\le 2\sqrt{2}$. In addition, notice that the constellation determined by~\eqref{eq:x1-k-even} has two properties: (a) $|{\mathcal Y}_{\rm opt}^{(1)}|=2^K$; (b) $y_1-y_2=2(1-j) z$ for any two distinct points $y_1$ and $y_2$ in ${\mathcal Y}_{\rm opt}^{(1)}$, where $z$ is some complex integer. As a consequence, $d_{\min}({\mathcal Y})=2\sqrt{2}$. Let us now check whether or not such a pair of ${\mathcal X}_{\rm opt}^{(1)}$ and ${\mathcal Y}_{\rm opt}^{(1)}$ constitutes a UFCP. Suppose that there exist $x, {\widetilde x}\in{\mathcal X}$ and $y, {\widetilde y}\in{\mathcal Y}$ such that $x{\widetilde y}={\widetilde x} y$, then, $({\widetilde y}-y)x=({\widetilde x}-x) y$. If $x={\widetilde x}$, then, $y={\widetilde y}$, since $x\ne 0$. If $x\ne{\widetilde x}$, then,  $x-{\widetilde x}=\pm(1-j)$. Combining this with the fact $({\widetilde y}-y)=2(1-j)z$, we have $2 z=\pm y$, which is impossible. Thus, $x={\widetilde x}$ and $y={\widetilde y}$. In other words,  a pair of the constellations ${\mathcal X}_{\rm opt}^{(1)}$ and ${\mathcal Y}_{\rm opt}^{(1)}$ constitutes a UFCP. Therefore, in this case,  ${\mathcal Y}_{\rm opt}^{(1)}$ is optimal. 
\item $K>5$ is odd. Following almost the same discussion as the possibility of the even $K$ exceeding 2, we can also arrive at the fact that the constellation given by~\eqref{eq:x1-k-odd} is still optimal. 
\end{enumerate}
 \begin{figure}[h]
\centering
\includegraphics[width=10cm]{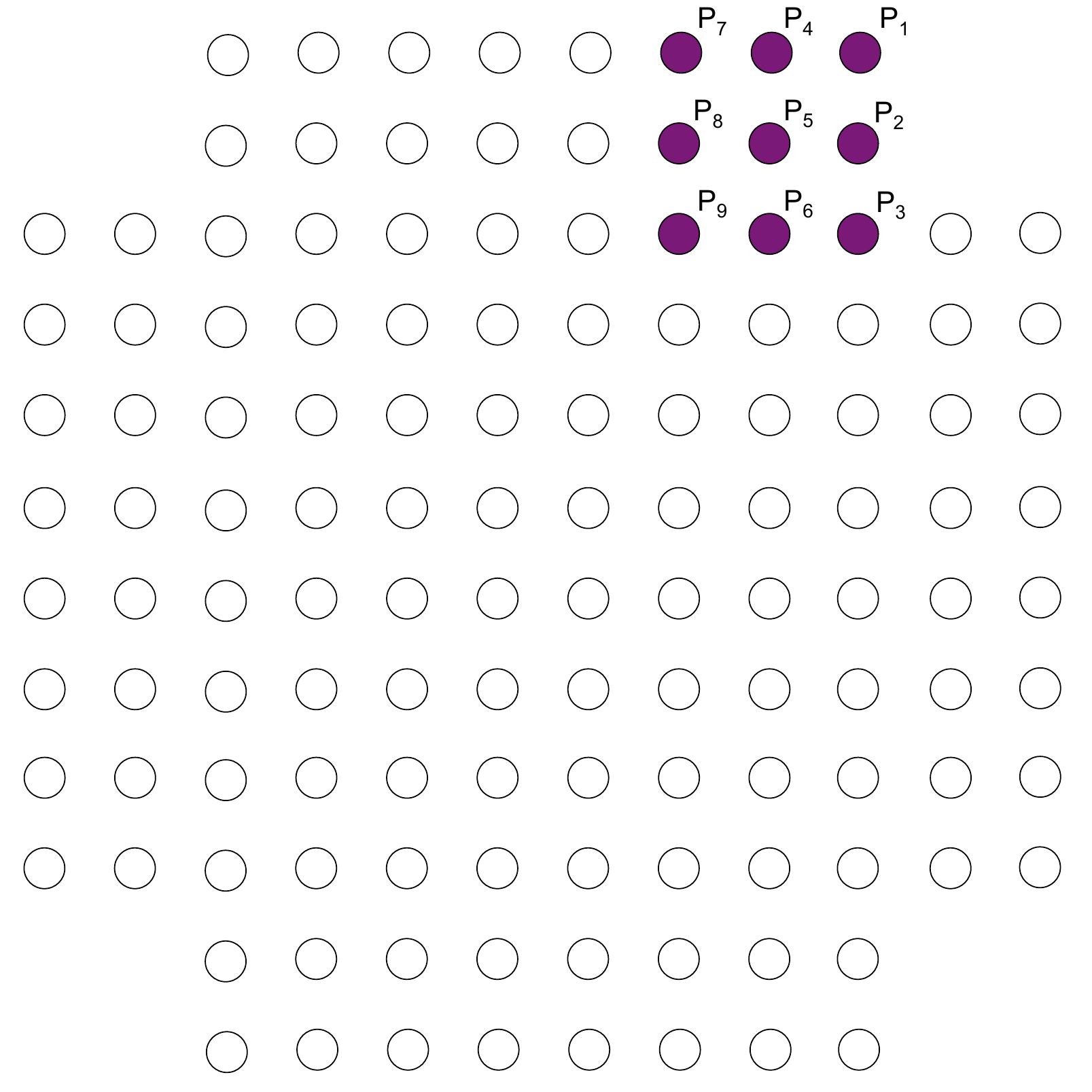}
\caption{Nine corner points in the cross QAM constellation \label{fig:qam-corner}}
\end{figure} {\bf Case 2}: $|{\mathcal X}|=4$. In this case, possibilities for $K=3$ and 5 can be verified directly by calculation. In addition, since the possibility for even $K$ greater than 2 is similar to that for odd $K$ exceeding 5, here we only provide a proof for the situation when $K$ is an odd number greater than 5.  By Proposition~\ref{pro:decom-group}, ${\mathcal Z}={\mathcal Z}_1\cup {\mathcal Z}_{-1}\cup{\mathcal Z}_{j}\cup{\mathcal Z}_{-j}$ with ${\mathcal Z}_{x_k}\cap{\mathcal Z}_{x_\ell}={\mathbf\Phi}$ for $k\ne\ell =1, 2, 3, 4$. Actually, ${\mathcal Z}_{x_k}=x_k^*{\mathcal Y}$, since $x_k\in{\mathcal X}=\{\pm 1, \pm j\}$. Thus, we have $d_{\min}({\mathcal Z}_{x_\ell})=d_{\min}({\mathcal Y})$.    
Let $P_i$ for $i=1, \cdots, 9$ be nine points in the first quadrant  around the corner of ${\mathcal Z}$ shown in Fig.~\ref{fig:qam-corner}. 
Using the pigeonhole   
principle, there exists one Goup, say, ${\mathcal Z}_{x_{k_0}}$, including at least three of these nine points, say, $P_{i_1}, P_{i_2}$ and $P_{i_3}$. Among these three points, if there exist two of them lying either in the same row or in the same column, then,    $d_{\min}({\mathcal Y})=d_{\min}({\mathcal Z}_{x_{k_0}})\le 4$. Otherwise, these three points locate in different rows and different columns and thus, $d_{\min}({\mathcal Y})=d_{\min}({\mathcal Z}_{x_{k_0}})\le 2\sqrt{2}$. Therefore, in any case, we can always have  $d_{\min}({\mathcal Y})\le 4$. On the other hand, notice that the constellation determined by~\eqref{eq:x2-k-odd} possesses two features: (a) $|{\mathcal Y}^{(2)}_{\rm opt}|=2^K$; (b)  $y_1-y_2=4z$ for any two distinct points $y_1$ and $y_2$ in ${\mathcal Y}_{\rm opt}^{(1)}$, where $z$ is some complex integer. Hence, we have $d_{\min}({\mathcal Y})=4$. Let us now examine whether such a pair of the constellations  ${\mathcal X}_{\rm opt}^{(2)}$ and ${\mathcal Y}_{\rm opt}^{(2)}$ forms a UFCP. Suppose that   there exist $x, {\widetilde x}\in{\mathcal X}$ and $y, {\widetilde y}\in{\mathcal Y}$ such that $x{\widetilde y}={\widetilde x} y$, then, we have
\begin{eqnarray}
({\widetilde y}-y)x=({\widetilde x}-x) y.\label{eq:proof-x2-odd} 
\end{eqnarray}
If $x={\widetilde x}$, then, $y={\widetilde y}$, since $x\ne 0$. If $x\ne{\widetilde x}$, then,  $x-{\widetilde x}=\pm 2, \pm 2j, \pm 1\pm j$. No matter whatever situation occurs, once we have substituted $({\widetilde y}-y)=4 z$ into~\eqref{eq:proof-x2-odd}, we can always obtain $y=2 z_0$ for some complex integer $z_0$, where we have used the fact that $2=(1+j)(1-j)$. This implies that $y$ is an even number,  which is impossible. Thus, $x={\widetilde x}$ and $y={\widetilde y}$ and ${\mathcal X}_{\rm opt}^{(1)}$ and ${\mathcal Y}_{\rm opt}^{(1)}$ indeed constitute a UFCP. Therefore, in this case,  ${\mathcal Y}_{\rm opt}^{(1)}$ is optimal. This completes the proof of Proposition~\ref{pro:ufcp-qam}.~\hfill$\Box$
 \subsection{Proof of Lemma~\ref{lem:energy-ineq}}\label{appendix:lem:energy-ineq}
 Since the proof of~\eqref{eq:energy-ineq2} is much similar to that of~\eqref{eq:energy-ineq1}, we only provide a proof for~\eqref{eq:energy-ineq1}, which can be fulfilled by considering the following possibilities:
\begin{enumerate}
 \item $v=3$. In this case, Lemma~\ref{lem:energy} tells us that $E_2=10$ and $E_{21}=2$. Hence, the inequality~\eqref{eq:energy-ineq1} is reduced to
 \begin{eqnarray}\label{eq:prof-q3}
 8+E_{11}\le E_1.
 \end{eqnarray}
 If $u=3$, $E_1=10$, $E_{11}=2$, so $E_1-E_{11}=8$, inequality~\eqref{eq:prof-q3} is hold.
 If $u$ is an odd number exceeding 3, then, by Lemma~\ref{lem:energy}, we have  $E_1=(2^{\frac{u-1}{2}}-1)^2+(3\times 2^{\frac{u-3}{2}}-1)^2$ and $E_{11}=(2^{\frac{u-1}{2}}-3)^2+(3\times 2^{\frac{u-3}{2}}-1)^2$. Thus, we obtain $E_1-E_{11}=(2^{\frac{u-1}{2}}-1)^2-(2^{\frac{u-1}{2}}-3)^2=4\times 2^{\frac{u-1}{2}}-8 \ge 8$, since $2^{\frac{u-1}{2}}\ge 4$. Hence, in this case, the inequality~\eqref{eq:prof-q3} is also true. If $u$ is an even integer, then, by Lemma~\ref{lem:energy} again, we have $E_1=2 (2^{\frac{u}{2}}-1)^2$ and $E_{11}=(2^{\frac{u}{2}}-1)^2+(2^{\frac{u}{2}}-3)^2$ so that $E_1-E_{11}=(2^{\frac{u}{2}}-1)^2-(2^{\frac{u}{2}}-3)^2=4\times 2^{\frac{u}{2}}-8\ge 8$, since $u\ge 4$. Hence, the inequality~\eqref{eq:prof-q3} is still true.
 \item $v$ is an odd number exceeding 3. In this case, Lemma~\ref{lem:energy} gives us that  $E_2=(2^{\frac{v-1}{2}}-1)^2+(3\times 2^{\frac{v-3}{2}}-1)^2$ and $E_{21}=(2^{\frac{v-1}{2}}-3)^2+(3\times 2^{\frac{v-3}{2}}-1)^2$ so that $E_2-E_{21}=(2^{\frac{v-1}{2}}-1)^2-(2^{\frac{v-1}{2}}-3)^2=4\times (2^{\frac{v-1}{2}}-2)$. On the other hand, if $u$ is an odd number not less than $v$, then, Lemma~\ref{lem:energy} gives us that  $E_1=(2^{\frac{p-1}{2}}-1)^2+(3\times 2^{\frac{u-3}{2}}-1)^2$ and $E_{11}=(2^{\frac{u-1}{2}}-3)^2+(3\times 2^{\frac{u-3}{2}}-1)^2$ so that $E_1-E_{11}=(2^{\frac{u-1}{2}}-1)^2-(2^{\frac{u-1}{2}}-3)^2=4\times (2^{\frac{u-1}{2}}-2)\ge 4\times (2^{\frac{v-1}{2}}-2)$, since the exponential function $2^t$ is increasing and $u\ge v$. Therefore, in this case, we have $E_1-E_{11}\ge E_2-E_{21}$, which is equivalent to the fact that $E_2+E_{11}\le E_1+E_{21}$. If $u$ is an even number not less than $v$, then, using Lemma~\ref{lem:energy} again yields $E_1=2 (2^{\frac{u}{2}}-1)^2$ and $E_{11}=(2^{\frac{u}{2}}-1)^2+(2^{\frac{u}{2}}-3)^2$. Hence, we have $E_1-E_{11}=(2^{\frac{u}{2}}-1)^2-(2^{\frac{u}{2}}-3)^2=4\times (2^{\frac{u}{2}}-2)\ge 4\times (2^{\frac{v-1}{2}}-2)=E_2-E_{21}$. Hence, in this case, the inequality~\eqref{eq:energy-ineq1} holds.
 \item $v$ is an even number greater than or equal to 2.  Similarly, by Lemma~\ref{lem:energy}, we can have $E_2-E_{21}=4\times (2^{\frac{v}{2}}-2)$ and
  \begin{eqnarray}
E_1-E_{11}=\Bigg\{
  \begin{array}{ll}
 4\times (2^{\frac{u}{2}}-2)\ge E_2-E_{21}   &~\mathrm{if~}u~{\rm is~even~and~not ~less~than~2},\nonumber\\
  8  &~\mathrm{if~}u=3,\\
4\times (2^{\frac{u-1}{2}}-2)\ge E_2-E_{21}  &~ \mathrm{if~} u~{\rm is~an~odd~number~exceeding~3}.\nonumber
  \end{array}
\end{eqnarray}
Notice that when $u=3$, $v$ must be 2, since $u\ge v\ge 2$. Then, in this case, $E_2-E_{21}=2-2=0$. Therefore, we can always obtain $E_1-E_{11}\ge E_2-E_{21}$.
\end{enumerate}
This completes the proof of Lemma~\ref{lem:energy-ineq}.~\hfill$\Box$ 
 \subsection{Proof of Lemma~\ref{lem:energy-min}}\label{appendix:lem:energy-min}
 Here, we only give the proof of~\eqref{eq:energy-min2}, since the proofs of the other inequalities are much similar. To do that,  let us consider the following situations.
\begin{enumerate}
 \item $w$ is an even number not less than 4. In this case, since $u+v=w$, both $u$ and $v$ are either even or odd. If both $u$ and $v$ are even, then,  Lemma~\ref{lem:energy} provides us with $E_1=2 (2^{\frac{u}{2}}-1)^2$ and $E_{21}=E_{22}=(2^{\frac{v}{2}}-1)^2+(2^{\frac{v}{2}}-3)^2$. Then, we have $E_1+E_{21}=2 (2^{\frac{u}{2}}-1)^2+(2^{\frac{v}{2}}-1)^2+(2^{\frac{v}{2}}-3)^2=2 (2^{\frac{u}{2}}-1)^2+(2^{\frac{w-u}{2}}-1)^2+(2^{\frac{w-u}{2}}-3)^2$. Since $u\ge v$, $w>u\ge w/2$. Now, consider the following function in terms of variable $t$:
 \begin{eqnarray}
 f(t)=2 (2^{\frac{t}{2}}-1)^2+(2^{\frac{w-t}{2}}-1)^2+(2^{\frac{w-t}{2}}-3)^2\qquad\textrm{for } t\ge w/2.\nonumber
  \end{eqnarray}
Then, the first order derivative of $f(t)$ with respect to $t$ is given by
  \begin{eqnarray}
 f'(t)=2 (2^{\frac{t}{2}}-1)2^{\frac{t}{2}}\ln 2-(2^{\frac{w-t}{2}}-1)2^{\frac{w-t}{2}}\ln 2-(2^{\frac{w-t}{2}}-3) 2^{\frac{w-t}{2}}\ln 2\qquad\textrm{for } w>t\ge w/2.\nonumber
  \end{eqnarray} 
Since the exponential function $2^{\frac{t}{2}}$ is increasing and $t\ge w-t$, $(2^{\frac{t}{2}}-1)2^{\frac{t}{2}}\ge (2^{\frac{w-t}{2}}-3) 2^{\frac{w-t}{2}}$ and thus, $f'(t)\ge 0$, showing that $f(t)$ is an increasing function. Therefore, we have 
$f(p)\ge f(r/2)$, i.e., $E_1+E_{21}\ge \widetilde{E}_1+\widetilde{E}_{21}$. Consequently, the inequality~\eqref{eq:energy-min2} holds in this case. Similarly, if both $p$ and $q$ are odd, we can also prove that~\eqref{eq:energy-min2} still holds.      
 \item $w$ is an odd number exceeding 4. Since $u+v=w$, either $u$ is even and $v$ is odd or $u$ is odd and $v$ is even. If $u$ is even and $v$ is odd, by Lemma~\ref{lem:energy}, we obtain $E_1=2 (2^{\frac{u}{2}}-1)^2$ $E_{21}=(2^{\frac{v-1}{2}}-3)^2+(3\times 2^{\frac{v-3}{2}}-1)^2$ and hence, $E_1+E_{21}=2 (2^{\frac{u}{2}}-1)^2+(2^{\frac{v-1}{2}}-3)^2+(3\times 2^{\frac{v-3}{2}}-1)^2=2 (2^{\frac{u}{2}}-1)^2+(2^{\frac{w-u-1}{2}}-3)^2+(3\times 2^{\frac{w-u-3}{2}}-1)^2$, where $w>u\ge (w+1)/2$. This leads us to considering a function:
 \begin{eqnarray}
 g(t)=2 (2^{\frac{t}{2}}-1)^2+(2^{\frac{w-t-1}{2}}-3)^2+(3\times 2^{\frac{w-t-3}{2}}-1)^2\qquad\textrm{for } w>t\ge (w+1)/2.\nonumber
  \end{eqnarray}
 The first order derivative of $g(t)$ is 
    \begin{eqnarray}
 g'(t)=2 (2^{\frac{t}{2}}-1)2^{\frac{t}{2}}\ln 2-(2^{\frac{w-t-1}{2}}-3)2^{\frac{w-t-1}{2}}\ln 2-3\times (3\times 2^{\frac{w-t-3}{2}}-1) 2^{\frac{w-t-3}{2}}\ln 2\nonumber
  \end{eqnarray}  
  for $w>t\ge (w+1)/2$. Since $2^{\frac{t}{2}}$ is increasing and $t\ge w-t$, we arrive at the fact that $\frac{1}{2}\times (2^{\frac{t}{2}}-1)2^{\frac{t}{2}}\ge\frac{1}{2}\times (2^{\frac{w-t}{2}}-1)2^{\frac{w-t}{2}}\ge (2^{\frac{w-t-1}{2}}-3)2^{\frac{w-t-1}{2}}$. In addition, notice that  $(2-\frac{1}{2})\times (2^{\frac{t}{2}}-1)2^{\frac{t}{2}}>(2-\frac{1}{2})\times(2^{\frac{w-t}{2}}-1) 2^{\frac{w-t}{2}}=3\sqrt{2}\times(2\sqrt{2}\times 2^{\frac{w-t-3}{2}}-1) 2^{\frac{w-t-3}{2}}>3\times(3\times 2^{\frac{w-t-3}{2}}-1) 2^{\frac{w-t-3}{2}}$ because of the fact that $2^{\frac{w-t-3}{2}}\ge 1$. As a result, $g'(t)\ge 0$ and $g(t)$ is increasing. Hence, $g(u)\ge g((w+1)/2)$, i.e., $E_1+E_{21}\ge \widetilde{E}_1+\widetilde{E}_{21}$. This can be also proved to be true if $u$ is odd and $v$ is even. Analogously, we can prove that   $E_1+E_{22}\ge \widetilde{E}_1+\widetilde{E}_{22}$. Therefore, the inequality~\eqref{eq:energy-min2} holds.   
  \end{enumerate} 
This completes the proof of Lemma~\ref{lem:energy-min}.~\hfill$\Box$ 
\subsection{Proof of Lemma~\ref{lem:energy-ineq-delta2}}\label{appendix:lem:energy-ineq-delta2}
We prove Lemma~\ref{lem:energy-ineq-delta2} by considering the following four different cases for $w$.
\begin{enumerate}
 \item $w=4\ell$, where $\ell$ is a positive integer not less than 1. In this case, by the definition of ${\widetilde u}$ and ${\widetilde v}$ in~\eqref{eq:def-pq}, we have  $\widetilde{u}=\widetilde{v}=w/2=2\ell$ and thus, $\widetilde{E}_1=\widetilde{E}_{2}$, $\widetilde{E}_{11}=\widetilde{E}_{22}$, which implies that $\widetilde{E}_1-\widetilde{E}_{11}=\widetilde{E}_2-\widetilde{E}_{22}$.
 \item $w=4\ell+1$. Then, \eqref{eq:def-pq} tells us that $\widetilde{u}=(w+1)/2=2\ell+1$ is odd, whereas $\widetilde{v}=(w-1)/2=2\ell$ is even. Particularly when $\ell=1$, ${\widetilde u}=3$ and ${\widetilde v}=2$. In this specular case, $\widetilde{E}_1-\widetilde{E}_{11}=10-2=6$ and $\widetilde{E}_2-\widetilde{E}_{22}=2-2=0$, resulting in $\widetilde{E}_1-\widetilde{E}_{11}>\widetilde{E}_2-\widetilde{E}_{22}$. If $\ell>1$, then,  by Lemma~\ref{lem:energy}, we have
 $\widetilde{E}_1-\widetilde{E}_{11}=4\times2^{\frac{\widetilde{u}-1}{2}}-8$ and $
 \widetilde{E}_2-\widetilde{E}_{22}=(2^{\frac{\widetilde{v}}{2}}-1)^2-(2^{\frac{\widetilde{v}}{2}}-3)^2=4\times2^{\frac{\widetilde{v}}{2}}-8$. 
Since
$\frac{\widetilde{u}-1}{2}=\frac{\widetilde{v}}{2}=\ell$, in this case we obtain $\widetilde{E}_1-\widetilde{E}_{11}=\widetilde{E}_2-\widetilde{E}_{22}$.
\item $w=4\ell+2$. From the definition of ${\widetilde u}$ and ${\widetilde v}$ given in~\eqref{eq:def-pq} we know that $\widetilde{u}=\widetilde{v}=w/2=2\ell+1$. This means that both ${\widetilde u}$ and ${\widetilde v}$ are odd. Then, Lemma~\ref{lem:energy} gives us
$\widetilde{E}_1-\widetilde{E}_{11}=(2^{\frac{\widetilde{u}-1}{2}}-1)^2-(2^{\frac{\widetilde{u}-1}{2}}-3)^2=4\times2^{\frac{\widetilde{u}-1}{2}}-8$ and
$\widetilde{E}_2-\widetilde{E}_{22}=(3\times 2^{\frac{\widetilde{v}-3}{2}}-1)^2-(3\times2^{\frac{\widetilde{v}-3}{2}}-3)^2=12\times2^{\frac{\widetilde{v}-3}{2}}-8$.
Since
$12\times2^{\frac{\widetilde{v}-3}{2}}=6\times2^{\frac{\widetilde{u}-1}{2}}>4\times2^{\frac{\widetilde{u}-1}{2}}$,
implying $\widetilde{E}_1-\widetilde{E}_{11}<\widetilde{E}_2-\widetilde{E}_{22}$.
\item $w=4\ell+3$. In this case, $\widetilde{u}=(w+1)/2=2\ell+2$ is even and $\widetilde{v}=(w-1)/2=2\ell+1$ is odd. Specially for $\ell=1$, we have ${\widetilde u}=4, {\widetilde v}=3$ and thus, $\widetilde{E}_1-\widetilde{E}_{11}=18-10=8$ and $\widetilde{E}_2-\widetilde{E}_{22}=10-2=8$. Hence, we have $\widetilde{E}_1-\widetilde{E}_{11}=\widetilde{E}_2-\widetilde{E}_{22}$ in this particular case. If $\ell>1$, then, by Lemma~\ref{lem:energy} again, we can attain that
$\widetilde{E}_1-\widetilde{E}_{11}=(2^{\frac{\widetilde{u}}{2}}-1)^2-(2^{\frac{\widetilde{u}}{2}}-3)^2=4\times2^{\frac{\widetilde{u}}{2}}-8$ and
$\widetilde{E}_2-\widetilde{E}_{22}=12\times2^{\frac{\widetilde{v}-3}{2}}-8$. Because of the fact that 
$4\times2^{\frac{\widetilde{u}}{2}}=4\times2^{\frac{\widetilde{v}+1}{2}}=16\times2^{\frac{\widetilde{v}-3}{2}}>12\times2^{\frac{\widetilde{v}-3}{2}}$,
we have $\widetilde{E}_1-\widetilde{E}_{11}>\widetilde{E}_2-\widetilde{E}_{22}$.
  \end{enumerate}
Now, summing up all the above results ends the proof of Lemma~\ref{lem:energy-ineq-delta2}.~\hfill$\Box$
 \subsection{Proof of Theorem~\ref{th:solu}}\label{appendix:th:solu}
 We prove Theorem~\ref{th:solu} by considering the following situations:
\begin{enumerate}
 \item $r=4$. In this case, we know from Properties~\ref{pro:delta0}, \ref{pro:delta1} and~\ref{pro:delta2} that 
 \begin{eqnarray}
 G(\widetilde{\mathcal X}^{(0)},\widetilde{\mathcal Y}^{(0)}_1, \widetilde{\mathcal Y}^{(0)}_2)&=&\frac{1}{4},\nonumber\\
 G(\widetilde{\mathcal X}^{(1)},\widetilde{\mathcal Y}^{(1)}_1, \widetilde{\mathcal Y}^{(1)}_2)&=&\frac{1}{6},\nonumber\\
 G(\widetilde{\mathcal X}^{(2)},\widetilde{\mathcal Y}^{(2)}_1, \widetilde{\mathcal Y}^{(2)}_2)&\le&\frac{2}{(\sqrt{5}+\sqrt{3})^2}<\frac{1}{4}.\nonumber 
  \end{eqnarray}
 Therefore, the optimal coding gain is $1/4, \widetilde{\delta}=0, \widetilde{p}=\widetilde{q}=2$, and $\widetilde{\mathcal X}$ and $\widetilde{\mathcal Y}$ are the 4-QAM constellation.  
 \item $r=10$. Similarly, Properties~\ref{pro:delta0}, \ref{pro:delta1} and~\ref{pro:delta2} tell us that  
 \begin{eqnarray}
 G(\widetilde{\mathcal X}^{(0)},\widetilde{\mathcal Y}^{(0)}_1, \widetilde{\mathcal Y}^{(0)}_2)&=&\frac{1}{(\sqrt{15}+\sqrt{17})^2},\nonumber\\
 G(\widetilde{\mathcal X}^{(1)},\widetilde{\mathcal Y}^{(1)}_1, \widetilde{\mathcal Y}^{(1)}_2)&=&\frac{1}{66},\nonumber\\
 G(\widetilde{\mathcal X}^{(2)},\widetilde{\mathcal Y}^{(2)}_1, \widetilde{\mathcal Y}^{(2)}_2)&\le&\frac{1}{86}.\nonumber 
  \end{eqnarray}
 Hence, the optimal coding gain is $\frac{1}{(\sqrt{15}+\sqrt{17})^2}, \widetilde{\delta}=0, \widetilde{p}=\widetilde{q}=5$, and $\widetilde{\mathcal X}$ and $\widetilde{\mathcal Y}$ are the cross 32-QAM constellation.   
 \item $r$ is an even integer exceeding 4 and not equal to 10. In this case, we consider the following two possibilities: 
 \begin{enumerate}
 \item If $r=6$, then,  Properties~\ref{pro:delta0}, \ref{pro:delta1} and~\ref{pro:delta2} give us  
  \begin{eqnarray}
 G(\widetilde{\mathcal X}^{(0)},\widetilde{\mathcal Y}^{(0)}_1, \widetilde{\mathcal Y}^{(0)}_2)&=&\frac{1}{\big(\sqrt{3}+\sqrt{5}\,\big)^2},\nonumber\\
  G(\widetilde{\mathcal X}^{(1)},\widetilde{\mathcal Y}^{(1)}_1, \widetilde{\mathcal Y}^{(1)}_2)&=&\frac{2}{\big(\sqrt{5}+\sqrt{7}\,\big)^2},\nonumber\\  
  G(\widetilde{\mathcal X}^{(2)},\widetilde{\mathcal Y}^{(2)}_1, \widetilde{\mathcal Y}^{(2)}_2)&\le&\frac{1}{14}.\nonumber 
  \end{eqnarray} 
  Thus, in this case, we have $\widetilde\delta=1, \widetilde{p}=4, \widetilde{q}=3$.
 \item If $r\ne 6$, then, we have from  Properties~\ref{pro:delta0}, \ref{pro:delta1} and~\ref{pro:delta2} that
   \begin{eqnarray}
 G(\widetilde{\mathcal X}^{(0)},\widetilde{\mathcal Y}^{(0)}_1, \widetilde{\mathcal Y}^{(0)}_2)&=&\frac{4}{\Big(\sqrt{\widetilde{E}^{(0)}_1+\widetilde{E}^{(0)}_2}+\sqrt{\widetilde{E}^{(0)}_1+\widetilde{E}^{(0)}_{21}}\,\Big)^2},\nonumber\\
  G(\widetilde{\mathcal X}^{(1)},\widetilde{\mathcal Y}^{(1)}_1, \widetilde{\mathcal Y}^{(1)}_2)&=& \frac{8}{\Big(\sqrt{\widetilde{E}^{(1)}_1+\widetilde{E}^{(1)}_{21}}+\sqrt{\widetilde{E}^{(1)}_1+\widetilde{E}^{(1)}_{22}}\,\Big)^2},\nonumber\\  
  G(\widetilde{\mathcal X}^{(2)},\widetilde{\mathcal Y}^{(2)}_1, \widetilde{\mathcal Y}^{(2)}_2)&\le&\frac{8}{\Big(\sqrt{\widetilde{E}_1^{(2)}+\widetilde{E}_{22}^{(2)}}+\sqrt{\widetilde{E}_2^{(2)}+\widetilde{E}_{12}^{(2)}}\Big)^2}.\nonumber 
  \end{eqnarray} 
  Recall the following facts:
  \begin{enumerate}
  \item $\widetilde{E}_1^{(2)}=\widetilde{E}_2^{(2)}, \widetilde{E}_{12}^{(2)}=\widetilde{E}_{22}^{(2)}$, $\widetilde{E}_1^{(2)}$ is the largest energy among all the points in the $2^{r/2+1}$-ary cross QAM constellation and $\widetilde{E}_{12}^{(2)}$ is the energy of the second neighbor of the point with the largest energy in the constellation.
  \item $\widetilde{E}_1^{(1)}$ is the largest energy among all the points in the $2^{r/2+1}$-ary cross QAM constellation, while $\widetilde{E}_{21}^{(1)}$ and $\widetilde{E}_{22}^{(1)}$  are the respective energies of the first and second neighbors of the point with the largest energy in the $2^{r/2}$-ary cross QAM constellation.  
  \item $\widetilde{E}_1^{(0)}=\widetilde{E}_2^{(0)}$, $\widetilde{E}_1^{(0)}$ is the largest energy among all the points in the $2^{r/2}$-ary cross QAM constellation and $\widetilde{E}_{21}^{(0)}$ is the energy of the first neighbor of the point with the largest energy in the constellation.  
  \end{enumerate}
Hence, we obtain $\widetilde{E}_{12}^{(2)}>\widetilde{E}_{21}^{(1)}>\widetilde{E}_{22}^{(1)}$ and as a result,   $G(\widetilde{\mathcal X}^{(1)},\widetilde{\mathcal Y}^{(1)}_1, \widetilde{\mathcal Y}^{(1)}_2)> G(\widetilde{\mathcal X}^{(2)},\widetilde{\mathcal Y}^{(2)}_1, \widetilde{\mathcal Y}^{(2)}_2)$. In addition, since
\begin{eqnarray}
\frac{ G(\widetilde{\mathcal X}^{(0)},\widetilde{\mathcal Y}^{(0)}_1, \widetilde{\mathcal Y}^{(0)}_2)}{G(\widetilde{\mathcal X}^{(1)},\widetilde{\mathcal Y}^{(1)}_1, \widetilde{\mathcal Y}^{(1)}_2)}=\left(\frac{\sqrt{\widetilde{E}^{(1)}_1+\widetilde{E}^{(1)}_{21}}+\sqrt{\widetilde{E}^{(1)}_1+\widetilde{E}^{(1)}_{22}}}{\sqrt{2\widetilde{E}^{(0)}_1+2\widetilde{E}^{(0)}_2}+\sqrt{2\widetilde{E}^{(0)}_1+2\widetilde{E}^{(0)}_{21}}}\right)^2\le  1,
\end{eqnarray}
where we have used the facts that $\widetilde{E}_{21}^{(0)}=\widetilde{E}_{21}^{(1)}\ge \widetilde{E}_{22}^{(1)}$ and that $2\widetilde{E}^{(0)}_1\ge \widetilde{E}^{(1)}_1$ if $r\ge 6$, we attain $G(\widetilde{\mathcal X}^{(1)},\widetilde{\mathcal Y}^{(1)}_1, \widetilde{\mathcal Y}^{(1)}_2)> G(\widetilde{\mathcal X}^{(0)},\widetilde{\mathcal Y}^{(0)}_1, \widetilde{\mathcal Y}^{(0)}_2)$. Therefore, in this case, the optimal coding gain is $G(\widetilde{\mathcal X},\widetilde{\mathcal Y}_1, \widetilde{\mathcal Y}_2)$ is $G(\widetilde{\mathcal X}^{(1)},\widetilde{\mathcal Y}^{(1)}_1, \widetilde{\mathcal Y}^{(1)}_2)$.
  \end{enumerate}
 \item $r$ is an odd integer exceeding 4. There are three cases which need to be considered: a) $r=5$, b) $r=9$ and c) $r\ne 5, 9$. However, since the discussions on Cases a) and b) are much similar to the previous cases $r=6$ and $r=10$, we only consider Case c).  In this situation, Properties~\ref{pro:delta0}, \ref{pro:delta1} and~\ref{pro:delta2} provide us with 
 \begin{eqnarray}
 G(\widetilde{\mathcal X}^{(0)},\widetilde{\mathcal Y}^{(0)}_1, \widetilde{\mathcal Y}^{(0)}_2)&=&\frac{4}{\Big(\sqrt{\widetilde{E}^{(0)}_1+\widetilde{E}^{(0)}_2}+\sqrt{\widetilde{E}^{(0)}_1+\widetilde{E}^{(0)}_{21}}\,\Big)^2},\nonumber\\
  G(\widetilde{\mathcal X}^{(1)},\widetilde{\mathcal Y}^{(1)}_1, \widetilde{\mathcal Y}^{(1)}_2)&=& \frac{8}{\Big(\sqrt{\widetilde{E}^{(1)}_1+\widetilde{E}^{(1)}_{21}}+\sqrt{\widetilde{E}^{(1)}_{11}+\widetilde{E}^{(1)}_2}\,\Big)^2},\nonumber\\  
G(\widetilde{\mathcal X}^{(2)},\widetilde{\mathcal Y}^{(2)}_1, \widetilde{\mathcal Y}^{(2)}_2)&\le&\frac{8}{\Big(\sqrt{\widetilde{E}_{11}^{(2)}+\widetilde{E}_2^{(2)}}+\sqrt{\widetilde{E}_{12}^{(2)}+\widetilde{E}_2^{(2)}}\Big)^2}.\nonumber
    \end{eqnarray} 
 Notice the following facts:
  \begin{enumerate}
  \item $\widetilde{E}_2^{(2)}$ is the largest energy among all the points in the $2^{(r+1)/2}$-ary cross QAM constellation, $\widetilde{E}_{11}^{(2)}$ and $\widetilde{E}_{12}^{(2)}$ are the energies of the first and second neighbors of the points with the largest energies in the $2^{(r+3)/2}$-ary cross QAM constellations.
   \item $\widetilde{E}_1^{(1)}=\widetilde{E}_1^{(2)}, \widetilde{E}_{11}^{(1)}=\widetilde{E}_{21}^{(1)}$,  $\widetilde{E}_1^{(1)}$ is the largest energy among all the points in the $2^{(r+1)/2}$-ary cross QAM constellation and $\widetilde{E}_{11}^{(1)}$ is the energy of the first neighbor of the point with the largest energy in the constellation.
  \item $\widetilde{E}_1^{(0)}$ is the largest energy among all the points in the $2^{(r+1)/2}$-ary cross QAM constellation, $\widetilde{E}_2^{(0)}$ is the largest energy among all the points in the $2^{(r-1)/2}$-ary cross QAM constellation  and $\widetilde{E}_{21}^{(0)}$ is the energy of the first neighbor of the point with the largest energy in the $2^{(r-1)/2}$-ary cross QAM constellation.
  \end{enumerate}
Now, following the way similar to the previous case where $r\ge 8$ is even but not equal to~10, we can prove that $G(\widetilde{\mathcal X}^{(1)},\widetilde{\mathcal Y}^{(1)}_1, \widetilde{\mathcal Y}^{(1)}_2)$ is still the optimal coding gain.
\end{enumerate} 
This completes the proof of Theorem~\ref{th:solu}.~\hfill$\Box$

\bibliographystyle{ieeetr}
\bibliography{tzzt}

\end{document}